\documentclass[11pt,a4paper]{article}
\usepackage{amssymb,bbm}

\font\got=eufm10 at 12pt
\font\sgot=eufm10 at 9pt

\newtheorem{theorem}{Theorem}[section] 
\newtheorem{lemma}[theorem]{Lemma}
\newtheorem{proposition}[theorem]{Proposition}
\newtheorem{corollary}[theorem]{Corollary}
\newtheorem{definition}[theorem]{Definition}
\newtheorem{remark}[theorem]{Remark}
\newtheorem{example}[theorem]{Example}
\newtheorem{convention}[theorem]{Convention}

\def\1{{\mathbbm{1}}}
\def\A{{\mathcal A}}
\def\Ad{{\mathop{\rm Ad}}}

\def\be{\begin{equation}}
\def\build#1_#2^#3{\mathrel{\mathop{\kern 0pt#1}\limits_{#2}^{#3}}}
\def\C{{\mathcal {C}}}

\def\dom{{\mathop{\rm Dom}}}
\def\E{{\mathbb{E}}}

\def\ee{\end{equation}}

\def\epsilon{\varepsilon}
\def\F{{\mathbb{F}}}
\def\f{{\mathcal F}}
\def\G{{\mathbb{G}}}
\def\GE{G^{\E^+}}

\def\LG{{\hbox{\got g}}}
\def\lg{{\hbox{\sgot g}}}
\def\J{{\mathcal{J}}}
\def\lra{\longrightarrow}
\def\M{{\mathcal M}}

\def\oo{{\hbox{\got{o}}}}
\def\ov{\overline}

\def\pf{\noindent \textbf{Proof -- }}
\def\ppf#1{\noindent \textbf{Proof of Proposition \ref{#1} -- }}

\def\qed{\hfill\hbox{\vrule\vbox to 2mm{\hrule width 2mm\vfill\hrule}\vrule}
  \\}
\def\R{{\mathbb{R}}}

\def\un{\underline}
\def\V{{\mathbb{V}}}

\def\wG{{\widetilde{G}}}
\def\wGE{{\widetilde G^{\E^+}}}

\def\YM{{\rm\scriptscriptstyle YM}}

\def\z{{\hbox{\got z}}}

\title{Large deviations for the Yang-Mills measure on a compact surface}

\author{Thierry L\'evy \\ CNRS, D\'epartement de Math\'ematiques et
Applications \\ \'Ecole Normale Sup\'erieure \\ 45, rue d'Ulm, F-75230
Paris Cedex 05 \\ levy@dma.ens.fr \and James R. Norris \\ Statistical
Laboratory \\ Center for Mathematical Sciences \\ Wilberforce Road,
Cambridge, CB3 0WB, UK \\ j.r.norris@statslab.cam.ac.uk }

\begin{document}
\maketitle

\begin{abstract}
We prove the first mathematical result relating the Yang-Mills measure on a compact surface and the
Yang-Mills energy. We show that, at the small volume limit, the scaled Yang-Mills measures satisfy a
large deviation principle with a rate function which is expressed in a simple and natural way in
terms of the Yang-Mills energy. 
\end{abstract}

\section*{Introduction}

The Yang-Mills measure is the distribution of a stochastic process indexed by paths on
a smooth manifold and with values in a Lie group. This process is at a heuristical level
the random holonomy process of a random connection on a principal bundle over the manifold,
the connection being distributed according to the Gibbs measure of the Yang-Mills energy. This is
usually expressed by the following inspiring but meaningless formula, where $S$ is the Yang-Mills
energy, $T$ a positive real parameter, $Z_{T}$ a normalization constant and $P_{T}$ the
Yang-Mills measure:
$$ dP_{T}(\omega) = \frac{1}{Z_{T}} e^{-\frac{1}{2T} S(\omega)} \; d\omega.$$

In the case of a compact two-dimensional base manifold and compact structure group, two different
constructions of the Yang-Mills measure are known. The first one proceeds by an
infinite-dimensional approach \cite{Driver,Sengupta_AMS} and the second one by a finite-dimensional
approach \cite{Levy_AMS}. Neither of these constructions involve the Yang-Mills energy more
than at an informal level, as a guide for the intuition. Both begin from the specifications
given by physicists of certain characterizing properties of the distribution of the stochastic
process, namely its finite-dimensional distributions, which physicists have of course derived from
the Yang-Mills energy, but by some non-rigorous arguments. These finite-dimensional distributions
seem to have been first described by A. Migdal in \cite{Migdal} and they are also discussed in
\cite{Witten}. The papers \cite{Migdal} and \cite{Witten} have thus so far played the role of a
touchstone for the constructions of the Yang-Mills measure. Nevertheless, a rigorous link between
the measure and the energy was still lacking.

That the constructions of the Yang-Mills measure do not incorporate explicitly the Yang-Mills energy
is not more surprising for example than the fact that one can and usually does construct Brownian
motion without referring to the Sobolev $H^1$ norm. However, essential links between the
Wiener measure and the  $H^1$ norm are attested for instance by Cameron-Martin's
quasi-invariance theorem and Schilder's large deviation principle. 

The point of this paper is to relate at a mathematical level the Yang-Mills measure in two
dimensions and the Yang-Mills energy by a large deviation principle. We consider a
base space which is a compact surface $M$ with or without boundary and a structure group which is
any compact connected Lie group $G$. We choose a principal $G$-bundle $P$ over $M$ and consider the
space $H^1\A(P)$ of $H^1$ connections on $P$, which is the most natural space of connections with
finite Yang-Mills energy. If $M$ has a boundary, we consider only those connections which satisfy
certain boundary conditions. Then, up to gauge transformations, we embed the space $H^1\A(P)$ of $H^1$
connections into the canonical space of the random holonomy process and define a
natural non-negative functional $I^\YM$ on this canonical space by extending the Yang-Mills energy by
$+\infty$ outside the range of the embedding. The Yang-Mills measure, denoted by $P_{T}$, depends on
a positive parameter $T$ which is the total area of $M$. The main result of this paper says that, as
$T$ tends to $0$ and for every measurable subset $A$ of the canonical space of the process, one has 
$$ -\inf_{x\in A^\circ} I^\YM(x) \leq \liminf_{T\to 0} T \log P_{T}(A) \leq \limsup_{T\to 0} T
\log P_{T}(A)
\leq -\inf_{x\in \overline{A}} I^\YM(x),$$
where $A^\circ$ and $\ov{A}$ denote respectively the interior and the closure of $A$ with respect
to the product topology on the canonical space. 

The paper is divided into four sections. In the first one, we give a
precise statement of the two main results, one for the case where $M$ has a boundary and one for
the case where it is closed. For this, we recall how the Yang-Mills measure is
constructed in both cases, what Sobolev connections are and then explain how the rate functions
for the large deviation principles are defined. 

In the second section, we collect the technical properties of $H^1$ connections which we will need at
various stages of the proof. In particular, we study the holonomy that they determine, the way they
are transformed by gauge transformations, and check that they satisfy a classical energy
inequality.

In the third section, we prove the large deviation principles. The starting point is the
classical short-time estimate of the heat kernel on a compact Riemannian manifold, that we apply
to the group $G$. Then, we build a large deviation principle for the random holonomy process by
following step by step its construction described in $\cite{Levy_AMS,Levy_PTRF}$. 

However, proving rather abstractly the existence of a large deviation principle is not enough and we
must, at each step, identify the rate function in terms of the Yang-Mills energy. Apart from standard
results from the theory of large deviations, this relies mainly on three technical tools. The first
one is the energy inequality mentioned above and of which we give a complete proof in the appendix.
The second one is a beautiful compactness theorem of
K. Uhlenbeck that we recall at the end of the second section. The third one is, as far as we know,
original, and asserts the existence of a connection of minimal energy with prescribed holonomy along
the edges of a graph on $M$.  The proof that such a minimizer exists occupies the fourth and last
section of the paper. 

\section{The large deviation principles}
\subsection{The Yang-Mills measures}
\subsubsection{The space of paths} 

Let $M$ be an oriented compact connected surface, possibly with boundary. Let $\sigma$ be a volume 2-form on $M$
consistent with the orientation. We will often identify $\sigma$ with its density which is a Borel measure on $M$. For
technical purposes, let us assume that $M$ is endowed with a Riemannian metric whose Riemannian
volume is the density of $\sigma$. 

By an {\it edge} on $M$ we mean a segment of a smooth oriented 1-dimensional submanifold. If $e$ is
an edge, we call {\it inverse} of $e$ and denote by $e^{-1}$ the edge obtained by reversing the
orientation of $e$. We also denote respectively by $\underline e$ and $\overline e$ the starting and
finishing point of $e$. Let $e_1,\ldots,e_n$ be $n$ edges. If, for each $i$ between $1$ and $n-1$,
one has $\overline{e_i}=\underline{e_{i+1}}$, then one can form the concatenation $e_1\ldots e_n$. If
moreover $f_1,\ldots,f_m$ are also edges which can be concatenated, we declare $e_1\ldots e_n$
equivalent to $f_1\ldots f_m$ if and only if there exists a continuous mapping $c:[0,1] \lra M$ and
two finite sequences $0=t_0<t_1<\ldots<t_n=1$ and $0=s_0<s_1<\ldots < s_m=1$ of real numbers such
that, for each $i=1\ldots n$, the restriction of $c$ to the interval $[t_{i-1},t_i]$ is a smooth
embedding of image $e_i$ and, for each $j=1\ldots m$, the restriction of $c$ to the interval $[t_{j-1},t_j]$ is a smooth
embedding of image $f_j$. By a {\it path} we mean an equivalence class of finite concatenations of
edges. We denote the set of paths by $PM$. Loops, starting and finishing points of
paths, their concatenation, are defined in the obvious way. We say that a loop $e_{1}\ldots e_{n}$ is
{\it simple} if the vertices $\un{e_{1}},\ldots,\un{e_{n}}$ are pairwise distinct. If $c$ is a
path, we denote respectively by $\underline{c}$ and $\overline{c}$ its starting and finishing point.

Let $l_1$ and $l_2$ be two loops. We say that $l_1$ and $l_2$ are cyclically equivalent if there
exist two paths $c$ and $d$ in $PM$ such that $l_1 = cd$ and $l_2=dc$. We call {\it cycle} an
equivalence class of loops for this relation. Informally, a cycle is a loop on which one has
forgotten the starting point. We say that a cycle is {\it simple} if its representatives are simple
loops.

For any two paths $c$ and $c'$, denote by $\ell(c)$ and $\ell(c')$ their respective lengths and
set $d_{\infty}(c,c')=\inf \sup_{t\in[0,1]} d(c(t),c'(t))$, where $d$ is the Riemannian distance
on $M$ and the infimum is taken over all continuous parametrizations of $c$ and $c'$ by the interval
$[0,1]$. The function $(c,c')\mapsto d_{\ell}(c,c')=d_{\infty}(c,c')+|\ell(c)-\ell(c')|$
is a distance on $PM$ (see \cite{Levy_AMS}, Lemmas 2.22, 2.23 and Remark 2.24) which depends on the
Riemannian metric on $M$. However, since $M$ is compact, the topology induced by $d_{\ell}$ on $PM$
is independent of the metric. In this paper, we simply say that a sequence of paths {\it
converges} to indicate that it converges in the topology induced by $d_{\ell}$. We will sometimes use
a stronger  notion of convergence, saying that a sequence $(c_n)_{n\geq 0}$
{\it converges} to $c$ {\it with fixed endpoints} if $c_n$ converges to $c$ and $c_n$ and $c$
share the same starting points and the same finishing points. 

\subsubsection{The measurable space}

Let $G$ be a connected compact Lie group endowed with a bi-invariant metric $\gamma$ of total
volume 1.

Consider a subset $J\subset PM$. We say that a function $f:J\lra G$ is {\it multiplicative} if the
following properties hold. Firstly, for all $c_{1}, c_{2}$ belonging to $J$ such that
$\overline{c_{1}}=\underline{c_{2}}$ and $c_{1}c_{2}\in J$, one has $f(c_{1}c_{2}) =
f(c_{2})f(c_{1})$. Secondly, if both $c$ and $c^{-1}$ belong to $J$, then $f(c^{-1})=f(c)^{-1}$. The
set of multiplicative functions from $J$ to $G$ is denoted by $\M(J,G)$.

For each path $c\in PM$, denote by $H_c : \M(PM,G) \lra G$ the evaluation at $c$ defined by
$H_c(f)=f(c)$. Let $\C$ be the cylinder $\sigma$-field on $\M(PM,G)$, that is, the smallest $\sigma$-field with respect
to which all the mappings $H_c$, $c\in PM$ are measurable. The Yang-Mills measures\footnote{There
is indeed a whole family of Yang-Mills measures, indexed by the fundamental group of $G$ if $M$ is
closed, or by one conjugacy class of $G$ for each connected component of the boundary of $M$ if it
is not empty, and in all cases by a positive real parameter which we call temperature and which is
really a scaling factor for the volume form $\sigma$ on $M$.} are probability measures on the
measurable space $(\M(PM,G),\C)$. 

Consider again a subset $J\subset PM$. Let $U$ be a subset of $M$ such that the endpoints of every
path of $J$ belong to $U$. The group $\f(U,G)$ of all $G$-valued functions on $U$ acts on $\M(J,G)$
according to the following rule: 
$$
(j\cdot f)(c)=j(\overline{c})^{-1} f(c) j(\underline{c}),
$$
which is inspired by the way a gauge transformation affects the holonomy of a connection. In
particular, the group $\f(M,G)$ acts on the measurable space $(\M(PM,G),\C)$. This action is
measurable and all the probability measures which we shall consider are invariant under this
action. This fact has important implications which we shall discuss later. 

\subsubsection{Graphs}

In order to characterize the different instances of the Yang-Mills measure that we are considering
here, we need to describe some of their finite dimensional marginals. This involves putting graphs on
$M$ and associating to each of them a probability measure on a finite product of copies of $G$. More
details can be found in \cite{Levy_AMS}, Sections 1.1 and 1.4, and in \cite{Levy_PTRF}, from which
what follows is inspired.

\begin{definition}
A {\sl graph} is a triple $\G=(\V,\E,\F)$, where

1. $\E$ is a finite collection of edges stable by inversion and such  that two distinct edges are
either inverse of each other or intersect, if at all, only at some of their endpoints. Moreover,
if two edges share the same starting point, then their angle at this point is different from 0 modulo
$2\pi$.

2. $\V$ is the set of endpoints of the elements of $\E$.

3. $\F$ is the set of the closures of the connected components of $M \backslash \bigcup_{e\in\E}
e$.

4. Each open connected component of $M \backslash \bigcup_{e\in\E} e$  is diffeomorphic to the
open unit disk of $\R^2$.

5. The boundary of $M$ is contained in $\bigcup_{e\in\E} e$.

\end{definition}

The elements of $\V,\E,\F$ are respectively called vertices, edges and faces of $\G$. We call {\it
open faces} the connected components of $M \backslash \bigcup_{e\in\E} e$. Beware that an open face
could be strictly contained in the interior of its closure. In fact, we make a further assumption
that makes this impossible.

We denote by $\E^*$ the set of paths that can be represented by a concatenation of elements of
$\E$. For example, each face of a graph has a boundary which is a cycle in $\E^*$. We say that a
graph is {\it simple} if the boundary of each one of its faces is a simple cycle. In this paper,
we shall always assume that the graphs are simple.

We will use the mapping $L:\E\lra \F \cup \{\varnothing \}$ defined by the fact that, for each edge
$e$, $L(e)$ is the unique face of $\G$ which $e$ bounds with positive orientation, in other words the
face located on the left of $e$. If $e\subset \partial M$ and $M$ is on the right of $e$, we set
$L(e)=\varnothing$.

Finally, let an {\it unoriented edge} of $\G$ be a pair $\{e,e^{-1}\}$ where $e \in \E$. We
call {\it orientation} of $\G$ a subset $\E^+$ of $\E$ which contains exactly one element of each unoriented edge. 

\begin{lemma}\label{lem:orientation} 
Let $\G$ be a graph on $M$. There exists an orientation $\E^+$ of $\G$ such
that, depending on whether $M$ has a boundary or not, one of the two following properties holds:

1. If $M$ is closed, then for each face $F$, there exists $e\in\E^+$ such that $F=L(e^{-1})$.

2. If $M$ has a boundary, then for each $e\in\E^+$ such that $e\subset \partial M$, one has $L(e)\neq
\varnothing$.
\end{lemma}

\pf In the case with boundary, the assertion is obvious. In the closed case, it is proved in
\cite{Levy_PTRF} by using a spanning tree in the dual graph to the fat graph induced by $\G$. \qed

We will always assume that the graphs that we consider are oriented in a way which satisfies the
relevant one of these two properties.

\subsubsection{The discrete measures}

Choose once for all in this section a simple graph $\G$ and a positive real number $T$. Choose an
orientation $\E^+=\{e_{1},\ldots,e_{r}\}$ of $\G$. We want to describe the discrete Yang-Mills measure
at temperature $T$ associated to $\G$. If $M$ is closed, it is a Borel probability measure on $\GE$ 
absolutely continuous with respect to the Haar measure. If $M$ has a boundary, it is supported by a
subset of $\GE$ which depends on some boundary conditions one has to specify.

Choose $c\in\E^*$. It can be written $c=e_{i_{1}}^{\epsilon_{1}}\ldots e_{i_{n}}^{\epsilon_{n}}$
for some $i_{1},\ldots,i_{n}\in{1,\ldots,r}$ and $\epsilon_{1},\ldots,\epsilon_{n}=\pm 1$. Define
$h_{c}:\GE \lra G$, the discrete holonomy along $c$, by setting
$h_{c}(g_{1},\ldots,g_{r})=g_{i_{n}}^{\epsilon_{n}}\ldots
g_{i_{1}}^{\epsilon_{1}}$. If $F$ is a face of $\G$, the mapping $h_{\partial F}$ is ill-defined
because the cycle $\partial F$ lacks a base point. However, let $\Ad$ denote the action of $G$ on
itself by conjugation: $\Ad(x)y=xyx^{-1}$. Then, if $c$ is a loop which represents $\partial F$,
the composition of $h_{c}$ with the canonical projection $G\lra G/\Ad$ does not depend on the choice
of $c$. We denote this composed mapping by $h_{\partial F}:\GE \lra G/\Ad$. 

Finally, let $p$ be the fundamental solution of the heat equation on $G$, that is, the unique
smooth function  $p : (0,\infty) \times G \lra (0,\infty)$, $(t,g) \mapsto p_t(g)$, which is a
solution of the equation $\left(\frac{1}{2}\Delta - \partial_t \right)p=0$ and satisfies, for every
continuous function $f$ on $G$,  the initial condition $\int_G f(g) p_t(g) \; dg \build{\lra}_{t\to
0}^{} f(1)$, where $1$ is the unit of $G$. For all $t>0$, the function $p_{t}$ is invariant by
conjugation on $G$, so that if $F$ is a face of $\G$, the function $p_{t} \circ h_{\partial F}
:\GE \lra (0,+\infty)$ is well defined.

We can now define the discrete Yang-Mills measure under the assumptions
that $M$ is closed and that $G$ is simply connected.

\begin{definition}[Closed surface, simply connected group] \label{def:closed-sc}
Assume that $G$ is simply connected. The discrete Yang-Mills measure associated to $\G$ at
temperature $T$ is the Borel probability measure $P^\G_{T}$ on $\GE$ defined by 
$$
dP^\G_{T}(g)=\frac{1}{Z^\G_{T}} \prod_{F\in\F} p_{T\sigma(F)}(h_{\partial F}(g)) \; dg,
$$
where $dg$ is the Haar measure on $\GE$.
\end{definition} 

Let us drop the assumption that $G$ is simply connected. In order to define the measure, we need a
few more definitions. 

Let  $\pi:\wG \lra G$  be a universal covering of $G$ and set $\Pi=\pi^{-1}(1)$. Recall that $\Pi$
is a discrete central subgroup of $\wG$ canonically isomorphic to the fundamental group of $G$.  If
$M$ is closed, then principal $G$-bundles over $M$ are classified up to bundle isomorphism \footnote{This classical fact is proved in \cite{Steenrod}, \S\S29--34. In the case where
$G=S^1$, a more accessible proof can be found in \cite{Morita}, Section 6.2. Appendix A of
\cite{Lawson_Michelsohn} is also worth reading. Finally, a good general reference on principal
bundles and smooth connections is \cite{Kobayashi_Nomizu_I}.} by $\Pi$. On the other hand, if $M$ has
a non-empty boundary, then all principal $G$-bundles over $M$ are trivial. Let us assume for the
moment that $M$ is closed. If $P$ is a principal $G$-bundle over $M$, we denote by $\oo(P)$ the
element of $\Pi$ which represents the isomorphism class of $P$.

If $c$ is a path in $\E^*$ and if $F\in \F$ is a face, then the definitions of the mappings
$h_{c}:\GE\lra G$ and $h_{\partial F} : \GE \lra G/\Ad$ still make sense when $G$ is replaced by
$\wG$. When we use the new mappings thus defined, we put a superscript to their names to indicate in
which group the mappings take their values, writing for example $h_{\partial F}^{\wG}$.

Recall that $G$ is endowed with a Riemannian metric $\gamma$. Let us endow $\wG$ with the Riemannian
metric $\pi^*\gamma$ and the corresponding Riemannian volume\footnote{This metric and this invariant
measure are not normalized: the total volume of $\wG$ is equal to the cardinality of $\Pi$, which is
finite if and only if $G$ is semi-simple.}.  Let $\tilde p$ be the fundamental solution of the heat
equation on $\wG$. It is a smooth positive function on $(0,+\infty)\times\wG$, and for each $t>0$,
$\tilde p_{t}$ is invariant by conjugation. The functions $p$ and $\tilde p$  are related by the
equality $p_{t}(x)=\sum_{\pi(\tilde x)=x} \tilde p_{t}(\tilde x)$ which holds for all $t>0$ and $x\in
G$.

For each $z\in \Pi$, define a subset $\displaystyle  \Pi^\F_{z}$ of $\wG^\F$ by $\displaystyle 
\Pi^\F_{z} =\{z_{\F}=(z_{F})_{F\in\F} \in \Pi^\F | \prod_{F\in\F} z_{F}=z \}.$ A proof of the
following result can be found in \cite{Levy_PTRF}, Proposition 2.4 and Lemma 2.7.

\begin{proposition} \label{prop:D}
Choose $z\in\Pi$. Let $g$ be an element of $\GE$. Let $\tilde g \in
\wG^{\E^+}$ be a lift of $g$. Then the number
$$
\sum_{z_{\F}\in \Pi^\F_{z}} \prod_{F\in\F} \tilde p_{T\sigma(F)}(h^{\wG}_{\partial F}(\tilde
g) z_{F})
$$
is finite and does not depend on the choice of $\tilde g$. We denote it by $D^\G_{T,z}(g)$.
Moreover, the function $D^\G_{T,z}$ is bounded on $\GE$.
\end{proposition}

This proposition provides us with a positive function on $\GE$ which is the density of the
discrete Yang-Mills measure. 

\begin{definition}[Closed surface, the general case] \label{def:closed}
Choose $T>0$ and $z\in \Pi$. The discrete
Yang-Mills measure at temperature $T$ associated to $\G$ and to the isomorphism class of $G$-bundles
corresponding to $z$ is the Borel probability measure $P^\G_{T,z}$ on $\GE$ defined by
$$
dP^\G_{T,z} (g) =\frac{1}{Z^\G_{T,z}} D^\G_{T,z}(g) \; dg.
$$
\end{definition}

Finally, the case where $M$ has a boundary requires also a few specific definitions. Whether or
not $G$ is simply connected does not matter any more since any $G$-bundle over $M$ is trivial.
On the other hand, it is natural to put constraints on the measure and to insist that the discrete
holonomy along each boundary component belong to some fixed conjugacy class in $G$. 

Let $X$ be a conjugacy class in $G$. For each $n\geq 1$, we are interested in the subset of
$G^n$ defined by the equation $g_{n}\ldots g_{1}\in X$. This subset is an orbit of the action
of $G^n$ on itself defined by $(x_{1},\ldots,x_{n})\cdot(g_{1},\ldots,g_{n})=
(x_{1}g_{1}x_{n}^{-1},x_{2}g_{2}x_{1}^{-1}, \ldots,x_{n}g_{n}x_{n-1}^{-1})$. As such, it
carries a natural measure which is the image of the Haar measure on $G^n$. This measure, which we
denote by $\nu^n_{X}$ or simply $\nu_{X}$, can alternatively be characterized by the fact that, for
every continuous function $f$ on $G^n$, 
$$\nu_{X}^n(f)=\lim_{t\to 0}\int_{G^{n+1}} f(g_{1},\ldots,g_{n}) 
p_{t}(g_{n}\ldots g_{1}yx^{-1}y^{-1}) \; dy dg_{1}\ldots dg_{n}.$$ 
The main properties of $\nu^n_{X}$ are the fact that the relation $g_{n}\ldots g_{1}\in X$
holds $\nu^n_{X}$-almost surely and its invariance under circular permutation of the factors in
$G^n$. 

Assume now that $\partial M$ has $p$ connected components $N_{1},\ldots,N_{p}$. For each
$i=1,\ldots,p$, set $\E^+_{N_{i}}=\{e\in\E^+ | e\subset N_{i}\}$. Set also $\E^+_{int}=\{e \in \E^+ |
e \not\subset \partial M \}$, so that $\E^+$ is the disjoint union of the subsets we have just
defined. 

Let $N$ be a component of $\partial M$. It is also the image of a cycle in $\E^*$. Let
$e_{i_{1}}\ldots e_{i_{n}}$ be a loop representing this cycle. Then, if $X$ is a conjugacy class in
$G$, the measure $\nu_{X}^n(g_{i_{1}},\ldots,g_{i_{n}})$ is well defined on $G^{\E^+_{N}}$ and
does not depend on the choice of the loop representing $\partial N$. We denote it by $\nu_{X}^N$.

Now if we choose $p$ conjugacy classes $X_{1},\ldots,X_{p}$ in $G$, then we can define the measure
$\nu_{X_{1}}^{N_{1}}\otimes \ldots \otimes \nu_{X_{p}}^{N_{p}}$ on $G^{\E^+_{N_{1}}} \times \ldots
\times G^{\E^+_{N_{p}}}$. Finally, let us denote by $dg_{int}$ the Haar measure on the remaining factors,
namely on $G^{\E^+_{int}}$.

\begin{definition}[Surface with boundary] \label{def:boundary}
Choose $p$ conjugacy classes $X_{1},\ldots,X_{p}$ in $G$. 
The discrete Yang-Mills measure at temperature $T$ associated to $\G$ with boundary conditions
$X_{1},\ldots,X_{p}$ along $N_{1},\ldots,N_{p}$ respectively is the Borel probability measure
$P^\G_{T;X_{1},\ldots,X_{p}}$ on $\GE$ defined by
$$
dP^\G_{T;X_{1},\ldots,X_{p}}(g) = \frac{1}{Z^\G_{T;X_{1},\ldots,X_{p}}} \prod_{F\in\F}
p_{T\sigma(F)}(h_{\partial F}(g)) \; d\nu_{X_{1}}^{N_{1}}\ldots d\nu_{X_{p}}^{N_{p}}dg_{int}.
$$
\end{definition}

Whether or not $M$ has a boundary, $\GE = \M(\E^+,G)$, so that the group $\f(M,G)$ acts on $\GE$. The
proof of the following proposition can be found in \cite{Levy_AMS} and \cite{Levy_PTRF}. 

\begin{proposition} \label{prop:Z}
1. The probability measures of Definitions \ref{def:closed-sc},
\ref{def:closed} and \ref{def:boundary} are all invariant under the action of the group $\f(M,G)$.

2. When $M$ is closed and $G$ is simply connected, then the measures $P^\G_{T}$ and $P^\G_{T,1}$
are identical for all $T>0$.

3. None of the normalization constants which appear in the definitions above depend on $\G$. They
depend only on the genus of $M$, denoted here by $g$, its total area $\sigma(M)$ and, when $M$ has
a boundary, on $X_{1},\ldots,X_{p}$. Their values are listed below. We use the notation
$[a,b]=aba^{-1}b^{-1}$ for $a,b \in G$. Recall also that, when $a,b\in G$ and $\tilde a,\tilde b
\in \wG$ satisfy $\pi(\tilde a)=a$ and $\pi(\tilde b)=b$, then $[\tilde a,\tilde b]$ depends only
on $a$ and $b$. We denote it by $[\widetilde{a,b}]$. Finally, we choose $x_{1}\in
X_{1},\ldots,x_{p}\in X_{p}$. 
\begin{eqnarray*}
Z_{T}&=&\int_{G^{2g}} p_{T\sigma(M)}([a_{1},b_{1}]\ldots [a_{g},b_{g}]) \; da_{1}db_{1}\ldots
da_{g}db_{g}. \\
Z_{T,z}&=&\int_{G^{2g}} \tilde p_{T\sigma(M)}([\widetilde{a_{1},b_{1}}]\ldots
[\widetilde{a_{g},b_{g}}]z) \; da_{1}db_{1}\ldots da_{g}db_{g}. \\
Z_{T;X_{1},\ldots,X_{p}}&=&\int_{G^{2g+p}} p_{T\sigma(M)}([a_{1},b_{1}]\ldots
[a_{g},b_{g}]c_{1}x_{1}c_{1}^{-1}\ldots c_{p}x_{p}c_{p}^{-1})\; da_{1}db_{1}\ldots
da_{g}db_{g} \; dc_{1}\ldots dc_{p}.
\end{eqnarray*}
\end{proposition}

\subsubsection{The Yang-Mills measures}

The discrete measures we have just defined provide us with many, but not all, finite dimensional
marginals of the Yang-Mills measure. For example, they do not allow us to write down the joint
distribution of $H_{c}$ and $H_{c'}$ if $c$ and $c'$ are two paths which intersect so often that
$M\backslash (c \cup c')$ has infinitely many connected components. A continuity requirement
fills this gap. 

We say that a sequence of $G$-valued random variables $(H_{n})_{n\geq 0}$ defined under a
probability $P$ converges in probability to $H$ if, for all $\epsilon>0$, $P(d(H_{n},H)>\epsilon)
\build\lra_{n\to\infty}^{} 0$, where $d$ denotes the Riemannian distance on $G$.
We can now state the theorem of existence and uniqueness of the Yang-Mills measure
(\cite{Levy_AMS}, Theorem 2.58). 
 
\begin{theorem}[Closed surface] Choose $T>0$ and $z\in\Pi$. There exists a unique probability measure
$P_{T,z}$ on $(\M(PM,G),\C)$ such that the two following properties hold:

1. For every graph $\G=(\V,\E,\F)$ on $M$, with orientation  $\E^+=\{e_1,\ldots,e_r\}$,
the law of $(H_{e_1},\ldots,H_{e_r})$ under $P_{T,z}$ is equal to $P^\G_{T,z}$.

2. Whenever $c$ belongs to $PM$ and $(c_n)_{n\geq 0}$ is a sequence of  $PM$ converging to $c$ 
with fixed endpoints, $(H_{c_{n}})_{n\geq 0}$ converges in probability to $H_{c}$.

Moreover, the measure $P_{T,z}$ is invariant under the action of $\f(M,G)$.
\end{theorem}

\begin{theorem}[Surface with boundary] Choose $T>0$. Let $X_{1},\ldots,X_{p}$ be $p$ conjugacy
classes in $G$. There exists a unique probability measure $P_{T;X_{1},\ldots,X_{p}}$ on
$(\M(PM,G),\C)$ such that the two following properties hold:

1. For every graph $\G=(\V,\E,\F)$ on $M$, with orientation  $\E^+=\{e_1,\ldots,e_r\}$,
the law of $(H_{e_1},\ldots,H_{e_r})$ under $P_{T;X_{1},\ldots,X_{p}}$ is equal to
$P^\G_{T;X_{1},\ldots,X_{p}}$.

2. Whenever $c$ belongs to $PM$ and $(c_n)_{n\geq 0}$ is a sequence of  $PM$ converging to $c$ 
with fixed endpoints, $(H_{c_{n}})_{n\geq 0}$ converges in probability to $H_{c}$.

Moreover, the measure $P_{T;X_{1},\ldots,X_{p}}$ is invariant under the action of $\f(M,G)$.
\end{theorem}

\subsection{Sobolev connections and the Yang-Mills energy}

In this paper, we shall often be dealing with connections on principal $G$-bundles which are not
smooth nor even continuous, but belong to some larger Sobolev spaces. We spend this paragraph
introducing them carefully. Let us fix a principal $G$-bundle $P$.

We call {\it local section} of $P$ a pair $(U,s)$, where $U$ is an open subset of $M$ and $s$ is a
smooth section of $P$ over $U$. We often denote such a pair simply by $s$ and use the notation
$\dom(s)=U$ when the domain has to be made explicit.

Let $\LG$ denote the Lie algebra of $G$. The Riemannian metric $\gamma$ on $G$ determines a scalar
product on $\LG$ invariant by the adjoint action of $G$. Let $\| \cdot \|$ denote the corresponding
Euclidean norm on $\LG$. 

\begin{definition} \label{def:Sobolev connections}
Choose $k\geq 1$ and $p\in[1,+\infty]$. A $W^{k,p}$ connection on $P$ is the
data, for each local section $s$ of $P$, of a $W^{k,p}$ $\LG$-valued 1-form $\omega_{s}$ on
$\dom(s)$. These locally defined forms must satisfy the following compatibility condition. Let $s$
and $s'$ be any two local sections of $P$. Let $\psi$ be the unique smooth $G$-valued function on
$\dom(s)\cap \dom(s')$ such that, on this domain, $s'=s\psi$. Then $\omega_{s'}=\Ad(\psi^{-1})
\omega_{s} + \psi^{-1} d\psi$. 
\end{definition}

\begin{remark} \label{rmk:def connection covering} A $W^{k,p}$ connection on $P$ is completely
specified by the data of the locally defined $1$-forms associated to a family of local sections of
$P$ whose domains cover $M$.
\end{remark}

We denote the set of $W^{k,p}$ connections on $P$ by $W^{k,p}\A(P)$. We shall in particular
consider $W^{s,2}$ connections, which we also call $H^s$ connections. The
norms on the spaces $W^{k,p}\A(P)$ depend on the Riemannian metric on $M$, but since $M$ is compact,
different metrics give rise to equivalent norms. Hence, the spaces themselves and their topologies
are intrinsically attached to the differentiable structure of $M$.

\begin{example} On the trivial bundle $P=M\times G$, one can choose a smooth global section and
identify $W^{k,p}$ connections with $W^{k,p}$ $\LG$-valued 1-forms on $M$. 
\end{example}

In the present context, the most important space of connections is $H^1\A(P)$. It will turn out that
this space plays to some extent the role of a Cameron-Martin space for the Yang-Mills measure. 

Let $\omega$ be a $H^1$ connection on some bundle $P$. Choose a local section $s$ of $P$ over the
domain of a coordinate chart of $M$, with coordinates $(x,y)$. Then one can write
$\omega_{s}=\omega_{s,1} dx + \omega_{s,2} dy$, where $\omega_{s,1}$ and $\omega_{s,2}$ are $H^1$
$\LG$-valued functions. Since $M$ is compact and two-dimensional, Sobolev embeddings imply that
$\omega_{s,1}$ and $\omega_{s,2}$ are also $L^p$ functions for every finite $p$. In particular,
they are $L^4$ and the formula 
\begin{equation}\label{eqn:def courbure}
\Omega_{s}=(\partial_{x}\omega_{s,2} - \partial_{y}\omega_{s,1} + [\omega_{s,1},\omega_{s,2}]) dx\wedge
dy
\end{equation}
defines locally an $L^2$ $\LG$-valued 2-form\footnote{This argument remains valid if the dimension of
$M$ is 3 or 4.}. As in the smooth
case, if $s'$ is another local section and $\psi$ is the smooth $G$-valued function such that $s'=s\psi$
on $\dom(s)\cap\dom(s')$, then $\Omega_{s'}=\Ad(\psi^{-1})\Omega_{s}$. 

Let $\Ad(P)$ denote the vector bundle associated with $P$ by the adjoint action of $G$ on
$\LG$\footnote{Take $m\in M$. Let $P_{m}$ denote the fibre of $P$ over $m$. Then the fibre of
$\Ad(P)$ over $m$ is the vector space of all mappings $\varphi:P_{m}\lra \LG$ such that, for all
$p\in P_{m}$ and all $g\in G$, $\varphi(pg)=\Ad(g^{-1})\varphi(p)$.  Thus this fibre is isomorphic
to $\LG$, though not canonically. Still, if $\varphi$ belongs to the fibre, then $\| \varphi
(p)\|$ does not depend on $p \in P_{m}$ and can safely be denoted by $\|\varphi\|$. In other
words, the scalar product on $\LG$ induces a metric on $\Ad(P)$.}.
The relation between $\Omega_{s}$ and $\Omega_{s'}$ stated above shows that the locally defined 2-forms
$\Omega_{s}$ build up into an $L^2$ $\Ad(P)$-valued 2-form on $M$, which is the curvature of
$\omega$ and is denoted by $\Omega$.

Let us consider the unique $L^2$ section $*\Omega$ of $\Ad(P)$ such that $\Omega = *\Omega \sigma$.
Then the Yang-Mills energy of $\omega$ is defined by the formula 
$$ S(\omega)=\int_{M} \|*\Omega\|^2 \; d\sigma. $$
We have just defined a functional $S: H^1\A(P)\lra [0,+\infty)$. If one multiplies the volume form
$\sigma$ by a positive real number $T$, then $\|*\Omega \|^2$ is multiplied by $\frac{1}{T^2}$ and
the energy $S$ is multiplied by $\frac{1}{T}$. Thus, the large deviation principle is concerned with
the asymptotic behaviour of the Yang-Mills measure as the area of $M$ is scaled by a factor which
tends to zero.

\begin{example} Let us assume that $P=M\times G$. Then the curvature of an $H^1$ connection
$\omega$ can be identified with an $L^2$ $\LG$-valued 2-form $\Omega$ on $M$, defined in local 
coordinates by (\ref{eqn:def courbure}). Then $\Omega$ can be written in a unique way as
$\Omega=*\Omega \sigma$, where $*\Omega$ is a square-integrable $\LG$-valued function on $M$. The
energy $S(\omega)$ is then nothing but the square of the $L^2$ norm of this function.   
\end{example}

\subsection{The rate functions}

In order to define the rate functions of the large deviation principles satisfied by the
Yang-Mills measure, we need to understand some properties of the holonomy induced by $H^1$
connections. This is a bit technical and explaining it completely now would distract us from our
goal which is to state the main theorems. Thus, we present here only the main ideas and postpone the
technical aspects until Section \ref{section:sobolev connections}.

Assume that $P$ is a trivial $G$-bundle over $M$. Identify $H^1$ connections on $P$ with $H^1$
$\LG$-valued 1-forms on $M$. Let $\omega$ be such a connection. The crucial property of $\omega$
as far as holonomy is concerned is the fact that it admits a trace along any smooth
$1$-dimensional submanifold of $M$, which is an $L^2$ $\LG$-valued function along this submanifold.
Thus, if a path $c\in PM$ is parametrized by the interval $[0,1]$, it makes sense to solve the
differential equation $\dot a_{t} a_{t}^{-1} = -\omega(\dot c(t)) , \; a_{0}=1,$
where the unknown function is $a:[0,1]\lra G$. If we set $f(c)=a_{1}$ and do this for each element
of $PM$, we get an element $f$ of the space $\M(PM,G)$. This element is called the holonomy of
$\omega$.

If $M$ has a boundary and if $N$ is a component of $\partial M$, then the holonomy of $\omega$
along $N$ is well defined as a conjugacy class in $G$, namely that of $f(c)$ if $c$ is a simple
loop which represents the cycle $N$. If $\partial M$ has $p$ components $N_{1},\ldots,N_{p}$ and
if $X_{1},\ldots,X_{p}$ are $p$ conjugacy classes in $G$, then we denote by
$H^1\A_{X_{1},\ldots,X_{p}}(P)$ the set of $H^1$ connections on $P$ whose holonomy along $N_{i}$
is equal to $X_{i}$ for each $i=1,\ldots,p$. 

When $P$ is not trivial, an $H^1$ connection induces only a class in $\M(PM,G)$ modulo the 
action of $\f(M,G)$. This class still contains a lot of information about $\omega$. For example, if
two connections, not even necessarily defined on the same bundle, induce the same class in
$\M(PM,G)$, then they have the same energy. 

Let us say that an $H^1$ connection $\omega$ on some $G$-bundle over $M$ and an element $f$ 
of $\M(PM,G)$ {\it agree} if they determine the same class, that is if $f$ belongs to the class
induced by $\omega$. Let us denote this by $\omega \sim f$. The claim made a few lines above
implies that, given $f$ in $\M(PM,G)$, if there exists an $H^1$ connection $\omega$ which agrees with
$f$, then $S(\omega)$ depends only on $f$, not on $\omega$.

Recall that, if $M$ is closed and $P$ is a principal $G$-bundle over $M$, then $\oo(P)$ denotes
the element of $\Pi$ which corresponds to the isomorphism class of $P$ among all principal $G$-bundles. 

\begin{definition}[Rate functions] 1. Assume that $M$ is closed. Let $z$ be an element of $\Pi$. Let
$P$ be a principal $G$-bundle over $M$ such that $\oo(P)=z$. Define the function $I^{\YM}_{z}:\M(PM,G)\lra [0,+\infty]$ by
$$ I^{\YM}_{z}(f)=\cases{\frac{1}{2}  S(\omega) \;\; {\rm if}\; \exists\;\omega\in H^1\A(P), \;
f\sim\omega, \cr +\infty \;\;\;{\rm otherwise.}} $$
2. Assume that $M$ has a boundary with $p$ components $N_{1},\ldots,N_{p}$. Let $P$ be a
$G$-bundle over $M$. Define for every choice of $p$ conjugacy classes $X_{1},\ldots,X_{p}$ in $G$ the
function $I^{\YM}_{X_{1},\ldots,X_{p}}:\M(PM,G)\lra [0,+\infty]$ by
$$ I^{\YM}_{X_{1},\ldots,X_{p}}(f)=\cases{\frac{1}{2} S(\omega)  \;\; {\rm if}\; \exists \; \omega
\in H^1\A_{X_{1},\ldots,X_{p}}(P), \; f\sim\omega, \cr +\infty \;\;\;{\rm otherwise.}} $$
\end{definition}

\subsection{Statement of the main results}

\begin{theorem}[Surface with boundary] \label{thm:b}
Let $N_{1},\ldots,N_{p}$ be the connected components of
$\partial M$. Let $X_{1},\ldots,X_{p}$ be $p$ conjugacy classes in $G$. The family
$(P_{T;X_{1},\ldots,X_{p}})_{T>0}$ of probability measures on $(\M(PM,G),\C)$ satisfies, as $T$
tends to 0, a large deviation principle of speed $T$ with rate function
$I^{\YM}_{X_{1},\ldots,X_{p}}$.
\end{theorem}

\begin{theorem}[Closed surface] \label{thm:c}
Let $z$ be an element of $\Pi$. The family $(P_{T,z})_{T>0}$ of
probability measures on $(\M(PM,G),\C)$ satisfies, as $T$ tends to 0, a large deviation principle 
of speed $T$ with rate function $I_z^{\YM}$. 
\end{theorem}

A good general reference on large deviation principles is \cite{DZ}. For the sake of clarity, let
us explain the meaning of the second theorem. The space $\M(PM,G)$ is endowed with the trace 
of the product topology on $G^{PM}$. That the function $I^{\YM}_{z}$ is a rate
function\footnote{Since $\M(PM,G)$ is a compact space, it does not tell much to say that it is
also a {\it good} rate function.} means that it
is lower semi-continuous on $\M(PM,G)$. Then, the large deviation principle asserts that, for each
measurable subset $A\in\C$ of $\M(PM,G)$, with closure $\overline{A}$ and interior $A^\circ$,  the
following inequalities hold:
$$-\inf_{f\in A^\circ} I^z_{YM}(f) \leq \liminf_{T\to 0} T \log P_{T,z}(A) \leq \limsup_{T\to 0} 
T \log P_{T,z}(A) \leq -\inf_{f\in \overline{A}} I^z_{YM}(f).$$

\section{Holonomy of $H^1$ connections}
\label{section:sobolev connections}
\subsection{Local results}
\label{subsec:local results}

Let $P$ be a principal $G$-bundle over $M$. Let $(U,s)$ be a local section of $P$. An $H^1$ connection
$\omega$ determines by definition a $\LG$-valued 1-form $\omega_{s}$ on $U$. In this first section,
we will focus on what can be done with this single locally defined 1-form. We use the notation
$\Omega^1(U)$ and $H^1\Omega^1(U)$ respectively for the spaces of smooth and $H^1$ real-valued 1-forms
on $U$. We put a subscript  $\LG$ to indicate $\LG$-valued forms or functions.

The covering $\pi:\wG \lra G$ induces an isomorphism of Lie algebras through which we identify $\LG$
with the Lie algebra of $\wG$. Most of the results we are about to prove hold in $\wG$ as well as
in $G$. Hence we decide, until the end of Section \ref{subsec:local results}, that $G$ is any Lie
group with Lie algebra $\LG$. Such a group is the direct product of a compact Lie group and a
group isomorphic to $\R^m$ for some $m\geq 0$.

\subsubsection{Holonomy}

Let $e$ be an edge contained in $U$. Let us choose a smooth parametrization $e:[0,1]\lra M$ of
$e$. If $\omega_{0}$ belongs to $\Omega^1_{\lg}(U)$, then $e^*\omega_{0}$ is a smooth
1-form on $[0,1]$ with values in $\LG$. We identify this 1-form with the $\LG$-valued function
$t\mapsto e^*\omega_{0}(\partial_{t})=\omega_{0}(\dot e_{t})$. 

If now $(\omega_{n})_{n\geq 0}$ is a sequence of smooth 1-forms which converges in
$H^1\Omega^1_{\lg}(U)$ to an $H^1$ 1-form $\omega$, then a classical result (\cite{Adams}, Theorem
5.22) asserts that the sequence $(e^*\omega_{n})_{n\geq 0}$ of $\LG$-valued functions on
$[0,1]$ converges in $H^{1/2}_{\lg}([0,1])$ to a limit which depends only on
$\omega$. This limit is denoted by $e^*\omega$ and it is called the {\it trace} of $\omega$ along the
parametrized edge $e$. Moreover, the mapping $e^*:H^1\Omega^1_{\lg}(U) \lra
H^{1/2}_{\lg}([0,1])$ is continuous. By composing this mapping with the
compact embedding $H^{1/2}_{\lg}\hookrightarrow L^2_{\lg}$, we get a compact linear mapping
$e^*:H^1\Omega^1_{\lg}(U) \lra L^2_{\lg}([0,1])$. We define the holonomy of $\omega$ along $e$ by
rolling the curve $e^*\omega$ onto $G$. For this, consider the space of $G$-valued functions on
$[0,1]$ 
$$H^1_\bullet([0,1];G) = \{ a \in H^1([0,1];G) | a_0=1\}.$$
Although this space is not a vector space, let us call $H^1$ {\sl norm} of one of its elements $a$
the number $ \| a \|_{H^1_\bullet}^2 = \int_0^1
\| \dot a_t a_{t}^{-1} \|^2 \; dt$. The following result is classical and explains what is meant by
rolling a curve in $\LG$ onto $G$.

\begin{proposition} The rolling map $R:L^2_{\lg}([0,1]) \lra H^1_{\bullet}([0,1];G)$ which assigns
to each $\alpha \in L^2_{\lg}([0,1])$ the unique element $a\in H^1_{\bullet}([0,1];G)$ such
that
\begin{equation}\label{eqn:EDO}
\dot a_{t}a_{t}^{-1}=-\alpha_{t} \; a.e. \; t\in [0,1]
\end{equation}
is a norm-preserving homeomorphism.
\end{proposition}

Let $\omega$ and $e$ be given as above. According to this proposition, we can consider the $H^1$
$G$-valued function $R(e^*\omega)$. It is straightforward to check that if we change the
parametrization of $e$, replacing $e$ by $e\circ \varphi$ for some diffeomorphism $\varphi$ of
$[0,1]$, then $R(e^*\omega)$ is replaced by $R((e\circ\varphi)^*\omega)=R(e^*\omega)\circ\varphi$.
In particular, the element $R(e^*\omega)(1)$ of $G$ is independent of the parametrization of $e$.
We denote this element by $\langle \omega,e\rangle$ or $\langle \omega, e \rangle_{G}$ if there is
any ambiguity.

Choose $t\in(0,1)$. Consider the two edges $e_{1}=e_{|[0,t]}$ and $e_{2}=e_{|[t,1]}$. Then it
follows from (\ref{eqn:EDO}) that $\langle \omega,e\rangle=\langle \omega,e_{2}\rangle \langle
\omega, e_{1}\rangle$. This relation implies the following: if the concatenations of edges
$e_{1}\ldots e_{n}$ and $f_{1}\ldots f_{n}$ are equivalent, then the products $\langle
\omega,e_{n}\rangle\ldots \langle \omega,e_{1}\rangle$ and $\langle
\omega,f_{n}\rangle\ldots \langle \omega,f_{1}\rangle$ are equal. Thus, if $c\in PM$ is a path
contained in $U$, then $\langle \omega,c\rangle$ is well defined. We call it the holonomy of
$\omega$ along $c$ and denote it sometimes by $\langle \omega, c \rangle_{G}$. We use the notation
$PU$ for the set of paths contained in $U$.

\begin{proposition} \label{prop:hol local}
Let $\omega$ be an element of $H^1\Omega^1_{\lg}(U)$. 

1. The mapping $\langle \omega,\cdot\rangle :PU \lra G$ belongs to $\M(PU,G)$. 

2. Let $(\omega_{n})_{n\geq 1}$ be a sequence of elements of the Hilbert space $H^1\Omega^1_{\lg}(U)$
which converges weakly to $\omega$. Then, for each $c\in PU$, $\langle \omega_{n},c\rangle$
converges to $\langle \omega, c\rangle$. 
\end{proposition}

\pf The first statement is straightforward. The second one is a consequence of the compactness of
the mapping $c^*$. Indeed, such a compact mapping sends a weakly convergent sequence to a strongly
convergent one. Hence, $R(c^*\omega_{n})$ converges in the $H^1$ topology to $R(c^*\omega)$, in
particular uniformly, and the result holds. \qed

\begin{lemma} \label{lemma:equality}
Let $\omega$ and $\omega'$ two elements of $H^1\Omega^1_{\lg}(U)$. Then the
functions $\langle \omega,\cdot \rangle$ and $\langle \omega',\cdot \rangle$ are equal on $PU$ if and
only if $\omega=\omega'$.
\end{lemma}

\pf Write locally $\omega=\alpha dx + \beta dy$ and $\omega'=\alpha' dx + \beta' dy$, where $\alpha,
\alpha', \beta, \beta'$ belong to $H^1_{\lg}(U)$. Since $\omega$ and $\omega'$ have the
same holonomy along every vertical segment in $U$, the forms $\beta$ and $\beta'$ have the same
trace along every vertical segment. Thus, by Fubini's theorem, the integral over any rectangle of the
difference $\beta-\beta'$ is equal to 0. Hence, $\beta=\beta'$ on $U$. The same argument with
horizontal segments shows that $\alpha=\alpha'$, hence $\omega=\omega'$. \qed 

\subsubsection{Energy inequality and the continuity of the holonomy}

We keep considering an element $\omega$ of $H^1\Omega^1_{\lg}(U)$. Let $\Omega$ be the element
of $L^2\Omega^2_{\lg}(U)$ defined in local coordinates by (\ref{eqn:def courbure}). Let
$*\Omega$ denote the unique element of $L^2_{\lg}(U)$ such that $\Omega=*\Omega\sigma$.
If $V\subset U$ is an open subdomain of $U$, let us define the Yang-Mills energy of $\omega$ on $V$
by $S_{V}(\omega)=\int_{V}\|*\Omega\|^2\; d\sigma$. For each $x\in G$, we denote by $\rho(x)$ the
Riemannian distance in $G$ between $x$ and the unit element.The following result was proved by A.
Sengupta in \cite{Sengupta_IE}. 

\begin{proposition}[Energy inequality]  \label{ineg fine} Let $l$ be a simple loop in $U$ which bounds a domain $V$
diffeomorphic to a disk. Assume that $\omega$ is smooth. Then one has the inequality
\begin{equation}
\label{E}
\rho(\langle\omega,l\rangle_{G})^2 \leq \sigma(V) S_V(\omega). 
\end{equation}
\end{proposition}

For the sake of completeness, we give a proof of this inequality in the appendix (Corollary
\ref{true ei}).

\begin{proposition}\label{energieh1} Proposition \ref{ineg fine} is still true 
under the weaker assumption that $\omega$ belongs to $H^1\Omega^1_{\lg}(U)$.
\end{proposition}

\pf Let us approximate $\omega$ in $H^1$ norm by a sequence $(\omega_{n})_{n\geq 0}$ of smooth 1-forms.
On one hand, the functional $S_{V}:H^1\Omega^1_{\lg}(U)\lra [0,+\infty)$ is continuous because the
embedding $H^1_{\lg}(U)\hookrightarrow L^4_{\lg}(U)$ is continuous. On the other hand, by
Proposition \ref{prop:hol local}, $\langle \omega_{n},l\rangle_{G}$ converges to $\langle
\omega,l\rangle_{G}$. Hence, inequality (\ref{E}) passes to the limit. \qed

This energy inequality is essential. For example, it suffices to imply the following continuity
result.

\begin{proposition} \label{cont fep}
Consider $\omega \in H^1\Omega^1_{\lg}(U)$. The multiplicative mapping $\langle \omega, \cdot
\rangle : PU \lra G$ is continuous with fixed endpoints. This means that, whenever $c_n$ converges
to $c$ with fixed endpoints,  $\langle \omega,c_n\rangle$ converges to $\langle \omega,c\rangle$. 
\end{proposition}

We prefer to state and prove the following slightly more general result. 

\begin{proposition} \label{continuite} Let $U$ be an open domain contained in the interior of $M$.
Let $f$ be an element of $\M(PU,G)$. Assume that there exists a constant $K$ such that, for every
sub-domain $V$ of $U$ diffeomorphic to an open disk and bounded by a simple loop $\partial V$, one
has the inequality 
\begin{equation} \label{Eg}
\rho(f(\partial V))^2 \leq K \sigma(V).  
\end{equation}
Then $f$ is continuous with fixed endpoints.
\end{proposition}

\pf The arguments for this proof are spread in the sections 2.4 to 2.6 of \cite{Levy_AMS}, but
this result was not stated there. We give here a full sketch of proof and refer the reader to
\cite{Levy_AMS} for the details. 

Let $f\in \M(PM,G)$ satisfy (\ref{Eg}). From now on in this proof, all references are to be found
in \cite{Levy_AMS}. The first step is to use an isoperimetric inequality on $M$
(Proposition 2.15), which holds locally, to deduce the existence of a new constant,
still denoted by $K$, such that, for any simple loop $l$ of sufficiently small length $\ell(l)$,
$\rho(f(l))\leq K \ell(l)$.

Since $f$ is multiplicative, this inequality can be extended to short loops with finite self-intersection, defined in
Definition 2.11. It holds in particular for piecewise geodesic loops. Hence, for
every short enough piecewise geodesic loop $\zeta$, one has $\rho(f(\zeta))\leq K \ell(\zeta)$.

Now, let $c$ be an edge. It is possible to find a sequence of piecewise geodesic paths
$(\alpha_n)_{n\geq 1}$ converging to $c$ with fixed endpoints and such that, for all $n$,
$\alpha_n^{-1} c$ is a simple loop bounding a domain diffeomorphic to a disk, whose area
tends to zero as $n$ tends to infinity. Hence, $f(\alpha_n)$ tends to $f(c)$. This is explained
in Section 2.5.3. 

Then, the arguments of the proofs of Lemma 2.36 and Proposition 2.35 show that,
whenever $(\zeta_n)$ is a sequence of piecewise geodesic paths converging to $c$ with fixed
endpoints, $f(\zeta_n)$ tends to $f(c)$. This is the main step of the proof. It involves cutting each
$\zeta_n$ in three parts, two short loops based at the endpoints of $c$ and one path with the same
endpoints as $c$ and staying in a tubular neighbourhood of $c$. The two short loops do not contribute
asymptotically to $f(\zeta_n)$ and the path in the tubular neighbourhood is shown to have a value
under $f$ close to that of $c$, by comparing it with an appropriate term in the first approximating
sequence $(\alpha_n)$.

After extending the result to the case of a piecewise embedded path $c$ (Section 2.6.4), one
concludes that $f$ is continuous with fixed endpoints (Proposition 2.42). \qed

\subsubsection{Gauge transformations}

Let $k\geq 1$ be an integer and $p\geq 2$ be a real number or $+\infty$. In what follows, we consider
$W^{k,p}$ connections. Since $M$ is compact, they are in particular $H^1$, so that the results of the
preceding sections apply.

The group $W^{k+1,p}(U;G)$ of $W^{k+1,p}$ $G$-valued functions on $U$ acts on
$W^{k,p}\Omega^1_{\lg}(U)$ as follows. If $j$ belongs to $W^{k+1,p}(U;G)$ and $\omega$ to
$W^{k,p}\Omega^1_{\lg}(U)$, then
$$ j\cdot \omega = \Ad(j^{-1})\omega + j^{-1} dj. $$

Since $M$ is 2-dimensional, the Sobolev embedding $W^{k+1,p}(U)\hookrightarrow C^{k-1}(U)$ holds and
$W^{k+1,p}$ functions can be evaluated at any point. If $c$ belongs to $PU$, then the relation 
\begin{equation}
\label{eqn:local transfo holonomie}
\langle j\cdot \omega ,c\rangle = j(\overline{c})^{-1}\langle \omega,c\rangle
j(\underline{c})
\end{equation}
is classical if all objects are smooth and easy to check in our setting. It fits with the action
of $\f(M,G)$ on $\M(PU,G)$.

We need to establish a result which allows us to determine when two elements of
$H^1\Omega^1_{\lg}(U)$ differ by the action of an element of $H^2(U;G)$ and how regular this
element is.

\begin{lemma} \label{lem:loops to paths}
Let $V\subset U$ be a connected subset of $U$, not necessarily open, such that any two points of $V$
can be joined by a path contained in $V$. Let $m$ be a point of $V$.

Let $\omega$ and $\omega'$  be two elements of $H^1\Omega^1_{\lg}(U)$. Assume that there exists an
element $x$ of $G$ such that, for each loop $l$ based at $m$ and contained in $V$, one has $\langle
\omega' , l \rangle = \Ad(x^{-1})\langle \omega, l \rangle.$

Then there exists $j\in \f(V,G)$ such that $j(m)=x$ and the multiplicative functions $j\cdot \langle
\omega, \cdot \rangle$ and $\langle \omega',\cdot \rangle$ coincide on the set $PV$ of paths
contained in $V$.
\end{lemma} 

\pf The mapping $j:V \lra G$ must satisfy the following condition: for each path starting at $m$
and finishing at some point $n$, $j(n)=\langle \omega, c\rangle j(m)\langle \omega',c
\rangle^{-1}$. 

Set $j(m)=x$. Then the assumption on $\omega$ and $\omega'$ shows that, for each $n\in V$, the value
of $j(n)$ determined by the equation above does not depend on the the choice of the path $c$
joining $m$ to $n$. This defines an element of $\f(V,G)$ which satisfies the required property. \qed

\begin{proposition} \label{prop:local jauge}
Assume that $U$ is connected and that its boundary is locally the graph of
a Lipschitz function. Endow $M$ with an auxiliary Riemannian metric and let $P_{g}U$ be the set of
paths contained in $U$ which are piecewise geodesic for this metric.  Let $k\geq 1$ be an integer and
$p\in [2,+\infty]$ be a real number or $\infty$. Let $\omega$ and $\omega'$ be two elements of
$W^{k,p}\Omega^1_{\lg}(U)$. Assume that some
element $j\in \f(U,G)$ satisfies, for each path $c\in P_{g}U$, the relation
\begin{equation}\label{eqn:carac j}
\langle\omega',c\rangle = j(\ov{c})^{-1}\langle \omega,c \rangle j(\un{c}).
\end{equation}

Then $j\in W^{k+1,p}(U;G)$ and $j\cdot \omega=\omega'$.
\end{proposition}

Since for every smooth open domain $V$ whose closure is contained in $U$, one has 
$C^\infty(\overline{V})=\cap_{k\geq 1} W^{k,2}(V)$, the proposition implies in particular that, if
$\omega$ and $\omega'$ are smooth on $U$, then $j$ is smooth on $U$. \\

\pf We prove this result by induction on $k$. Proving that $j$ belongs to $W^{1,p}$ is a non-trivial
step of the proof since we assume no regularity at all a priori on $j$. However, (\ref{eqn:carac
j}) allows us to prove that $j$ is continuous and indeed $W^{1,q}$ for all $q<\infty$.

We will use the fact that every point of $U$ has a neighbourhood diffeomorphic to $(-1,1)^2$ in
such a way that all horizontal and all vertical segments are geodesic. To construct such a
neighbourhood, choose two geodesic segments $\gamma_{h}$ and $\gamma_{v}$ which cross orthogonally
at $m$. Let $T_{h}$ and $T_{v}$ be two tubular neighbourhoods of $\gamma_{h}$ and $\gamma_{v}$
respectively (see \cite{Gray} for a definition of tubular neighbourhoods and Fermi or normal
coordinates). Let $(x_{h},r_{h})$ and $(x_{v},r_{v})$ respectively be normal coordinates on these
tubes, such that $r_{h}$ is the distance to $\gamma_{h}$ and $r_{v}$ the distance to $\gamma_{v}$.
The mapping $T_{h}\cap T_{v}\lra \R^2$ which sends a point $n$ to $(x_{h}(n),x_{v}(n))$ is smooth
and its differential at $m$ is invertible. On a small neighbourhood of $m$, this mapping is a
coordinate chart such that all segments parallel to the axes are geodesic.

For more convenience in dealing with Sobolev spaces, we will also assume that $G$ is a subgroup of a
vector space of matrices. This puts no further restriction on $G$, since it is by our assumptions the
direct product of a compact group and a group isomorphic to $(\R^m,+)$. In this way, $\LG$ is 
a subspace of the same vector space of matrices. So, we consider that all forms and functions on
$U$ are matrix-valued.

$\bullet$ Let us prove that $j$ is continuous. Let $m$ be a point of $U$. Let us restrict
ourselves to a neighbourhood of $m$ as above.

For each $x\in(-1,1)$, let $h_{x}$ denote the horizontal segment joining $m$ to $(x,0)$. For each
$(x,y)\in(-1,1)^2$, let $v_{x,y}$ denote the vertical segment joining $(x,0)$ to $(x,y)$. Finally,
for each $(x,y)\in(-1,1)^2$, set $c_{x,y}=h_{x}v_{x,y}$. This is a piecewise geodesic path and
this allows us to write 
$$j(x,y)=\langle \omega , c_{x,y}\rangle j(0,0) \langle \omega', c_{x,y}\rangle^{-1}.$$
From this equality we deduce that 
\begin{eqnarray*}
d_{G}(j(0,0),j(x,y)) &\leq& \rho(\langle \omega,c_{x,y}\rangle) + \rho(\langle \omega',
c_{x,y}\rangle) \\
&\leq &\rho(\langle \omega,h_{x}\rangle) + \rho(\langle \omega,v_{x,y}\rangle) +
\rho(\langle \omega',h_{x}\rangle) + \rho(\langle \omega',v_{x,y}\rangle).
\end{eqnarray*}
Assume that $0\leq x \leq \frac{1}{2}$. Then $x\mapsto \langle \omega,h_{x}\rangle =
R(h_{\frac{1}{2}}^*\omega)(x)$ is an absolutely continuous function of $x$. Now let
$\varphi:(-1,1)^2\lra [0,1]$ be a smooth cutoff function such that $\varphi(x,y)=1$ if
$\max(|x|,|y|)\leq \frac{1}{2}$ and $\varphi(x,y)=0$ if $\max(|x|,|y|)\geq \frac{3}{4}$. Then,
if $0\leq x,y \leq \frac{1}{2}$, then $\langle \omega , v_{x,y} \rangle = \langle \varphi\omega ,
v_{x,y} \rangle $. Let us write $\omega=\omega_{1}\; dx + \omega_{2}\; dy$. Then, by an elementary
estimate (\cite{Adams}, Lemma 5.7), there exists a constant $C$ independent of $y$ such that
$$\int_{0}^y \|\omega_{2}(x,t)\|^2 \; dt \leq C \int_{(-1,1)\times [0,y)}
\|\omega_{2}(s,t)\|^2 + \| \partial_{x}\omega_{2}(s,t)\|^2 \; ds dt .$$
Since $\omega$ belongs to $H^1$, the right hand side tends to $0$ as $y$ tends to $0$. Thus, the
$L^2$ norm of $v_{x,y}^*\omega$ tends to $0$ as $y$ tends to $0$, uniformly in $x$. This implies
that $\rho(\langle \omega,v_{x,y}\rangle)$ tends to $0$ as $(x,y)$ tends to $(0,0)$.
The same arguments applied to $\omega'$ finish the proof that $j$ is continuous at $m$. 

$\bullet$ Let us prove that $j$ belongs to $W^{1,q}(U)$ for all $q<\infty$. We begin by proving
that $j$ admits a weak derivative and for this we restrict again to a neighbourhood diffeomorphic
to $(-1,1)^2$ of some point $m$.
For each $(x,y) \in (-1,1)^2$, we have 
$$ j(x,y)=\langle \omega , v_{x,y}\rangle j(x,0) \langle\omega', v_{x,y}\rangle^{-1}.$$
Hence, for any fixed $x$, the map $y \mapsto j(x,y)$, as a product of two 
$H^1$ functions of one variable, belongs to $H^1$. Let us compute its derivative:
\begin{eqnarray*}
\partial_y j(x,y) &=& -\omega(\partial_y) \langle \omega, v_{x,y}\rangle
j(x,0) \langle \omega',v_{x,y}\rangle^{-1} +  \langle \omega,v_{x,y}\rangle j(x,0) \langle
\omega',v_{x,y}\rangle^{-1} \omega'(\partial_y)\\
&=& j(x,y) \omega'(\partial_y) - \omega(\partial_y)j(x,y).
\end{eqnarray*}

This shows that $j$ is absolutely continuous along every vertical segment and admits there an
almost everywhere derivative which is the trace of the function $j \omega'( \partial_y)-\omega(
\partial_y) j$. Hence, $j$ admits a weak partial derivative with respect to $y$, namely
$j\omega'-\omega j$ evaluated on the vector field $\partial_y$. A similar statement holds in the
direction of $y$ and $j$ admits the weak differential $dj=j\omega'-\omega j$.
 
Since $j$ is continuous, this weak differential belongs to $L^q_{loc}\Omega^1(U)$ for all $q<\infty$.
Hence, $j\in W^{1,q}_{loc}(U)$ for all $q<\infty$. If $G$ is compact, $j$ is in fact bounded, so that
$dj \in L^q\Omega^1(U)$ and $j\in W^{1,q}(U)$ for all $q<\infty$. If $G$ is not compact, it is the
direct product of a compact group by $\R^m$ for some integer $m\geq 1$. The $\LG$-valued 1-forms and
the action of $G$-valued functions on them split between the compact part of $G$ and the part
isomorphic to $\R^m$. It is thus enough to check that $j\in W^{1,q}(U)$ when $G=\R^m$. 

In this case, on has $dj=\omega'-\omega$, which belongs to $H^1(U)$. Set $f=*d*(\omega'-\omega) \in
L^2(U)$. Let $\nu$ be the outer normal vector field along the boundary of $U$ and let
$g=(\omega' - \omega)(\nu) \in H^{1/2}( \partial U)$. Then, up to an additive
constant, $j$ is the unique solution to the inhomogeneous Neumann problem 
$$\cases{\Delta j= f \;\; {\rm on }\; U \cr
\partial_{\nu} j = g \;\; {\rm on }\; \partial U.}$$
Hence, according to standard results on elliptic boundary value problems, $j$ belongs to $H^2(U)$.
This implies that $j$ admits a continuous extension on the closure of $U$. In particular, it is
bounded on $U$. Finally, for all $q<\infty$, $dj \in L^q\Omega^1(U)$, so that $j\in W^{1,q}(U)$.

$\bullet$ Let us prove that $j$ belongs to $W^{k+1,p}(U)$. In a first step, let $r\geq 2$ be such
that $\omega$ and $\omega'$ belong to $W^{1,r}\Omega^1(U)$. By using Leibnitz's rule, it is easy
to check that the product of a function of $W^{1,r}(U)$ by a function of $C^0(\overline{U})\cap
W^{1,2r}(U)$ belongs to $W^{1,r}(U)$. Hence, $dj\in W^{1,r}\Omega^1(U)$ and $j\in W^{2,r}(U)$. If
$k=1$, the proof is finished. 

In general, we use a simple iteration argument. Set $W^{l,\infty-}(U)=\cap_{2\leq
p<\infty} W^{l,p}(U)$. We use the fact that $W^{k,p}(U)$ is stable by multiplication as soon as
$kp>2$.
In particular, if $\omega$ and $\omega'$ belong to $W^{l,\infty-}$ for some $l\geq 1$, and if $j$
belongs to $W^{l,\infty-}$, then $dj$ belongs to $W^{l,\infty-}\Omega^1(U)$ and $j$ belongs
actually to $W^{l+1,\infty-}(U)$. Now we use the Sobolev embedding $W^{l,q}(U) \hookrightarrow W^{l-1,\infty-}$,
which is valid for all $q\geq 2$ and all $l\geq 1$. By this embedding, $\omega$
and $\omega'$ belong to $W^{l,\infty-}(U)$ for $l=0,\ldots, k-1$. Assume that $k>1$. Since
$\omega$ and $\omega'$ belong to $W^{1,\infty-}(U)$, we
have proved above that $j$ belongs to $W^{1,\infty-}(U)$ and we conclude by iteration that $j\in
W^{k,\infty-}(U)$. In particular, $j$ and hence $dj$ belong to $W^{k,p}(U)$. Finally, $j\in
W^{k+1,p}(U)$.

The fact that $j\cdot \omega=\omega'$ is obvious. \qed

\subsection{Global results}

Let $P$ be a principal $G$-bundle over $M$. Let $\omega$ be an element of $H^1\A(P)$. In this second
section, we explain how the holonomy of the locally defined 1-forms $\omega_{s}$ indexed by local
sections of $P$ fit together into a global object. 

\subsubsection{Holonomy}

For each point $m$ of $M$, we denote by $P_{m}$ the fibre of $P$ over $m$. Let $m$ and $n$ be two
points of $M$. Recall that a mapping $ \tau : P_{m} \lra P_{n}$ is said to be {\it $G$-equivariant}
if, for all $p\in P_{m}$ and all $g\in G$, one has $\tau(pg)=\tau(p)g$.

\begin{definition} \label{def:holonomie}
A {\sl holonomy} or {\sl parallel transport} on $P$ is a collection of mappings
$(\tau_{c},c\in PM)$ indexed by $PM$ with the following properties:

1. For each $c\in PM$, $\tau_{c}$ is a $G$-equivariant mapping from $P_{\underline{c}}$ to
$P_{\overline{c}}$.

2. The collection $(\tau_{c},c\in PM)$ is multiplicative, that is, for each path $c$,
$\tau_{c^{-1}}=\tau_{c}^{-1}$ and, if $c_{1}$ and $c_{2}$ are two paths such that
$\overline{c_{1}}=\underline{c_{2}}$, then $\tau_{c_{1}c_{2}}=\tau_{c_{2}}\circ \tau_{c_{1}}$.

The set of holonomies on $P$ is denoted by $\mathcal{T}(P)$.
\end{definition}

\begin{remark} \label{rmk:equiv}
Let $m$ and $n$ be two points of $M$. Since $G$ acts transitively on $P_{m}$, a $G$-equivariant
mapping $\tau:P_{m}\lra P_{n}$ is determined by the image of any single point of $P_{m}$. If a
point $p$ is chosen in $P_{m}$ and a point $q$ in $P_{n}$, there exists a unique element $g$ of
$G$ such that $\tau(p)=qg$. Conversely, for each $g\in G$ there exists a unique $G$-equivariant
mapping $\tau: P_{m}\lra P_{n}$ such that $\tau(p)=qg$. This one-to-one correspondence defines a
natural topology on the set of equivariant mappings from $P_{m}$ to $P_{n}$.
\end{remark}

\begin{proposition} \label{prop:h1 hol} Every element of $H^1\A(P)$ determines a holonomy on $P$. 
Moreover, if two elements of $H^1\A(P)$ determine the same holonomy, then they are equal.
\end{proposition}

\pf Let $\omega$ be an $H^1$ connection on $P$. Let $c$ be a path on $M$. Assume that there exists a
local section $(U,s)$ of $P$ such that $c$ is contained in the domain of $U$. Then define
$\tau^s_{c}$ to be the unique $G$-equivariant mapping from $P_{\underline{c}}$ to $P_{\overline{c}}$
such that $\tau_{c}(s(\underline{c}))=s(\overline{c})\langle \omega_{s},c\rangle$. The collection of
mappings $(\tau^s_{c},c\in PU)$ is clearly multiplicative in the sense of Definition
\ref{def:holonomie}. In particular, if $c=c_{1}c_{2}$, then
$\tau^s_{c}=\tau^s_{c_{2}}\tau^s_{c_{1}}$. 

If $s'$ is another local section of $P$ on a domain which contains $c$, then one checks easily by
using Definition \ref{def:Sobolev connections} and (\ref{eqn:local transfo
holonomie}) that the mappings $\tau^{s}_{c}$ and $\tau^{s'}_{c}$ are the same. Let us denote them
both by $\tau_{c}$.

Now pick any path $c$ in $PM$. Write $c$ as a concatenation of shorter paths $c=c_{1}\ldots c_{n}$
in such a way that each shorter path is contained in the domain of a local section of $P$. Then
the mapping $\tau_{c_{n}}\ldots \tau_{c_{1}}$ does not depend on the decomposition of $c$. Indeed,
if $c=c'_{1}\ldots c'_{m}$ is another decomposition of $c$, then there exists a third
decomposition $c=c''_{1}\ldots c''_{r}$ which is finer than the two other ones, in the sense that
each $c_{i}$ and each $c'_{j}$ can be written as a concatenation of some $c_{k}''$'s. Then the
multiplicativity stated above inside each domain of a local section implies that the two first
decompositions give rise to the same mapping $P_{\underline{c}}\lra P_{\overline{c}}$ as the third
decomposition. It is thus legitimate to call this mapping $\tau_{c}$. 

Two $H^1$ connections $\omega$ and $\omega'$ induce the same holonomy if and only if, for each
local section $(U,s)$ of $P$, the multiplicative functions $\langle \omega_{s},\cdot \rangle$ and
$\langle \omega'_{s},\cdot \rangle$ are equal on $PU$. According to Lemma \ref{lemma:equality}, this
is equivalent to the fact that $\omega_{s}=\omega'_{s}$ on the interior of $M$. If $M$ has a
boundary, this shows that $\omega=\omega'$ almost everywhere, hence $\omega=\omega'$. \qed
 
The proof of the following lemma is straightforward.

\begin{lemma} Let $\omega$ be an element of $H^1\A(P)$. Let $l$ and $l'$ be two loops on $M$,
based respectively at $m$ and $m'$. Assume that $l$ and $l'$ are cyclically equivalent, that is,
that they differ only by the choice of their base points. Choose $p\in P_{m}$ and $p'\in P_{m'}$.
Let $g$ and $g'$ be the elements of $G$ such that $\tau_{l}(p)=pg$ and $\tau_{l'}(p')=p'g'$. Then
$g$ and $g'$ are conjugate. 
\end{lemma}

If $c$ is a cycle on $M$, for example a component of $\partial M$ or the boundary of a face of a
graph on $M$, then this lemma allows us to define $\langle \omega, c \rangle$ as a {\it conjugacy
class} of $G$. In particular, if $\partial M$ has $p$ components $N_{1},\ldots,N_{p}$ and if $X_{1},
\ldots, X_{p}$ are $p$ conjugacy classes in $G$, then we set
$$ H^1\A_{X_{1},\ldots,X_{p}}(P)=\{\omega \in H^1\A(P) : \langle \omega, N_{1}\rangle = X_{1},
\ldots, \langle \omega, N_{p} \rangle =X_{p} \}. $$

 The space $H^1\A(P)$ is an affine space with direction $H^1\Omega^1(M)\otimes\Ad(P)$. We endow
$H^1\A(P)$ with the corresponding topology. Concretely, this topology is generated by the subsets
$\{\omega \;|\; \| \omega_{s} - \eta_{s} \|_{H^1} <\epsilon \}$, where $\epsilon$ runs over the
positive reals, $s$ over the local sections of $P$ and $\eta_{s}$ over
$H^1\Omega^1_{\lg}(\dom(s))$. This topology can be metrized by choosing a finite covering of $M$.
The weak topology on $H^1\A(P)$ is defined similarly. 
 
 \begin{proposition} Let $(\omega_{n})_{n\geq 0}$ be a sequence of $H^1$ connections on $P$ which
converges weakly to a connection $\omega_{\infty}$. For each $n$ with $0\leq n\leq \infty$, let
$(\tau^n_{c},c\in PM)$ be the holonomy induced by $\omega_{n}$. Then, for each path $c\in PM$, the
mappings $\tau^n_{c}$ converge to $\tau^\infty_{c}$. 
\end{proposition}

\pf If $c$ is contained in the domain of a local section of $P$, then the result is a direct
consequence of  Proposition \ref{prop:hol local}. If $c$ is not contained in the domain of a local
section, then we decompose it as a concatenation of shorter paths to which the local argument can
be applied. The result follows by multiplicativity of the holonomy induced by a connection. \qed
 
\subsubsection{Gauge transformations}

\begin{definition} A {\sl gauge transformation} on $P$ is a collection of mappings
$(\gamma_{m},m\in M)$ indexed by the points of $M$ such that, for each $m\in M$, $\gamma_{m}$
is a $G$-equivariant mapping of $P_{m}$ onto itself. The set of gauge transformations on $P$ is
denoted by $\J(P)$.
\end{definition}

Let $j=(\gamma_{m},m\in M)$ be a gauge transformation. Let $(U,s)$ be a local section of $P$.
Define $j_{s}$ as the unique $G$-valued function on $U$ such that, for each $m\in U$,
$\gamma_{m}(s(m))=s(m)j_{s}(m)$. It is easily checked that, if $s'$ is another local section of $P$
and if $s'=s\psi$ on $\dom(s)\cap\dom(s')$, then $j_{s'}=\psi^{-1}j_{s}\psi$. 

In fact, a gauge transformation is completely determined by the family of locally defined $G$-valued
mappings $j_{s}$, where $s$ runs over local sections of $P$. It can be defined as such a family
which satisfies the compatibility condition stated above. If in addition one puts
regularity conditions on the mappings $j_{s}$, this allows us to define Sobolev gauge
transformations. In particular, we shall consider the space $H^2\J(P)$ of $H^2$ gauge
transformations. 

The set $\J(P)$ and the space $H^2\J(P)$ are groups under pointwise composition. These groups act
respectively on $\mathcal{T}(P)$ and $H^1\A(P)$ as follows\footnote{These are actions on the right
but we denote them on the left. This will not cause any ambiguity.}.

Let $T=(\tau_{c},c\in PM)$ be a holonomy and $j=(\gamma_{m},m\in M)$ be a gauge
transformation. We define a new holonomy $j \cdot T=(j\cdot \tau_{c},c\in PM)$ by
setting, for each path $c$, $(j\cdot \tau)_{c}=\gamma_{\overline{c}}^{-1}\circ \tau_{c} 
\circ\gamma_{\underline{c}}$. 

On the other hand, let $\omega$ be an $H^1$ connection and $j$ an $H^2$ gauge transformation. Then
we define a new element $j\cdot \omega$ of $H^1\A(P)$ by setting, for each local section $s$ of
$P$, $(j\cdot\omega)_{s}=j_{s}\cdot \omega_{s}=\Ad(j_{s}^{-1})\omega_{s}+j_{s}^{-1} dj_{s}$. 

According to (\ref{eqn:carac j}), the mapping $H^1\A(P)\lra \mathcal{T}(P)$ which sends a connection to its
holonomy induces a mapping between quotient spaces:
\begin{equation}\label{eqn:left}
H^1\A(P)/H^2\J(P) \lra \mathcal{T}(P)/\J(P).
\end{equation}

Let us choose a reference point $p(m)$ in the fibre $P_{m}$ for each $m\in M$.  Then, according to
the Remark \ref{rmk:equiv}, a holonomy on $P$ determines an element of $\f(PM,G)$. It follows from
the multiplicativity of a holonomy that this function is actually multiplicative. Now it is easily
checked that changing the reference point in each fibre or changing the holonomy by the action of
a gauge transformation modifies the multiplicative function by the action of an element of
$\f(M,G)$. Hence, there is a second mapping:
\begin{equation}\label{eqn:right}
\mathcal{T}(P)/\J(P) \lra \M(PM,G)/\f(M,G).
\end{equation}
Combining (\ref{eqn:left}) with (\ref{eqn:right}), we get a mapping which we denote by
$\mathcal{H}_P$:
$$
\mathcal{H}_{P} : H^1\A(P)/H^2\J(P) \lra \M(PM,G)/\f(M,G).
$$

Recall that a function $f\in \M(PM,G)$ is said to be continuous with fixed endpoints if it is
sequentially continuous along sequences of paths converging with fixed endpoints. Observe that this
property is not affected by the action of $\f(M,G)$. Our main result in this section is the
following. 

\begin{proposition}\label{prop:1-1}
1. Let $P$ be a principal $G$-bundle over $M$. Then the mapping $\mathcal{H}_{P}$ is a one-to-one
mapping whose range contains only functions continuous with fixed endpoints. 

2. Let $Q$ be another $G$-bundle over $M$. If $P$ and $Q$ are isomorphic, and if $\varphi:P\lra Q$ is
a bundle isomorphism, then $\varphi$ induces an energy-preserving map $\varphi^*:H^1\A(Q)/H^2\J(Q)\lra
H^1\A(P)/H^2\J(P)$ which satisfies the relation $\mathcal{H}_{P}\circ \varphi^*=\mathcal{H}_{Q}$. 

If $P$ and $Q$ are not isomorphic, then the ranges of $\mathcal{H}_{P}$ and $\mathcal{H}_{Q}$ are
disjoint.
\end{proposition}

\begin{remark} \label{rem:loops}
It could be argued that it is very unpleasant to choose a reference point in each fibre of $P$
because, unless $P$ is trivial, this cannot be done in a continuous way. It is possible to avoid this
by choosing a point $m$ on $M$ and restricting oneself to the space $L_{m}M$ of loops based at $m$.
The action
of $\f(M,G)$ on $\M(L_{m}M,G)$ reduces to the diagonal action of $G$ by conjugation. Then, by
choosing only a reference point in $P_{m}$, one is able to construct a mapping
$\tilde\mathcal{H}_{P}: H^1\A(P)/H^2\J(P)\lra
\M(L_{m}M,G)/G$.  This point of view is exactly equivalent to ours since the spaces
$\M(PM,G)/\f(M,G)$ and $\M(L_{m}M,G)/G$ equipped with the traces of the product topologies are
canonically homeomorphic. Moreover, the canonical homeomorphism sends functions continuous with
fixed endpoints to continuous functions. Nevertheless, we choose to consider $PM$ instead of
$L_{m}M$ because it is easier to work with the Yang-Mills measure if it is defined on $\M(PM,G)$.
\end{remark}

\noindent \textbf{Proof of Proposition \ref{prop:1-1} -- } 1. Let $\omega$ be an $H^1$ connection.
For each $m\in M$, let $p(m)$ be a reference point in $P_{m}$. Let $f$ be the element of
$\M(PM,G)$ determined by $\omega$ and the set-theoretic section $m\mapsto p(m)$. Let us prove
that $f$ is continuous with fixed endpoints.

Let $(c_{n})_{n\geq 0}$ be a sequence of paths converging with fixed endpoints to $c$. Let us
assume that $c$ is contained in the domain of a local section $s$ of $P$. Then, for $n$ large
enough, $c_{n}$ is also contained in the domain of this local section.
Let us also assume that $s(\underline{c})=p(\underline{c})$ and $s(\overline{c})=p(\overline{c})$.
Since $G$ is connected, this causes no loss of generality. Now, for each $n$, we have
$f(c_{n})=\langle \omega_{s},c_{n}\rangle$ and the similar equality for $c$. By Proposition
\ref{cont fep}, this implies that $f(c_{n})$ tends to $f(c)$ as $n$ tends to infinity. 

If $c$ is not contained in the domain of a local section, let us decompose it as $c=c^{1}\ldots
c^{r}$ in such a way that, for each $k=1,\ldots,r$, $c^{k}$ is contained in the domain $U_{k}$ of
some local section $s_{k}$. We assume that, for each $k$, the section $s_{k}$ coincides with $p$ at
the endpoints of $c^{k}$. Let $R>0$ be such that every geodesic ball on $M$ of radius smaller than
$R$ is geodesically convex and such that, for each $k$, the $R$-neighbourhood of $c^{k}$, denoted
by $c^{k}_R=\{m\in M | d_{M}(m,c_{k})<R\}$ is contained in $U_{k}$. Then, for $n$ large enough,
$c_{n}$ is contained in $c^{1}_R\cup \ldots \cup c^{r}_R$. For such an $n$, decompose $c_{n}$ as
$c_{n}=c_{n}^1\ldots c_{n}^r$ in such a way that for each $k$,
$d_{M}(\underline{c}_{n}^k,\underline{c}^k)<R$ and $d_{M}(\overline{c}_{n}^k,\overline{c}^k)<R$.
Then, for each $k$, $c^k_{n}$ converges to $c^k$, but not with fixed endpoints. 

For each $n$ large enough and each $k=1,\ldots,r-1$, let $\zeta_{k,n}$ denote the geodesic segment
joining $\overline{c}^k$ to $\overline{c}^k_{n}$. Let $\zeta_{0,n}$ be the point $\underline{c}$
and $\zeta_{r,n}$ be the point $\overline{c}$. Then, for each $n$ and each $k=1,\ldots,r$, set
$\tilde c^k_{n} = \zeta_{k-1,n}c^k_{n}\zeta_{k,n}^{-1}$. For each $k$, $(\tilde c^k_{n})$ tends
to $c^k$ with fixed endpoints and, by multiplicativity, $f(c_{n})=f(c^1_{n}\ldots c^r_{n})=f(\tilde
c^1_{n}\ldots \tilde c^r_{n})=f(\tilde c^r_{n})\ldots f(\tilde c^1_{n})$. We have now reduced the
problem to the case of paths lying in the domain of a local section and find that $f(c_{n})$ tends
to $f(c^r)\ldots f(c^1)=f(c)$.

We prove now that $\mathcal{H}_{P}$ is injective. Let us consider two $H^1$ connections
$\omega$ and $\omega'$ on $P$ which are sent to the same class of $\M(PM,G)$ by the composed mapping
$H^1\A(P) \twoheadrightarrow H^1\A(P)/H^2\J(P) \build\lra_{}^{\mathcal{H}_{P}}\M(PM,G)/\f(M,G)$.
We claim that $\omega$ and $\omega'$ differ by the action of an element of $H^2\J(P)$.

Indeed, let $T=(\tau_{c},c\in PM)$ and $T'=(\tau'_{c},c\in PM)$ be the holonomies induced by $\omega$
and $\omega'$ respectively. There exists a gauge transformation $j=(\gamma_{m},m\in M)$ such that
$T'=j\cdot T$. Let $(U,s)$ be a local section of $P$. Then the equality$\langle \omega'_{s},\cdot 
\rangle= j_{s}\cdot \langle \omega_{s},\cdot \rangle$ of multiplicative function on $PU$ holds. By
Proposition \ref{prop:local jauge}, $j_{s}$ belongs to $H^2(U;G)$. Since this argument is valid for
each local section of $P$, we conclude that $j$ belongs to $H^2\J(P)$. Now, the connections $j\cdot\omega$
and $\omega'$ determine the same holonomy on $P$. According to Proposition \ref{prop:h1 hol}, this
implies that they are equal. 

2. If $P$ and $Q$ are isomorphic, the statement is straightforward. Let us assume that $P$ and $Q$
are not isomorphic. Then $M$ is necessarily closed. Let $f$ be an element of $\M(PM,G)$ whose
class modulo $\f(M,G)$ belongs to the range of $\mathcal{H}_{P}$. We claim that $\oo(P)$ can be
computed from $f$. Since $\oo(P)\neq \oo(Q)$, this implies that $f$ does not belong to the range
of $\mathcal{H}_{Q}$. 

Let $m$ be a point of $M$. Let $g$ denote the genus of $M$. Let $a_{1},b_{1},\ldots,a_{g},b_{g}$
be $2g$ loops based at $m$ which generate the fundamental group $\pi_{1}(M,m)$ with the single
relation $[b_{g}^{-1},a_{g}^{-1}]\ldots[b_{1}^{-1},a_{1}^{-1}]=1$. Let $L:[0,1]^2\lra M$ be a
smooth homotopy such that, for each $s\in [0,1]$, $L(s,\cdot)$ is a smooth loop based at $m$, 
$L(0,\cdot)$ is the constant loop at $m$ and  $L(1,\cdot)$ is the loop
$[b_{g}^{-1},a_{g}^{-1}]\ldots[b_{1}^{-1},a_{1}^{-1}]$. The mapping from $[0,1]$ to $G$ defined by
$s\mapsto f(L(s,\cdot))$ is a continuous path starting from $1$. Let $s\mapsto \tilde f(s)$ be the
lift starting at $1$ in $\wG$ of this curve. Recall that, if $x,y\in G$ and $\tilde x,\tilde
y\in\wG$, then $[\tilde x,\tilde y]=\tilde x \tilde y \tilde x^{-1} \tilde y ^{-1}$ depends only on
$x$ and $y$. We denote it by $[\widetilde{x,y}]$. Then $\oo(P)=\tilde f(1)
\left([\widetilde{f(a_{1}),f(b_{1})}]\ldots [\widetilde{f(a_{g}),f(b_{g})}]\right)^{-1}$
(\cite{Sengupta_AMS}, Theorem 3.9). \qed

To finish this section on Sobolev connections, let us state Uhlenbeck's compactness theorem in the
particular case that we are going to use. The original reference for this theorem is
\cite{Uhlenbeck}. For the case where $M$ has a boundary, and also for a more comprehensive and
detailed proof, we refer the reader to \cite{Wehrheim}. 

\begin{theorem}[Compactness theorem] \label{Uh}
Let $P$ be a principal $G$-bundle over $M$. Let $(\omega_n)_{n\geq 1}$ be a sequence of
connections in $H^1\A(P)$ such that $S(\omega_n)$ is uniformly bounded. Then there exists a
subsequence
$(\omega_{n_k})_{k\geq 1}$, a sequence $(j_k)_{k\geq 1}$ in $H^2\J(P)$ and an element $\omega$ of
$H^1\A(P)$ such that
\begin{enumerate}
\item $\displaystyle j_k \cdot \omega_{n_k} \rightharpoonup  \omega$ in $H^1\A(P)$,
\item $\displaystyle S(\omega)\leq \liminf_{k\to\infty} S(\omega_{n_k})$.
\end{enumerate}
\end{theorem}

\section{Large deviations for the Yang-Mills measures}

In this section, we prove Theorems \ref{thm:b} and \ref{thm:c}, except for the construction of a
connection with minimal energy with prescribed holonomy along the edges of a graph, which is the
object of Section \ref{section:construction}. We are going to follow a
route close to the one followed to construct the measure in \cite{Levy_AMS}. The starting point is
the classical short-time estimate of the heat kernel on a compact Riemannian manifold which we
apply to $G$, and, with a minor modification, to its possibly non-compact universal covering $\wG$.
A large deviation principle for the discrete Yang-Mills measures follows by elementary arguments.
An application of the contraction principle produces a large deviation principle for the
finite-dimensional distributions associated to families of paths which are contained in some
$\E^*$, where $\E$ is the set of edges of a graph. Just as in the construction
of the measure, it is not enough to take the projective limit of these discrete principles: we must
first obtain an large deviation principle for all finite dimensional marginals of the holonomy
process. For this, we use a standard result on exponential approximations of measures. In 
identifying the rate function at this stage in terms of the Yang-Mills measure, we make repeated
uses of Uhlenbeck's compactness theorem. Finally, Dawson-G\"artner's theorem yields the large deviation
principle for the whole process.

\subsection{The discrete Yang-Mills measures}

Let $N$ be a Riemannian manifold. Let us denote for all $t>0$ by $p_t(\cdot,\cdot)$ the heat kernel
on $N$, that is, the kernel of the operator $\exp{\frac{t\Delta}{2}}$ on the space of square-integrable
functions on $N$. The fundamental estimate is the following. The prototype of this result was proved
by Varadhan in \cite{Varadhan1,Varadhan2}. For the form given here the reader may consult
\cite{Norris} or \cite{Li}. 

\begin{theorem} \label{molcha}
Let $p_t(\cdot,\cdot)$ be the heat kernel on a compact Riemannian manifold $N$. Then,
uniformly for all $x,y \in N$, one has
$$\lim_{t\to 0} -2t \log p_t(x,y)=d(x,y)^2.$$
\end{theorem}

When $M$ has a boundary, we deduce directly from this theorem the large deviation principle for the
discrete Yang-Mills measure associated to a graph on $M$ with boundary conditions.

Let $N_{1},\ldots,N_{p}$ be the connected components of $\partial M$. Let $X_{1},\ldots,X_{p}$ be
$p$ conjugacy classes in $G$. Recall that, if $x\in G$, then $\rho(x)$ denotes the Riemannian
distance between $1$ and $x$.

\begin{proposition} Let $\G$ be a graph on $M$. The family of measures
$(P^\G_{T;X_{1},\ldots,X_{p}})_{T>0}$ on $G^{\E^+}$ satisfies a large deviation principle with good
rate function
$$
I^\E_{X_{1},\ldots,X_{p}}(g)=\cases{\displaystyle\sum_{F\in\F} \frac{\rho(h_{\partial
F}(g))^2}{2\sigma(F)} \;\; {\rm if}\;\; h_{N_{1}}(g)=X_{1},\ldots,h_{N_{p}}(g)=X_{p},\cr
+\infty \;\; {\rm otherwise}.}
$$
\end{proposition}

\pf Let $S$ denote the subset of all $g\in G^{\E^+}$ such that the boundary conditions
$h_{N_{1}}(g)=X_{1},\ldots,h_{N_{p}}(g)=X_{p}$ are satisfied. It is a closed subset of $G^{\E^+}$.
With the notation of Definition \ref{def:boundary} and by Theorem \ref{molcha}, we have, as
$T$ tends to 0, for all $g\in S$,
$$dP^\G_{T;X_{1},\ldots,T_{p}}(g)=\frac{1}{Z^\G_{T;X_{1},\ldots,X_{p}}}
e^{-\frac{1}{T}(I^\E_{X_{1},\ldots,X_{p}}(g)+o(1))}\; d\nu_{X_{1}}^{N_{1}}\ldots
d\nu_{X_{p}}^{N_{p}}dg_{int},$$
where $o(1)$ is uniform on $S$. 

According to Proposition \ref{prop:Z} and to a standard estimation of the supremum of the heat 
kernel (for instance Theorem V.4.3 of \cite{Varopoulos}, to which we will refer again later),
$Z^\G_{T;X_{1},\ldots,X_{p}}\leq
\|p_{T\sigma(M)}\|_{\infty}=O(T^{-\frac{\dim G}{2}})$ is negligible at the exponential scale. 
The large deviation principle on the subset $S$ follows now from the fact that the measure
$d\nu_{X_{1}}^{N_{1}}\ldots d\nu_{X_{p}}^{N_{p}}dg_{int}$ charges every open subset of $S$. Finally,
since $S$ is closed and supports the measures $P^\G_{T;X_{1},\ldots,X_{p}}$, the large deviation
principle holds on $G^{\E^+}$. \qed

When $M$ is closed, the Yang-Mills measures on $M$ are defined in terms of the heat kernel on
$\wG$ which may not be compact. However, this possible non-compactness is easy to deal with,
since it comes from the presence of a Euclidean direct factor $\R^m$. 

\begin{proposition} Uniformly for all $\tilde g, \tilde h\in \wG$, one has
$$\lim_{t\to 0} -2t \log \tilde p_{t}(\tilde g,\tilde h)=d_{\wG}(\tilde g,\tilde h)^2.$$
\end{proposition}

\pf Since $G$ is a compact group, its Lie algebra $\LG$ which is also that of $\wG$
can be written as $\LG=[\LG,\LG]\oplus \z(\LG)$, where $\z(\LG)$ is the center of $\LG$
(\cite{Duistermaat}, Theorem 3.6.2). Accordingly, $\wG=K\times
\R^m$, where $K$ is the subgroup of $\wG$ with Lie algebra $[\LG,\LG]$ and $m=\dim\z(\LG)$. The group
$K$ is compact and simply connected. Let $\Delta$, $\Delta_{K}$ and $\Delta_{\R^m}$ denote the
Laplace operators on $\wG$, $K$ and $\R^m$ respectively, where $K$ and $\R^m$ are endowed with the
induced metric. Observe that $\Delta_{K}$ and $\Delta_{\R^m}$ commute and, independently, that the
induced metric on $\R^m$ is a constant Euclidean metric. The scalar product on $\LG$ corresponding to
the Riemannian metric on $\wG$ is invariant under the adjoint action of $\wG$ on $\LG$, so that the
adjoint action of $\LG$ on itself is antisymmetric and $\z(\LG)\perp[\LG,\LG]$. This implies the
relations $\Delta=\Delta_{K}+\Delta_{\R^m}$ and $\exp{\frac{t\Delta}{2}}=
\exp{\frac{t\Delta_{K}}{2}}\exp{\frac{t\Delta_{\R^m}}{2}}$. Finally, for all $t>0$, $k,l\in K$ and
$x,y\in\R^m$, and with an obvious notation,
$$\tilde p_{t}\left((k,x),(l,y)\right)=p^K_{t}(k,l)p^{\R^m}_{t}(x,y).$$
On one hand, as $t$ tends to 0, $-2t\log p^K_{t}(k,l)$ tends, by Theorem \ref{molcha}, to
$d_{K}(k,l)^2$ uniformly. On the other hand, $-2t\log p^{\R^m}_{t}(x,y)=d_{\R^{m}}(x,y)^2-mt\log(2\pi
t)$. The result follows now from the identity $d_{\wG}((k,x),(l,y))^2=d_{K}(k,l)^2+d_{\R^m}(x,y)^2$.
\qed

Choose $z\in \Pi$. Let $\G$ be a graph on $M$. Recall the definition of the discrete measures
$P^\G_{T,z}$ given in Definition \ref{def:closed}. Let us also introduce, for $\tilde x\in \wG$, the
notation $\tilde \rho(\tilde x)$ for the Riemannian distance in $\wG$ between $\tilde x$ and the
unit element. 

\begin{proposition} \label{prop:ldp closed 1}
The family of probability measures $(P^\G_{T,z})_{T>0}$ on $\GE$ satisfies a
large deviation principle with rate function 
\begin{equation}\label{eqn:rate closed 1}
I^\E_{z}(g)=\min_{z_{\F}\in \Pi^\F_{z}} \sum_{F\in\F} \frac{\tilde \rho(h^\wG_{\partial F}(\tilde 
g)z_{F})^2}{2\sigma(F)},
\end{equation}
where $\tilde g\in \wG^{\E^+}$ satisfies $\pi(\tilde g)=g$.
\end{proposition}

\begin{remark} 1. The value of the rate function does not depend on the choice of $\tilde g$ for the
same reason as the number defined as $D^\G_{T,z}(g)$ in Proposition \ref{prop:D} does not. Less
obvious is the fact, which is part of the last proposition, that the minimum in (\ref{eqn:rate closed
1}) is attained for some $z_{\F}\in \Pi^\F_{z}$.\\
2. When $G$ is simply connected, the rate function takes the simpler form
$$ I^\E(g)=\sum_{F\in\F} \frac{\rho(h_{\partial F}(g))^2}{2\sigma(F)}. $$
In this case, Proposition \ref{prop:ldp closed 1} is a direct consequence of Theorem \ref{molcha}.
\end{remark}

\pf What makes this proof a bit more difficult than in the case with boundary is the possible
presence of an infinite sum in the density $D^\G_{T,z}$. We need to truncate this sum and
estimate the error we make. 

Let us choose a bounded measurable section $\GE\lra\wGE$ of $\pi$. Let us simply denote by $\tilde
g$ the image by this section of $g\in \GE$. Set $c=\sup \{\tilde \rho(h^\wG_{\partial F}(\tilde
g)) : g\in \GE, F\in\F\}$. Let also $s$ and $S$ be two real numbers such that $0<s<S$ and,
for each face $F$, $s<\sigma(F)<S$. Finally, let $C>0$ be such that, for all $t>0$, all $\tilde x,
\tilde y \in \wG$, $\tilde p_{t}(\tilde x, \tilde y) \leq C t^{-\frac{\dim G}{2}} \exp
-\frac{d_{\wG}(\tilde x,\tilde y)^2}{Ct}$. Such a constant exists by \cite{Varopoulos}, Theorem
V.4.3. 

For each integer $k\geq 0$, set $\Lambda_{k}=\{z_{\F}\in \Pi^\F_{z}| \forall F\in \F, \tilde
\rho(z_{F})<k\}$. Fix $g\in\GE$. Then
\begin{equation}\label{eqn:boites}
 D^\G_{T,z}(g)=\sum_{k=0}^\infty \left\{\sum_{z_{\F}\in \Lambda_{k+1}\backslash \Lambda_{k}}
\prod_{F\in\F} \tilde p_{T\sigma(F)}(h^\wG_{\partial F}(\tilde g)z_{F}) \right\}.
\end{equation}
If $z_{\F}$ belongs to $\Lambda_{k+1}\backslash\Lambda_{k}$, then there exists a 
face $F$ such that $\tilde\rho(h^\wG_{\partial F}(\tilde g)z_{F})\geq |k-c|$, so 
$$\prod_{F\in\F}\tilde p_{T\sigma(F)}(h^\wG_{\partial F}(\tilde g)z_{F}) \leq
\left(\frac{C}{(sT)^{\frac{d}{2}}}\right)^{|\F|} e^{-\frac{(k-c)^2}{CST}}.$$
Hence, if $L$ is a non-negative integer, the tail of $(\ref{eqn:boites})$ satisfies
$$ \sum_{k=L}^\infty \left\{\ldots\right\} \leq \frac{C}{T^\frac{d|\F|}{2}} \sum_{k\geq
L}  |\Lambda_{k+1}| e^{-\frac{(k-c)^2}{CT}}, $$
where $C$ denotes now a constant which varies from line to line. The set $\Pi\subset \wG\simeq
K\times \R^m$ is a sub-lattice of $Z(K)\times R$, where $Z(K)$ is the center of $K$, which is finite,
and $R$ is a discrete additive subgroup of $\R^m$. Hence, the cardinality $|\Lambda_{k}|$ is
dominated by a power of $k$. Thus there exists a rational function $Q$ of two variables such that
the tail of (\ref{eqn:boites}) satisfies
$$ \sum_{k=L}^\infty \left\{\ldots\right\} \leq Q(L,\sqrt{T}) e^{-\frac{(L-c)^2}{CT}}. $$
Hence, for each $L$, the density can be put in the form
$$ D^\G_{T,z}(g)=\sum_{z_{\F}\in \Lambda_{L}} \exp \left[-\frac{1}{T}\sum_{F\in\F}\frac{\tilde
\rho(h^\wG_{\partial F}(\tilde g)z_{F})^2}{2\sigma(F)}+o\left(\frac{1}{T}\right)\right] +
\epsilon(L,T)(g), $$
with $0\leq \epsilon(L,T)(g)\leq Q(L,\sqrt{T}) e^{-\frac{(L-c)^2}{CT}}$.

Now, with $g\in \GE$ still fixed, the function from $\Pi^\F_{z}$ to $\R^+$ which sends $z_{\F}$ to
$\sum_{F\in\F} \frac{\tilde \rho(h^\wG_{\partial F}(\tilde g)z_{F})^2}{2\sigma(F)}$ tends to
infinity as $z_{\F}$ tends to infinity. Thus, this function attains its infimum, on a subset
$M(g)$ of $\Pi^\F_{z}$ which may not be a singleton. Since $g\mapsto \tilde g$ is a bounded
mapping, the convergence of the sum above is uniform in $g$, so that
$\cup_{g\in\GE} M(g)$ is a bounded set and there exists a positive integer $L_{0}$ such that
$\cup_{g\in \GE} M(g)\subset \Lambda_{L_{0}}$. 

The large deviation principle can now be proved easily. As in the case where $M$ has a boundary,
Proposition \ref{prop:Z} and a classical estimate on the heat kernel imply that $Z^\G_{T,z}$ is
negligible at the exponential scale. Let $A\subset \GE$ be a Borel subset. Then, from the discussion
above we deduce that, for $L\geq L_{0}$, 
$$ \overline{\lim_{T\to 0}} T \log P^\G_{T,z}(A)\leq \max\left[\overline{\lim_{T\to 0}} T \log
\int_{A} \epsilon(L,T)(g)
\; dg , -\inf_{g\in\overline{A}} \min_{z_{\F}\in \Pi^\F_{z}} \sum_{F\in\F} \frac{\tilde 
\rho(h^\wG_{\partial F}(\tilde g)z_{F})^2}{2\sigma(F)} \right]. $$
Since $\overline{\lim}_{T\to 0}T \log \int_{A}\epsilon(L,T)(g)\; dg \leq -\frac{(L-c)^2}{C}$
tends to $-\infty$ as $L$ tends to infinity, the upper bound of the large deviation principle is
proved by taking $L$ large enough. A similar argument for the lower bound finishes the proof.\qed
 
We want to give an expression of the rate functions $I^\E_{X_{1},\ldots,X_{p}}$ and $I^\E_{z}$ in
terms of the Yang-Mills energy. For this, we need to establish a link between $H^1$ connections
and elements of $\GE$.

Let $J$ be a subset of $PM$, for example, the set of edges of a graph, or a set of loops. Any 
$H^1$ connection on some $G$-bundle $P$ over $M$ determines, via the mapping $\mathcal{H}_{P}$, an
element of $\M(J,G)/\f(M,G)$.
Now for every subset $K$ of $PM$, let $K^*$ denote the set of paths that can be constructed by
concatenating elements of $K$. There is a natural one-to-one correspondence between $\M(K,G)$ and
$\M(K^*,G)$. Hence, if $K$ is a subset of $PM$ such that $J\subset K^*$, then any function of
$\M(K,G)$ determines a function of $\M(J,G)$ and thus an element of $\M(J,G)/\f(M,G)$. The main
example of this situation
is the following: $K$ is the set of edges of a graph and $J$ is a set of paths in this graph.

\begin{definition} \label{agree}
Let $J$ and $K$ be two subsets of $PM$ such that $J\subset K^*$.
Two connections of $H^1\A$, or two functions of $\M(K,G)$, or one such
connection and one such function are said to {\sl agree up to gauge transformation on $J$}, or
simply to {\sl agree on $J$}, if they induce the same class of $\M(J,G)/\f(M,G)$. We denote this
relation by $\sim_J$.
\end{definition}

From now on, we will alternatively use two sets of assumption, corresponding to the cases with and
without boundary. Let us state them once for all and give them a name. 

\begin{convention}
1. (Boundary) means: Assume that $M$ has a non-empty boundary. Assume that
$N_{1},\ldots,N_{p}$ are the connected components of $\partial M$. Let $X_{1},\ldots,X_{p}$ be $p$
conjugacy classes of $G$. Let $P$ be a principal $G$-bundle over $M$.   

2. (Closed) means: Assume that $M$ is closed. Let $z$ be an element of $\Pi$. Let $P$ be a
principal $G$-bundle over $M$ such that $\oo(P)=z$.
\end{convention}

Let us state the main technical result of this paper. 

\begin{proposition} \label{prop:identify rate 1}
Let $\G$ be a graph on $M$. Let $g$ be an element of $\GE$. 

1. (Boundary) The following equality holds:
\begin{equation}\label{eqn:id rate b}
I^\E_{X_{1},\ldots,X_{p}}(g) =\frac{1}{2} \inf \{ S(\omega) : \omega \in
H^1\A_{X_{1},\ldots,X_{p}}(P), \omega \sim_{\E} g\}.
\end{equation}

2. (Closed) The following equality holds:
\begin{equation}\label{eqn:id rate c}
I^\E_{z}(g)=\frac{1}{2} \inf\{S(\omega) : \omega \in H^1\A(P), \omega\sim_{\E}g\}.
\end{equation}

In both cases, the infima are attained by an element of $W^{1,\infty}\A(P)$, hence continuous 
and Lipschitz on $M$, which is smooth outside $\bigcup_{e\in \E}e$.
\end{proposition}

In (\ref{eqn:id rate b}) and (\ref{eqn:id rate c}), the fact that the left hand side is smaller
than the right hand side is a simple consequence of the energy inequality (Proposition \ref{energieh1}).
Besides, the fact that the infima are attained by $H^1$ connections is a consequence of Uhlenbeck's
theorem (Theorem \ref{Uh}). Indeed, from a minimizing sequence one can extract a weakly convergent
one and closed constraints on the holonomy are stable by weak $H^1$ limits. Proving that the minimum
is equal to the left hand side is the difficult part. We do this by constructing an explicit
minimizer. This is
rather long and we postpone the construction until Section \ref{section:construction}. Let us state
the result here.
\begin{proposition} \label{prop:exists min}
Let $\G$ be a graph on $M$. Let $g$ be an element of $\GE$. 

1. (Boundary) Assume that $h_{N_{1}}(g)=X_{1}, \ldots, h_{N_{p}}(g)=X_{p}$. Then there exists a
connection $\omega\in W^{1,\infty}\A_{X_{1},\ldots,X_{p}}(P)$ which is smooth outside
$\bigcup_{e\in\E}e$ such that $\omega\sim_{\E} g$ and $S(\omega)=2I^\E_{X_{1},\ldots,X_{p}}(g)$.

2. (Closed) There exists a connection $\omega\in W^{1,\infty}\A(P)$ which is smooth outside
$\bigcup_{e\in \E}e$ such that $\omega\sim_{\E}g$ and $S(\omega)=2I^\E_{z}(g)$.
\end{proposition}

Let us  give briefly an idea of what a minimizing connection looks like. The key to the construction
is that minimizers of the energy on disks with prescribed holonomy along the boundary are well-known.
A connection on a face $F$ with holonomy $x$ along the boundary and minimal energy is
gauge-equivalent  to a connection of the form $X \lambda$, where $X$
is an element of $\LG$ of minimal norm such that $\exp(\sigma(F)X)=x$ and $\lambda$ is a smooth 1-form
such that $d\lambda = \sigma$. We construct a minimizing connection on $M$ essentially by taking
one such minimizer on each face and gluing them all together.\\

\ppf{prop:identify rate 1} 1. Let $\omega$ be an $H^1$ connection which satisfies the boundary
conditions and agrees with $g$ on $\E$. Then, the energy inequality (Proposition \ref{energieh1})
applied on each face of $\G$ implies that $S(\omega)\geq 2I^\E_{X_{1},\ldots,X_{p}}(g)$. The reverse
inequality follows from Proposition \ref{prop:exists min}.

2. Let $\omega$ be an $H^1$ connection on $P$ which agrees with $g$ on $\E$. It is not enough to
apply the energy inequality to $\omega$ in this case. Instead, let us choose for each face $F$ a
smooth section $s_{F}$ of $P$ over $F$. Let us choose a face $F$. The form $\omega_{F}$ belongs to
$H^1\Omega^1_{\lg}(F)$. Let us compute $\langle \omega_{F},\partial F\rangle_{\wG}$, which is a
conjugacy class in $\wG$ and which projects on the conjugacy class $\langle \omega,\partial F
\rangle$ of $G$. The energy inequality applied on each face with the structure group $\wG$ gives us 
$$ S(\omega)\geq \sum_{F\in\F} \frac{\tilde\rho(\langle \omega_{F},\partial
F\rangle_{\wG})^2}{\sigma(F)}. $$
We claim that the right hand side of this inequality is of the form $\sum_{F\in\F} \frac{\tilde
\rho(h^\wG_{\partial F}(\tilde g)z_{F})^2}{\sigma(F)}$ for some $\tilde g\in \wGE$ and some
$z_{\F}\in \Pi^\F_{z}$. 

Recall that, if $e$ is an edge, $L(e)$ is the face located on the left of $e$. For each
$e\in\E^+$, set $\tilde g_{e}=\langle \omega_{L(e)},e\rangle_{\wG}$. Then $\tilde g=(\tilde
g_{e},e\in \E^+)$ belongs to $\wGE$ and satisfies $\pi(\tilde g)=g$. Finally, for each face $F$,
set 
$$z_{F}=\prod_{e\in\E^+ : L(e^{-1})=F}\langle \omega_{L(e)} , e\rangle_{\wG} \langle
\omega_{L(e^{-1})},e^{-1}\rangle_{\wG}.$$ 
Then it follows from \cite{Levy_AMS}, Lemma 1.7, that
$z_{\F}=(z_{F},F\in\F)$ belongs to $\Pi^\F_{z}$. On the other hand, $z_{\F}$ is defined in such a
way that, for each face $F$, $h^\wG_{\partial F}(\tilde g)z_{F}=\langle \omega_{F},\partial
F\rangle_{\wG}$. Our claim is thus proved and it follows that $S(\omega)\geq 2I^\G_{z}(g)$.  The
reverse inequality follows as in the case with boundary from Proposition \ref{prop:exists min}. \qed

\subsection{Holonomy along a family of paths in a graph (Contraction principle)}

The contraction principle \cite[Thm. 4.2.1]{DZ} allows us to state a large deviation principle for
the law of the random holonomy along a finite set of paths in a graph.

\begin{proposition}\label{fdg}
 Let $\G$ be a graph on $M$. Let $J=\{c_{1},\ldots,c_{n}\}$ be a finite subset
of $\E^*$. 

1. (Boundary) The laws of $(H_{c_{1}},\ldots,H_{c_{n}})$
under $P^\G_{T;X_{1},\ldots,X_{p}}$ satisfy as $T$ tends to $0$ a large deviation principle on
$\M(J,G)$ with rate function
$$ I^J_{X_{1},\ldots,X_{p}}(g)=\frac{1}{2}\inf\{S(\omega): \omega\in H^1\A_{X_{1},\ldots,X_{p}}(P),
\omega \sim_{J} g\}.$$

2. (Closed) The laws of $(H_{c_{1}},\ldots,H_{c_{n}})$ under $P^\G_{T,z}$ satisfy as $T$ tends to
$0$ a large deviation principle on $\M(J,G)$ with rate function
$$ I^J_{z}(g)=\frac{1}{2}\inf\{S(\omega): \omega\in H^1\A(P), \omega\sim_{J}g\}.$$
\end{proposition}

\pf The proof is exactly the same whether or not $M$ has a boundary. We write the proof when $M$
is closed. Changing the names of the probabilities and rate functions gives the proof in the case
with boundary.

2. The mapping $h_J=(h_{c_1},\ldots,h_{c_n}):\GE \lra G^J\simeq G^n$ is
continuous. Hence, by contraction of the large deviation principle on $\GE$, the laws of
$(H_{c_1},\ldots,H_{c_n})$ under $P^\G_{T;z}$ satisfy a large deviation principle
on $G^J$ with rate function 
$$\tilde I^J_{z}(g)=\inf\{I^\E_{z}(k) : k\in \GE, h_J(k)=g\}.$$

We claim that this large deviation principle holds on the smaller space $\M(J,G)\subset G^J$.  
Indeed, $h_J(k)=g$ implies $g\in \M(J,G)$. Hence, the support of $\tilde I^J_{z}$ is contained in
the closed subset $\M(J,G)$ of $G^J$, which supports the laws of $(H_{c_1},\ldots,H_{c_n})$ under
$P^\G_{T,z}$. The claim follows by \cite{DZ}, Lemma 4.1.5. 

Now, on one hand, $h_J(k)=g$ implies $k\sim_J g$. On the other hand, $k\sim_J g$ implies that there
exists $j\in \f(M,G)$ such that $h_J(j\cdot k)=g$. Since $I^\E_{z}$ is gauge-invariant, we get the
following expression for $\tilde I^J_{z}$: 
\begin{eqnarray*}
\tilde I^J_{z}(g) &=& \inf\{I^\E_{z}(k):k\in \GE, k \sim_J g\}\\
&=& \frac{1}{2} \inf \{ S(\omega) : \omega\in H^1\A(P), \exists k \in \GE ,\omega \sim_\E k, k\sim_J g \}.
\end{eqnarray*}
It appears that $\tilde I^J_{z}(g)$ is the infimum of $S$ over a smaller set of connections than
$I^J_{z}(g)$, so that $\tilde I^J_{z} \geq I^J_{z}$.

Now, take $g\in G^J$ for which $I^J_{z}(g)<\infty$, that is, such that there exists an $H^1$
connection on $P$ which agrees with $g$ on $J$. For every $\alpha >0$, there exists a connection
$\omega_\alpha$ which agrees with $g$ on $J$ and such that $\frac{1}{2}S(\omega_\alpha) \leq
I^J_{z}(g)+\alpha$. This connection induces a certain class of $\GE/\f(M,G)$. Let $k_{\alpha}$ be
an element of this class. Then $\omega_{\alpha}$ agrees tautologically with $k_{\alpha}$ on $\E$ and
 $k_\alpha\sim_J g$. Hence, $\tilde I^J_{z}(g) \leq \frac{1}{2}S(\omega_\alpha) \leq
I^J_{z}(g)+\alpha$. By letting $\alpha$ tend to 0, we get $\tilde I^J_{z} \leq I^J_{z}$. \qed

For the last part of the proof, we could also have argued that the infimum defining the function
$I^J_{z}$ is attained, as a consequence of Uhlenbeck's compactness theorem. Then, a minimizer
agrees on $\E$ with a certain configuration in $\GE$  and the inequality $\tilde I^J_{z} \leq
I^J_{z}$ follows.

\subsection{Holonomy along arbitrary finite families of paths (Exponential approximation)}

The last step at the finite-dimensional level is to prove that Proposition \ref{fdg} holds for an
arbitrary finite subset $J$ of $PM$.  To do this, we use an exponential approximation result. 

Fix $J=\{p_1,\ldots,p_n\}$ an arbitrary finite subset of $PM$.

\begin{lemma} \label{niceapprox}
There exist $n$ sequences $(c^m_i)_{m\geq 1}$, $i=1\ldots n$, of paths such that the following
properties hold.

0. For each $i=1\ldots n$, $(c^m_{i})_{m\geq 1}$ converges to $p_{i}$ with fixed endpoints.

1. For all $m\geq 1$, there exists a graph $\G_m$ such that $c^m_1,\ldots,c^m_n$ belong to $\E_m^*$. 

2. For all $\delta >0$, 
$$\lim_{m\to \infty} \limsup_{T\to 0} \; T \log P_T \left[\max_{1\leq
i \leq n} d(H_{c_i^m},H_{p_i}) > \delta\right] = -\infty,$$
where $P_{T}$ stands either for $P_{T;X_{1},\ldots,X_{p}}$ or for $P_{T,z}$.

3. For every $H^1$ connection $\omega$, in $H^1\A_{X_{1},\ldots,X_{p}}(P)$ if $M$ has a boundary, or
in $H^1\A(P)$ if $M$ is closed,
$$\rho(\langle \omega, c^m_{i}p_{i}^{-1}\rangle) \leq C [S(\omega)
d_\ell(c^m_i,p_i)]^\frac{1}{2}$$
 for some constant $C$ independent of $\omega$.
\end{lemma}

\pf Let us begin with the case where $p_1,\ldots,p_n$ are edges. It is proved in
\cite[Section 2.5.3]{Levy_AMS} that we can find sequences of piecewise geodesic paths
$(c^m_i)_{m\geq 1}$, $i=1\ldots n$, converging to $p_1,\ldots,p_n$ with fixed endpoints, such that
$c^m_i p_i^{-1}$ bounds for each $i$ a domain diffeomorphic to a disk, of arbitrarily small area. Let
$\sigma_i^m$ denote this area. We assume that $\sigma^m_{i}<\frac{1}{2}\sigma(M)$ for each $i$ and
each $m$. If $M$ has a boundary, choose $x_{1}\in X_{1},\ldots,x_{p}\in X_{p}$.
Then, for each bounded non-negative measurable function $f$,
$E_{T;X_{1},\ldots,X_{p}}[f(d(H_{p_i},H_{c^m_i}))]$ is equal to
\begin{eqnarray*}
&& \hskip -1cm \frac{1}{Z_{T;X_{1},\ldots,X_{p}}}\int_{G^{2g+p+1}} f(x)
p_{T\sigma^m_{i}}(x) p_{T(\sigma(M)-\sigma^m_{i})}(x^{-1}\prod_{k=1}^g[a_{k},b_{k}]\prod_{l=1}^py_{l}^{-1}
x_{l}y_{l})\\
&&\hskip 8cm da_{1}db_{1}\ldots da_{g}db_{g}dy_{1}\ldots dy_{p}\;dx\\
&& \leq \frac{1}{Z_{T;X_{1},\ldots,X_{p}}}\|p_{\frac{T}{2}\sigma(M)}\|_{\infty} \int_{G} f(x) p_{T\sigma^m_{i}}(x)\; dx.
\end{eqnarray*}
If $M$ is closed, then $E_{T,z}[f(d(H_{p_i},H_{c^m_i}))]$ is equal to
\begin{eqnarray*}
&&\frac{1}{Z_{T,z}}\int_{G^{2g+1}}f(x) \sum_{w\in \Pi} \tilde p_{T\sigma^m_{i}}(\tilde x w) \tilde
p_{T(\sigma(M)-\sigma^m_{i})}(\tilde x^{-1} \prod_{k=1}^g[\widetilde{a_{k},b_{k}}]w^{-1}z)\;
da_{1}db_{1}\ldots da_{g}db_{g}\;dx \\
&&\hskip 1cm \leq \frac{1}{Z_{T,z}}\|\tilde p_{\frac{T}{2}\sigma(M)}\|_{\infty} \int_{G} f(x)
p_{T\sigma^m_{i}}(x)\; dx.
\end{eqnarray*}
In both cases, by Proposition \ref{prop:Z} and the estimate of the heat kernel given for example
in \cite{Varopoulos}, Theorem V.4.3, we get, for some constant $C$ depending only on $G$ and for
every $\delta >0$, 
$$P_T[d(H_{c^m_i},H_{p_i})>\delta] \leq C (T\sigma^m_i)^{-\frac{\dim G}{2}}
e^{\frac{C\mathrm{diam}(G)^2}{T}-\frac{\delta^2}{CT\sigma^m_i}}.$$
Hence, 
\begin{eqnarray*}
\limsup_{T\to 0} \; T \log P_T\left[\max_{1\leq i\leq n} d(H_{c_i^m},H_{p_i}) > \delta\right] &\leq &
\limsup_{T\to 0} \; T \log \sum_{i=1}^n P_T[d(H_{c^m_i},H_{p_i})>\delta] \\
&& \hskip -2.5cm = \max_{1\leq i \leq n} \limsup_{T\to 0} \; T \log
P_T[d(H_{c^m_i},H_{p_i})>\delta]\\
&& \hskip -2.5cm \leq C\mathrm{diam}(G)^2-\min_{1\leq i \leq n}
\frac{\delta^2}{C\sigma^m_i}.
\end{eqnarray*}
Since the sequences $(\sigma^m_i)_m$ converge to 0, the limit as $m$ tends to infinity of this expression is equal to $-\infty$. 

Let $\omega$ be an $H^1$ connection. By the energy inequality (Proposition \ref{ineg fine}),
$$\rho (\langle \omega, c^m_i p_i^{-1}\rangle)^2 \leq \sigma^m_i S(\omega).$$
Now, the domain bounded by $p_i^{-1} c^m_i$ is contained in a tube around $p_i$ of
width $d_\infty(p_i,c^m_i) \leq d_\ell(p_i, c^m_i)$. Thus, there exists a constant $K$, depending on
the paths $p_1,\ldots,p_n$, such that $\sigma^m_i \leq K d_\ell(p_i,c^m_i)$. Finally, we get
$$\rho (\langle \omega, c^m_i p_i^{-1}\rangle)^2 \leq K d_\ell(p_i,c^m_i) S(\omega)$$ 
and property 3 holds.

If the paths $p_1,\ldots,p_n$ are not edges, let us write them in some way as concatenations of
edges. Let $\{q_1,\ldots,q_r\}$ be the set of distinct edges that have been used in at least one of
the decompositions. We  apply the arguments above to this new set of paths. We find $r$ sequences
$(d^m_j)_{m\geq 1}$, $j=1\ldots r$ satisfying properties 0-3. Let us make the further assumption that
$\ell(d^m_j)\geq \ell(q_j)$ for all $j$ and $m$. 

Assume for instance that $p_1=q_{i_{1}} \ldots q_{i_{s}}$, where $1\leq i_{1},\ldots,i_{s} \leq r$.
Then, for all $m\geq 1$, 
$$d(H_{p_1},H_{d^m_{i_{1}} \ldots d^m_{i_{s}}}) \leq \sum_{j=1}^s d(H_{q_{i_{j}}}, H_{d^m_{i_{j}}}).$$
 Set $c^m_1=d^m_{i_{1}} \ldots d^m_{i_{s}}$ and define the others $c^m_i$, $i=2\ldots n$ in a similar
fashion. Let $N$ be the largest number of non necessarily distinct edges that it is necessary to
concatenate in order to get one of the paths $p_i$. We have 
$$P_T\left[\max_{1\leq i \leq n} d(H_{c_i^m},H_{p_i}) > \delta\right] \leq P_T\left[\max_{1\leq j
\leq r} d(H_{d^m_j},H_{q_j}) > \frac{\delta}{N}\right]$$
and property $2$ follows. 

Let us prove property $3$ for $i=1$. By applying the special case of property 3
that we have proved above to $q_{i_{1}},\ldots,q_{i_{s}}$, we find a constant $K'$ such that
\begin{eqnarray*} 
\rho(\langle \omega, c^m_{1}p_{1}^{-1}\rangle)^2 &\leq& \left(\sum_{j=1}^s \rho(\langle
\omega,d^m_{i_{j}} q_{i_{j}}^{-1}\rangle)\right)^2 \\
&\leq & K' S(\omega) s \sum_{j=1}^s d_\ell(d^m_{i_{j}},q_{i_{j}}). \\
\end{eqnarray*}
Since $\ell(q_i)\leq \ell(d^m_i)$, we have for each $j$ the inequality
$|\ell(d^m_{i_{j}})-\ell(q_{i_{j}})| \leq |\ell(c^m_i)-\ell(p_i)|$ and the last
term is bounded above by $K' s^2 S(\omega) d_\ell(c^m_1, p_1).$
Finally, $s\leq N$ and property 3 follows. \qed

\begin{proposition} Let $J=\{p_1,\ldots,p_n\}$ be a finite subset of $PM$. 

1. (Boundary) The laws of $(H_{p_1},\ldots,H_{p_n})$ under $P_{T;X_{1},\ldots,X_{p}}$, satisfy, as
$T$ tends to 0, a large deviation principle on $\M(J,G)$ with rate function 
$$ I^J_{X_{1},\ldots,X_{p}}(g)=\frac{1}{2}\inf\{ S(\omega) : \omega \in
H^1\A_{X_{1},\ldots,X_{p}}(P), \omega \sim_J g.\}. $$

2. (Closed) The laws of $(H_{p_1},\ldots,H_{p_n})$ under $P_{T,z}$, satisfy, as $T$ tends to 0, a
large deviation principle on $\M(J,G)$ with rate function
$$ I^J_{z}(g)=\frac{1}{2}\inf\{ S(\omega) : \omega \in H^1\A(P), \omega \sim_J g.\}. $$
\end{proposition}

\pf Here again, the proof is exactly the same with and without boundary. We drop the subscripts
that usually indicate in  which context we are.

Let $(c^m_i)_{m\geq 1}$, $i=1\ldots n$, be given by Lemma \ref{niceapprox}. For each $m\geq 1$,
denote by $J_m$ the set of paths $\{c^m_1,\ldots,c^m_n\}$. By Proposition \ref{fdg}, the laws of
$(H_{c^m_{1}},\ldots,H_{c^m_{r}})$ under $P_{T}$ satisfy a large deviation principle 
in $G^J$ with rate function $I^{J_{m}}$.

By a standard result on exponential approximations \cite[Theorem 4.2.16]{DZ}, property 2 of Lemma
\ref{niceapprox} ensures that the laws of $(H_{p_1},\ldots,H_{p_n})$ under $P_T$, $T>0$ satisfy a
large deviation principle on $G^J$ with rate function 
$$\hat I^J(g)=\sup_{\delta>0} \liminf_{m\to\infty} \inf_{h\in B(g,\delta)} I^{J_{m}}(h).$$

Here, $B(g,\delta)$ denotes the open ball of radius $\delta$ around $g$ in $G^J$. The proof is
completed by the next lemma, after noticing that the domain $\{I^J<+\infty\}$ is contained in the
closed subset $\M(J,G)$ of $G^J$. \qed

\begin{lemma} We keep the preceding notation. Then
$$\sup_{\delta>0} \liminf_{m\to\infty} \inf_{h\in B(g,\delta)}
\inf_{\omega \sim_{J_m} h} S(\omega) = \inf_{\omega\sim_J g} S(\omega).$$
In the case with boundary, the two last infima are taken over all $H^1$ connections which satisfy the
boundary conditions. In the closed case, they are taken over all $H^1$ defined on a principal
$G$-bundle over $M$ which belongs to the correct isomorphism class. 
\end{lemma} 

\pf  For each $m\geq 1$ and $\delta>0$, define the set 
$$O_{m,\delta}=\{ \omega \in H^1\A : \exists h \in B(g,\delta), \omega \sim_{J_m} h \}.$$
Since the holonomy along a fixed path depends continuously on the connection in the $H^1$ topology,
these are open subsets of $H^1\A_{X_{1},\ldots,X_{p}}(P)$ or $H^1\A(P)$.  

Now, let $\omega$ be an $H^1$ connection such that $\omega\sim_J g$. According to Proposition
\ref{prop:1-1}, the holonomy induced by $\omega$ is continuous  with fixed endpoints. Hence, for
every $\delta>0$, $\omega$ belongs to $O_{m,\delta}$ for $m$ large enough.

Choose $\alpha >0$ and an $H^1$ connection $\omega_0$  such that $\omega_0\sim_J g$ and such that
$S(\omega_0) \leq \inf\{S(\omega) : \omega\sim_J g\} + \alpha$. Choose $\delta >0$. By the
observation above, 
$$ \liminf_{m\to\infty} \inf_{O_{m,\delta}} S \leq S(\omega_0).$$
By letting $\alpha$, then $\delta$ tend to 0, we get the inequality $\hat I^J \leq I^J$.  

Assume that this inequality is strict. Then, for some $\alpha >0$, for all $\delta >0$, one has 
$$\liminf_{m\to \infty} \inf_{O_{m,\delta}} S \leq \inf_{\omega\sim_J g} S(\omega)- \alpha = s-\alpha,$$
where we have set $s=\inf_{\omega\sim_J g}S(\omega)$.  Let us fix $\delta>0$. We can construct an
increasing sequence of integers $(m_{k})_{k\geq 1}$ and a sequence of $H^1$ connections
$(\omega_{m_k})_{k\geq 1}$ with $\omega_{m_k} \in O_{m_k,\delta}$ and $\sup_k S(\omega_{m_k})
\leq s-\alpha/2$. From this sequence with bounded energy, we can, by Uhlenbeck's theorem, extract a
subsequence which is gauge-equivalent to a weakly convergent sequence of connections. Thus, there
exists a subsequence $(\omega_r)_{r\geq 1}$ of $(\omega_{m_k})_{k\geq 1}$, a sequence $(j_r)_{r\geq
1}$ in $H^2\J(P)$ and an $H^1$ connection $\omega$ such that $j_r \cdot \omega_r \rightharpoonup
\omega$ and $S(\omega) \leq s-\alpha/2$. Let $(N_r)_{r\geq 1}$ denote an increasing sequence such
that $\omega_r \in O_{N_r,\delta}$.

For every path $c$, the holonomy along $c$ of $j_r \cdot \omega_r$ converges to that of $\omega$.
This holds in particular for the paths $p_{1},\ldots,p_{n}$. Moreover, since $S(\omega_r)$ is
bounded independently of $r$, by property $3$ of Lemma \ref{niceapprox}, the
distance $\rho(\langle j_{r}\cdot\omega_{r}, c^{N_{r}}_{i}p_{i}^{-1}\rangle)$ tends
to 0 as $r$ tends to infinity. Hence, for each $i=1\ldots n$, the holonomy of $j_{r}\cdot \omega_{r}$
along $c^{N_{r}}_{i}$ converges as $r$ tends to infinity to the holonomy of $\omega$ along $p_{i}$.
Hence, $\omega \in O_{\infty,2\delta}$, where we take the convention $J_\infty=J$.

For every $\delta>0$, we are thus able to construct a connection $\omega$ in $O_{\infty,2\delta}$
such that $S(\omega) \leq s-\alpha/2$. For each $n\geq 1$, let us do this construction with
$\delta=1/2n$. This produces a sequence $(\omega_n)$, from which we may again extract a subsequence
gauge-equivalent to a weakly $H^1$ convergent one, with limit $\omega_*$. This limit satisfies both
$S(\omega_*) \leq s-\alpha/2$ and $\omega_* \sim_J g$. This contradicts the definition of $s$. \qed 

\subsection{The Yang-Mills measures (Projective limit)}

As explained in \cite{Levy_AMS}, Section 2.10.2, the probability space $(\M(PM,G),\C,P_T)$ is the
projective limit of the spaces $(\M(J,G),\C_J,P_T^J)$, where $J$ spans the set of finite subsets of
$PM$ and $P_T^J$ denotes the distribution of the holonomy along the paths of $J$ under $P_T$. A
straightforward application of Dawson-G\"artner's theorem (\cite{DZ}, Theorem 4.6.1) gives the
following result. 

\begin{proposition} 1. (Boundary) The probability measures $(P_{T;X_{1},\ldots,X_{p}})_{T>0}$
satisfy, as $T$ tends to 0, a large deviation principle on $\M(PM,G)$ with rate function
$$ \tilde I^\YM_{X_{1},\ldots,X_{p}}(f)=\sup_{J\subset PM, |J|<\infty}
I^J_{X_{1},\ldots,X_{p}}(f)=\frac{1}{2} \sup_{J\subset PM, |J|<\infty} \inf \{S(\omega):
\omega \in H^1\A_{X_{1},\ldots,X_{p}}(P), \omega\sim_{J}f \}. $$

2. (Closed) The probability measures $(P_{T,z})_{T>0}$ satisfy, as $T$ tends to 0, a large
deviation principle on $\M(PM,G)$ with rate function 
$$ \tilde I^\YM_{z}(f)=\sup_{J\subset PM, |J|<\infty} I^J_{z}(f)=\frac{1}{2} \sup_{J\subset PM,
|J|<\infty} \inf \{S(\omega): \omega \in H^1\A(P), \omega\sim_{J}f \}. $$
\end{proposition}

The proof of Theorems \ref{thm:b} and \ref{thm:c} will be complete after we have proved that
$\tilde I^\YM=I^\YM$ in both cases. Since the proof of this equality is the same with and without
boundary, we drop again the subscripts. Let us start with an easy lemma.

\begin{lemma} \label{lem:facile}
The inequality $\tilde I^\YM\leq I^\YM$ holds.
\end{lemma}

\pf Let $f\in \M(PM,G)$. If $I^\YM(f)=+\infty$, then it is certainly true that $\tilde I^\YM(f)\leq
I^\YM(f)$. Let us assume that $I^\YM(f)<\infty$. Then there exists an $H^1$ connection $\omega$
which agrees with $f$ on $PM$. In particular, it agrees with $f$ on $J$ for every subset $J$ of
$PM$. Hence, $\tilde I^\YM(f)\leq I^\YM(f)$. \qed

Let us define another function $\hat I^\YM:\M(PM,G)\lra [0,+\infty]$ by
$$ \hat I^\YM(f)=\sup_{\G \; {\rm graph}} I^\E(f). $$
Since $\hat I^\YM$ is a supremum over a smaller class of subsets of $PM$ than $\tilde I^\YM$,
the inequality $\hat I^\YM \leq \tilde I^\YM$ holds. According to Lemma \ref{lem:facile}, it is
enough to prove that $\hat I^\YM=I^\YM$, or even that $I^\YM\leq \hat I^\YM$.

\begin{proposition} The inequality $I^\YM\leq \hat I^\YM$ holds on $\M(PM,G)$.
\end{proposition}

\pf Consider $f\in \M(PM,G)$. Assume that $\hat I^\YM(f)<\infty$, otherwise there is nothing to
prove. Let $l$ be a simple loop in $PM$. There exists a graph, say $\G$, such that $l$ belongs to
$\E^*$. Assume that $l$ bounds a domain $V$ diffeomorphic to an open disk. Then we may assume that
$V$ is a face of $\G$. 

If $M$ has a boundary, then by definition of $\hat I^\YM$,
\begin{equation}\label{eqn:elle est cont}
\rho(f(l))^2=\rho(f(\partial V))^2\leq 2\hat I^\YM(f)\sigma(V).
\end{equation} 
If $M$ is closed, we can only say that there exists a lift $\widetilde{f(l)}$ of $f(l)$ to $\wG$ and
$z\in \Pi$ such that $\tilde\rho(\widetilde{f(l)}z)^2\leq 2\hat I^\YM(f) \sigma(V)$. However, notice
that, if $x\in G$ and $\tilde x\in \wG$ satisfy $\pi(\tilde x)=x$, then $\rho(x)=\min_{z\in \Pi}
\tilde\rho(\tilde x z)$. So, (\ref{eqn:elle est cont}) holds even when $M$ is closed. In both
cases, Proposition \ref{continuite} allows us to deduce that $f$ is continuous with fixed endpoints. 

Now let $m$ be a point in $M$. There exists a countable dense subset of the space
$L_{m}M$ of loops based at $m$. For example, consider a countable dense subset of $M$. Then the
set $\Lambda_{m}M$ of piecewise geodesic loops based at $m$ and joining a finite number of these
points is countable and dense in $L_{m}M$. Let $(\zeta_{n})_{n\geq 1}$ be a sequence which exhausts
$\Lambda_{m}M$. For each $n\geq 1$, there exists a graph, say $\G_{n}$, such that $\zeta_{i}\in
\E_{n}^*$ for each $i=1\ldots n$. 

For each $n\geq 1$, let $\omega_{n}$ be an $H^1$ connection which agrees with $f$ on $\G_{n}$ and
such that $S(\omega_{n})=2I^{\E_{n}}(f)$. Such a connection exists by Proposition \ref{prop:exists
min}. Uhlenbeck's compactness theorem allows us to extract a weakly convergent subsequence of
the sequence $(\omega_{n})_{n\geq 1}$, up to gauge transformations. The weak limit of this
subsequence has an energy at most equal to $2\hat I_{YM}(f)$ and it agrees with $f$ on
$\zeta_1,\ldots,\zeta_n$ for each $n$. Since $G$ is compact, this implies that $\omega$ and $f$ agree
on $\Lambda_{m}M$. Finally, since both are continuous on $L_{m}M$, they agree on $L_{m}M$ itself. As
pointed out in the remark \ref{rem:loops}, this is equivalent to saying that $\omega$ and $f$ agree
on $PM$. 
 
Since there exists an $H^1$ connection $\omega$ which agrees with $f$, $I^\YM(f)$ is finite. It is
equal to $\frac{1}{2}S(\omega)$ and we have observed that $\frac{1}{2}S(\omega) \leq \hat
I^\YM(f)$. The result is proved. \qed 

\section{Connections that minimize the Yang-Mills energy under holonomy constraints} 
\label{section:construction}
\subsection{The main result}

The purpose of this section is to prove Proposition \ref{prop:exists min}. We deduce it from the
next proposition, in which we assume that $M$ is closed. Once for all, let us choose a simple graph
$\G=(\V,\E,\F)$ on $M$. Let $\E^+$ be an orientation of $\V$, which satisfies
the properties explained in Lemma \ref{lem:orientation}. 

\begin{proposition} \label{prop:mini gene} Let $g=(g_{e})_{e\in\E^+}$ be an element of $\GE$. Let $\tilde
g=(\tilde g_{e})_{e\in\E^+}$ be an element of $\wGE$ such that $g=\pi(\tilde g)$. Let $z$ be an
element of $\Pi$. Let $z_{\F}=(z_{F})_{F\in\F}$ be an element of $\Pi^\F_{z}$. There exists a
principal
$G$-bundle $P$ over $M$ and a connection $\omega$ on $P$ such that the following properties hold.

1. $\oo(P)=z$.

2. $\omega$ belongs to $W^{1,\infty}\A(P)$ and, for each open face $F$ of $\G$, the
restriction of $\omega$ to $P_{|F}$ is smooth.

3. $\omega$ and $g$ agree on $\E$.

4. For each face $F$ of $\G$, $\displaystyle S_{F}(\omega)=\frac{\tilde \rho (h^\wG_{\partial
F}(\tilde g)z_{F})^2}{\sigma(F)}$.
\end{proposition}

Let us explain how this result implies Proposition \ref{prop:exists min}. \\

\ppf{prop:exists min} 1. Let $\overline{M}$ be a closure of $M$, that is, a closed surface in
which $M$ is embedded in such a way that $\overline{M}\backslash M$ is a disjoint
union of $p$ disks. Then $\G$ is still a simple graph on $\overline{M}$, it only has $p$ more faces.
Let $\tilde g\in \wGE$ be such that $\pi(\tilde g)=g$. For each face $F$ of $\G$ contained in $M$,
let $z_{F}$ be an element of $\Pi$ such that $\tilde \rho(h^\wG_{\partial F}(\tilde
g)z_{F})=\rho(h_{\partial F}(g))$. Such a $z_{F}$ exists because, for each $\tilde x \in \wG$,
$\rho(\pi(\tilde x))=\min_{z\in\Pi}\tilde \rho(\tilde x z)$. For the faces of $\G$ not
contained in $M$, choose $z_{F}$ arbitrarily. Then set $z=\prod_{F} z_{F}$. Proposition
\ref{prop:mini gene} produces a bundle $Q$ over $\overline{M}$ and a connection $\eta$ on $Q$.
A bundle $P$ is given by assumption over $M$. Since $M$ has a boundary, $P$ is trivial, and so is
the restriction of $Q$ to $M$. Let  $\varphi:P\lra Q_{|M}$ be a bundle isomorphism. Set
$\omega=\varphi^*\eta$.  Then, since $\omega$ agrees with $g$ on $\E$ and $g$ satisfies the
correct boundary conditions, $\omega$ belongs to $W^{1,\infty}\A_{X_{1},\ldots,X_{p}}(P)$. It is
smooth outside $\cup_{e\in \E} e$ because $\eta$ is. Finally, the choice of $(z_{F})_{F\in\F}$
guarantees that $S(\omega)$ is equal to $I^\E_{X_{1},\ldots,X_{p}}(g)$. 

2. In this case, the result is straightforward: it suffices to choose $\tilde g \in \wGE$ such
that $\pi(\tilde g)=g$ and then $z_{\F}$ which minimizes the right hand side of (\ref{eqn:rate closed
1}). \qed

\subsection{An open covering of $M$}

We begin by constructing an open covering of $M$ which is nicely adapted to $\G$. Recall
that $M$ is endowed with a Riemannian metric. This metric allows us to define tubular neighbourhoods
around embedded submanifolds of $M$, see for example \cite{Gray}. 

Up to this point, we have always called {\it faces} the closed faces of $\G$. In this section, we
change this convention and decide to call faces the {\it open faces} of $M$.

\begin{lemma} \label{recouvrement}
There exist two real numbers $R,L>0$ such that the following properties hold. 
\begin{enumerate}
\item The balls $B(v,R),v\in\V$ are diffeomorphic to disks and pairwise disjoint. Moreover, an edge
$e$ meets the ball $B(v,R)$ if and only if $v\in\{\un{e},\ov{e}\}$. In this case, $e\cap B(v,r)$ is
connected. 
\item Let $v\in \V$. Let $f_1,\ldots,f_k$ be the edges sharing $v$ as their starting point, indexed
in their cyclic order around $v$ induced by the orientation of $M$. It is possible to choose polar
normal coordinates $(r,\theta)$ in $B(v,R)$ such that there exist $k$ smooth functions
$\theta_1,\ldots,\theta_k:[0,R)\lra [0,2\pi) $ and $2k$ real numbers 
$0\leq \theta_1^-<\theta_1^+<\ldots<\theta_k^-<\theta_k^+<2\pi$, such that, for all $j=1\ldots k$,
$(r,\theta_j(r))$ is a parametrization of $f_j$ inside $B(v,R)$ and $\theta_j^- < \inf_{[0,R)}
\theta_j < \sup_{[0,R)} \theta_j < \theta_j^+.$ We call the sector $\{(r,\theta) : \theta_j^- <
\theta <\theta_j^+\}$ the angular sector of $f_j$ at $v$.

Moreover, let $(r,\theta)\mapsto \sigma(r,\theta)$ be the smooth density of the measure $\sigma$
with respect to $r dr d\theta$ on $B(v,R)$. Then any partial derivative of any order of $\sigma$ is
uniformly bounded on $D(0,R)$.

For each edge $e$, let $e^\circ$ denote the intersection of $e$ with the subset $M\backslash \bigcup_{v\in
\V} \overline{B(v,R/2)}$. Let $T_e$ denote the tubular neighbourhood of radius $L$ around $e^\circ$.

\item For each edge $e$, the tubular neighbourhood of radius $L$ around $e^\circ$ exists. It is
denoted by $T_{e}$. If $e,e'\in\E$ are not equal nor inverse of each other, then $T_{e}$ and $T_{e'}$ are disjoint.
\item Let $v$ be a vertex and $e$ an edge. The tube $T_e$ meets $B(v,R)$ only if $v$ is an endpoint
of $e$.  
\item Let $e$ be an edge. There exists a coordinate chart $T_{e} \lra (-3,3)\times (-1,1)$ with
coordinates $(x,y)$ such that $e\cap T_{e}=\{y=0\}$, $L(e)\cap T_{e}=\{y>0\}$, $B(\un{e},
3R/4)\cap T_{e}=\{x<-2\}$, $B(\un{e},R)\cap T_{e}=\{x<-1\}$, $B(\ov{e},R)\cap T_{e}=\{x>1\}$ and
$B(\ov{e},3R/4)\cap T_{e}=\{x>2\}$.

Moreover, let $(x,y)\mapsto \sigma(x,y)$ be the smooth density of the measure $\sigma$
with respect to $dxdy$ on $T_{e}$. Then any partial derivative of any order of $\sigma$ is uniformly
bounded on $(-3,3)\times (-1,1)$.
\end{enumerate}
\end{lemma}

\pf Let $R_{inj}$ be the injectivity radius of $M$, so that any ball of radius smaller than $R_{inj}$ is
diffeomorphic to a disk. Set $R_1=R_{inj} \wedge \frac{1}{2}\inf_{v\neq w} d(v,w)$, where the infimum
is taken over all pairs of distinct vertices. The balls $B(v,R), v\in \V$ are diffeomorphic to disks
and pairwise disjoint as soon as $R\leq R_1$.

Let us choose in each ball $B(v,R_1)$ a system of normal polar coordinates. This amounts to choosing
the initial speed of the geodesic of equation $\{\theta=0\}$. We choose it in such a way that no edge
starting from $v$ is tangent to this geodesic at $v$.

Let $v$ be a vertex and $e$ an edge such that $\un{e}=v$. Since $e$ is a segment of an embedded
submanifold, it can be parametrized near $v$ and inside $B(v,R_1)$ as $e(s)=(r(s),\theta(s))$, where
$s \geq 0$ and $e(0)=v$. Then $\dot r(0)>0$. Choose $s_v(e)>0$ such that $\dot r(s)>0$ for all $s \in
[0,s_v(e)]$. Set $R_2=\inf\{r(s_v(e)) : v\in \V, \un{e}=v\}$.  

Let $e$ be an edge. Set $\tilde e = e\cap (B(\underline{e},R_2)\cup B(\overline{e},R_2))^c$. Finally,
set $R_3=\frac{1}{2}\inf d(\tilde e, v)$, where the infimum is taken over all edges and all vertices.
Observe that $R_3 < R_2$.  Take $R\leq R_3$. Then the balls $B(v,R)$ satisfy the first point of the
lemma. Moreover, we can say that, if $e$ is incident to $v$, it crosses the circle of radius $R$
around $v$ transversally.

Let us consider a vertex $v$ and an edge $e$ incident at $v$. Once again, because $e$ is a segment of
an embedded submanifold, the local parametrization $(r(s),\theta(s))$ of $e$ defined above is such
that $s \mapsto \theta(s)$ can be extended by continuity at $s=0$. 

Let us denote by $f_1,\ldots,f_k$ be the edges starting at $v$, given in their cyclic order around
$v$, and $(r_1,\theta_1), \ldots, (r_k,\theta_k)$ their local parametrizations. We assume that the edges are
indexed in such a way that $0<\theta_1(0) < \ldots < \theta_k(0) <2\pi$. Set $\delta = \theta_1(0)
\wedge (2\pi - \theta_k(0)) \wedge \inf_{1\leq i \leq k-1} |\theta_{i+1}(0)-\theta_i(0)|$. Choose
$s'_v>0$ such that, for all $i=1\ldots k$ and all $s\in [0,s'_v]$, $|\theta_i(s)-\theta_i(0)|<
\delta/4$. Finally, set $R_4=\inf r_i(s'_v)$, where $v$ runs over $\V$ and $r_i$ is the local
parametrization of an edge incident to $v$. Then, any $R$ such that $0<R\leq R_4$ satisfies the two
first points of the lemma, except maybe for the boundedness condition on the derivatives of
$\sigma$. Let us choose $R=R_4/2$ to make sure that it holds. This allows us to define $e^\circ$ for
each edge $e$. In fact, let us temporarily consider $e^\bullet$ which is the larger portion of 
$e$ defined by $e^\bullet=e\cap (M\backslash \bigcup_{v\in \V} \overline{B(v,R/4)})$.

For each edge $e$, let $L_e$ be the largest width of a tube around $e^\bullet$. Set $L_1=\inf_e L_e
\wedge \frac{1}{2}\inf_{e\neq f} d(e^\bullet, f^\bullet)$. Then any positive $L$ smaller than $L_1$
satisfies the third point.  Now set $L_2=\frac{1}{2} \inf_{e,v} d(e^\bullet, B(v,R))$, where the
infimum runs over all pairs $(e,v)$ with $v\notin\{\un{e},\ov{e}\}$.  Any positive $L$ smaller than
$L_3=L_1\wedge L_2$ satisfies the fourth point.

Let $e$ be an edge. Let $T_e$ be a tube around $e^\bullet$, endowed with Fermi coordinates $(t,s)$,
such that $e^\bullet$ is defined by the equation $s=0$. Consider a curve $\gamma$ which crosses
$e^\bullet$ only once and transversally. Near its intersection point with $e^\bullet$, $\gamma$ can
be parametrized as $(t(\tau),s(\tau))$, with $\tau=0$ corresponding to the intersection point. Then,
$\dot s(0)\neq 0$. Hence, there exists a positive number $L'_e(\gamma)$ such that the portion of
$\gamma$ contained in a tube of width smaller than $L'_e(\gamma)$ is the graph in Fermi coordinates
of a smooth function $s\mapsto T(s)$, which crosses the boundary of the tube transversally. Then, the
domain $\{(s,t) : t \leq T(s)\}$ is diffeomorphic to a rectangle by a diffeomorphism which sends
$e^\bullet$ to a segment parallel to an edge.

By applying for each edge $e$ this argument to the circles of centers $\underline{e}$ and
$\overline{e}$ and radii $R$ and $3R/4$, we find a positive width $L'_e$ such that $L_4=L_3 \wedge
\inf_e L'_e$ satisfies the fifth and sixth points, excepts perhaps for the boundedness condition
of the derivatives of $\sigma$. For each edge $e$, the tube of radius $L=L_{4}/2$ around $e^\circ$ is
contained, as well as its closure, in the tube of radius $L$ around $e^\bullet$ and satisfies all the
required properties. \qed

Let $R$ and $L$ be two positive numbers given by this lemma. We define a collection of open
subsets of $M$ as follows. For each vertex $v$, set
$U_{v}=B(v,R)$. For each edge $e$, set $U_{e}=B({\un{e}},3R/4) \cup T_{e}\cup B(\ov{e},3R/4)$. For each face $F$, set
$U_{ \partial F}=\cup_{L(e)=F}U_{e}$ and $U_{F}=F\cup U_{\partial F}$. Finally, set
$U_{\G}=\cup_{F\in\F}U_{\partial F}=\cup_{e\in\E}U_{e}$. 

\subsection{Around the singularities}

The connections we are going to construct are singular on the subset $\cup_{e\in\E}e$ of $M$. This
set contains two kinds of points: the vertices of $\G$ and the points which are interior to an
edge. We treat these two cases separately. 

\begin{proposition} \label{prop:sommet}
Let $v$ be a vertex of $\G$. For each face $F$ such that $v$ lies on $\partial F$, let $X_{F}$
be an element of $\LG$. There exists $\omega \in W^{1,\infty}\Omega^1_{\lg}(U_{v})$ such that the
following properties hold.

1. If $e$ is an edge starting at $v$, then $\omega$ vanishes in all directions on
$U_{v}\cap e$.

2. If $F$ is a face such that $v$ lies on $\partial F$, then $\omega$ is smooth on $U_{v}\cap F$.
Moreover, there exists $\lambda \in \Omega^1(U_{v}\cap F)$ such  that $d\lambda=\sigma$ and
$\omega=X_{F}\lambda$ on $U_{v}\cap F$. In particular, the curvature of $\omega$ is equal to
$X_{F}\sigma$ on $U_{v}\cap F$.
\end{proposition}

\pf Let us denote, as in Lemma \ref{recouvrement}, by $f_1,\ldots,f_k$ be the edges starting at $v$,
given in their cyclic order around $v$, and $(r_1,\theta_1), \ldots, (r_k,\theta_k)$ their local
parametrizations. Pick an integer $i$ between $1$ and $k$. Set $X_i=X_F$, where $F$ is the face sitting between
$f_i$ and $f_{i+1}$, with the convention $f_{k+1}=f_1$. Let $\sigma(r,\theta)$ be the smooth positive
function on $D(0,R)$ such that $\sigma = \sigma(r,\theta) r dr d\theta$. Define, for $r \in [0,R)$,
$$ \beta_i(r)=X_i \int_{\theta_i(r)}^{\theta_{i+1}(r)} \sigma(r,\theta) \; d\theta $$
and
$$ \alpha_i(r)=\frac{2}{r^2} \int_0^r u \beta_i(u) \; du.$$
Let $\varphi_i:[0,2\pi) \lra [0,\infty)$ be a smooth nonnegative function
such that $\int_0^{2\pi} \varphi_i(\theta) \; d\theta = 1$ and such that the support of $\varphi_{i}$ 
is contained in $(\theta_i^+, \theta_{i+1}^-)$. Set
$$ a(r,\theta)=\varphi_i(\theta) \alpha_i(r) \; , \;\;\; \theta_i(r) \leq
\theta < \theta_{i+1}(r), $$
and define a function $b$ on $B(v,R)$ by setting $b(r,\theta_1(r))=0$ and 
$$ \frac{\partial b}{\partial \theta} = \sigma(r,\theta)X_i - \varphi_i(\theta) \beta_i(r) \; , \;\;\;
\theta_i(r) \leq \theta <\theta_{i+1}(r). $$

After doing this for each $i=1\ldots k$, we have defined two functions $a$ and $b$ on $U_{v}$.
Finally, set 
$$ \omega= \frac{r^2}{2} a(r,\theta) d\theta - rb(r,\theta)dr. $$
We claim that $\omega$ belongs to $W^{1,\infty}\A(U_{v})$. To see this, let us write $\omega$ in
Cartesian coordinates corresponding to our choice of polar coordinates:
$$ \omega= \left(-\frac{y}{2}a(r,\theta)-xb(r,\theta)\right) dx + \left(\frac{x}{2}a(r,\theta)-y
b(r,\theta)\right)dy. $$
Recall that the function $\sigma$ is bounded as well as its derivatives on $D(0,R)$. For each
$i=1\ldots k$, the function $\beta_i$ is continuous and bounded on $[0,R)$ and so is
$\alpha_i$. Both are also smooth on $(0,R)$. Hence, $a$ is bounded on $U_{v}$ and smooth except at
$v$. Observe that $\int_{\theta_i(r)}^{\theta_{i+1}(r)} \sigma(r,\theta) X_i - \varphi_i(\theta)
\beta_i(r)\; d\theta =0$, so that $b$ is continuous and bounded on $U_{v}-\{v\}$. It is also smooth
in $U_{v}$ outside $\G$. 

So far, we have proved that the components of $\omega$ belongs to $L^\infty$. Now, along any
segment in $U_{v}$ parallel to one of the coordinate axes and which does not contain $v$, $a$ is
smooth and $b$ is continuous and piecewise smooth with bounded derivative. In particular, both are
absolutely continuous and so are the components of $\omega$. Let us show that the almost-everywhere
defined derivatives of these components are uniformly bounded on $U_{v}$. For
example,
$$ \frac{\partial}{\partial x} (y a)(r,\theta) = \frac{2}{r^2}\int_0^r u\beta_i(u) \; du \left( -
\frac{y^2}{r^2} \varphi'(\theta)- \frac{2xy}{r^2} \varphi(\theta) \right) + \frac{2xy}{r^2}
\varphi_i(\theta)\beta_i(r), $$
so that
$$ \| \partial_x (ya) \|_{L^\infty} \leq 10 \sup_{1\leq i \leq k} \left( \| \beta_i \|_\infty (\|
\varphi_i \|_\infty + \| \varphi_i' \|_\infty)\right)< \infty. $$
A similar computation shows that the partial derivatives of $xa, ya, xb,yb$ are all uniformly bounded
on $U_{v}$. Finally, the components of $\omega$ are $L^\infty$, absolutely continuous along almost
all segments parallel to one of the coordinate axes inside $U_{v}$, with partial derivative belonging
to $L^\infty(U_{v})$. This implies that the components of $\omega$ belong to $W^{1,\infty}$.

The fact that $\omega$ vanishes along each edge starting at $v$ comes from the fact that $a$ vanishes
in the angular sector of each edge and $b(r,\theta_i(r))$ vanishes identically for each $i=1\ldots k$. 

Let $F$ be a face incident to $v$. By construction, $\omega$ is smooth inside $F$ and it can be
written as $X_F(v)$ times a real-valued $1$-form. It is readily checked that the differential of
this form is $\sigma$ itself: $a$ and $b$ have been designed for that purpose. The statement on the
curvature of $\omega$ follows immediately. \qed

\begin{proposition} \label{prop:arete}
Let $e$ be an edge of $\G$. For each face $F$ bounded by $e$, let $X_{F}$
be an element of $\LG$. There exists $\omega\in W^{1,\infty}\Omega^1_{\lg}(T_{e})$ such that the
following properties hold. 

1. The form $\omega$ vanishes in all directions on $T_{e}\cap e$.

2. If $F$ is a face bounded by $e$, then $\omega$ is smooth on $T_{e}\cap F$.
Moreover, there exists $\lambda \in \Omega^1(T_{e}\cap F)$ such  that $d\lambda=\sigma$ and
$\omega=X_{F}\sigma$ on $T_{e}\cap F$.

Let $\tilde g$ be an element of $\wG$. There exists a smooth function $j:T_{e} \lra \wG$ which is
identically equal to $1$
on $T_{e}\cap U_{\un{e}}$ and identically equal to $\tilde g^{-1}$ on $T_{e}\cap U_{\ov{e}}$. Then
the following properties hold.

3. $\langle j\cdot \omega, e^\circ \rangle_{\wG} = \tilde g$.

4. The form $j\cdot \omega$ vanishes in all directions on $T_{e}\cap(U_{\un{e}}\cup U_{\ov{e}})\cap
e$.

5. If $F$ is a face bounded by $e$, then the curvature of $j\cdot \omega$ is equal to $X_{F}\sigma$
on $T_{e}\cap F \cap U_{\un{e}}$, and it is equal to $\Ad(\tilde g)X_{F}\sigma $ on $T_{e}\cap F\cap
U_{\ov{e}}$.
\end{proposition}

\pf  Let $(x,y)\in (-3,3)\times (-1,1)$ be the local coordinates on $T_e$ given by Lemma
\ref{recouvrement}. Let
$\sigma(x,y)$ be the smooth positive function on $(-3,3)\times (-1,1)$ such that the equality $\sigma
= \sigma(x,y) dx dy$ holds. Set
$$\omega(x,y)=-(X_{L(e)} {\mathbf{1}}_{y\geq 0} + X_{L(e^{-1})} {\mathbf{1}}_{y\leq 0}) \left(
\int_0^y \sigma(x,t) \; dt \right) \; dx.$$
Since $(x,y)\mapsto \sigma(x,y)$ is bounded on $(-3,3)\times (-1,1)$ as well as its derivatives,
$\omega$ is Lipschitz, hence $W^{1,\infty}$. It is also smooth on $T_e$ outside $e$ and satisfies
property 2.

Let $\psi:(-3,3) \lra \wG$ be a smooth mapping such that $\psi(x)=1$ whenever $x\leq -1$ and
$\psi(x)=\tilde g^{-1}$ whenever $x\geq 1$. Finally, set $j(x,y)=\psi(x)$. Property 3, 4 and 5 are
straightforward.\qed

\subsection{On a neighbourhood of the graph}

We want to combine the two constructions presented above to get an element of $W^{1,\infty}
\Omega^1_{\lg}(U_{\G})$. For this, we need to choose a configuration in $\wGE$ and several elements
of $\LG$. To begin with, a configuration $\tilde g$ is given by assumption in Proposition \ref{prop:mini
gene}. Then, for each face $F\in\F$, let us choose a vertex $o(F)$ on the boundary of $F$ and call $\partial F$
the loop based at $o(F)$ going once around $F$ with positive orientation. If $v$ is a vertex on
$\partial F$ other than $o(F)$, denote by $\partial F_{o\to v}$ the portion of $\partial F$ going
from $o(F)$ to $v$. Let us decide that $\partial F_{o\to o}$ is the path $\partial F$ itself. For
each vertex $v$ on the boundary of $F$, set $\tilde x_F(v)=h^\wG_{\partial F_{o\to v}}(\tilde g)$. We
denote $\tilde x_F(o(F))=h^\wG_{\partial F}(\tilde g)$ simply by $\tilde x_F$. 

Choose an element $X_F \in \LG$ of minimal norm such that $\widetilde{\exp}(\sigma(F) X_F)=\tilde x_F
z_{F}$, where $\widetilde{\exp}:\LG\lra \wG$ is the exponential map. Observe that
$\tilde\rho(\tilde x_F z_{F})=\sigma(F) \| X_F \|$. Finally, for each $v$ on the boundary of $F$, set
$X_{F,v}= \Ad(\tilde x_{F}(v)) X_{F}$. By definition, the following compatibility condition is
satisfied for each edge $e$:
\begin{equation} \label{eqn:comp}
\forall e\in\E^+ ,\forall F\in\{L(e),L(e^{-1})\}, X_{F,\ov{e}}=\Ad(\tilde g_{e})X_{F,\un{e}}.
\end{equation}

Once these choices are made, Proposition \ref{prop:sommet} provides us with a collection of 1-forms
$(\omega_{v},v\in\V)$ and Proposition \ref{prop:arete} with a collection $(\omega_{e},e\in\E^+)$. We
prove now that it is possible to let a gauge transformation act on each form $\omega_{e}$ in such a
way that it coincides with $\omega_{\un{e}}$ and $\omega_{\ov{e}}$ on the
domains $B(\un{e},3R/4) \cap T_{e}$ and $B(\ov{e},3R/4)$ respectively. 

\begin{proposition} \label{prop:arete jauge}
Let $e\in \E^+$ be an edge. There exists $j_{e}\in W^{2,\infty}(T_{e};\wG)$ such that the following
properties hold.

1. The forms $j_{e}\cdot \omega_{e}$ and $\omega_{\un{e}}$ (resp. $\omega_{\ov{e}}$) coincide on
$B(\un{e},3R/4)\cap T_{e}$ (resp. $B(\ov{e},3R/4)\cap T_{e}$).

2. $j_{e}$ is smooth outside $e$ and identically equal to $1$ on $e$.
\end{proposition}

\pf Let $(x,y)\in (-3,3)\times (-1,1)$ be the local coordinates on $T_{e}$ given by Lemma
\ref{recouvrement}. The forms $\omega_{e}$ and $\omega_{\un{e}}$ are both defined on
$(-3,-1)\times (-1,1)$. We are going to apply Lemma \ref{lem:loops to paths} and Proposition
\ref{prop:local jauge} to these two forms on this domain.

First, let us choose on $M$ an auxiliary Riemannian metric for which $e$ is a geodesic. Let $l$ be a
loop based at $(-2,0)$ and contained in $(-3,-1)\times [0,1)$. Then, according to
Propositions \ref{prop:sommet} and \ref{prop:arete}, there exist two smooth 1-forms $\lambda$ and
$\lambda'$ such that $d\lambda=d\lambda'=\sigma$,  $\omega_{\un{e}}=X_{L(e),\un{e}}\lambda$
and $\omega_{e}=X_{L(e),\un{e}}\lambda'$ on $(-3,-1)\times [0,1)$. Since this domain is simply
connected, we conclude that $\langle \omega_{\un{e}},l\rangle_{\wG} =
\widetilde{\exp}(X_{L(e),\un{e}}\int_{l}\lambda)
= \widetilde{\exp}(X_{L(e),\un{e}}\int_{l}\lambda')=\langle \omega_{e},l\rangle_{\wG}$. 
Hence, Lemma \ref{lem:loops to paths} shows that there exists $j_{+}:(-3,-1)\times[0,1)\lra \wG$
which transforms the holonomy of $\omega$ into that of $\omega'$ on this domain.

 Replacing $L(e)$ by $L(e^{-1})$ and $(-3,-1)\times [0,1)$ by $(-3,-1)\times (-1,0]$, we find that
the equality $\langle \omega_{\un{e}},l\rangle_{\wG} =\langle \omega_{e},l\rangle_{\wG}$ holds also
if $l$ is contained in $(-3,-1)\times (-1,0]$. Lemma \ref{lem:loops to paths} gives similarly $j_{-}:(-3,-1)\times
[0,1) \lra \wG$. Moreover, $j_{+}(-2,0)=j_{-}(-2,0)=1$.

Finally, both $\omega$ and $\omega_{e}$ vanish in all directions on $e\cap U_{\un{e}}\cap T_{e}$, so
that if $c$ is a segment contained in $e$, both $c^*\omega_{e}$ and $c^*\omega_{\un{e}}$ are equal to
0. Hence, both $j_{+}$ and $j_{-}$ are identically equal to $1$ on $e$. They combine to give a
function $j_{\un{e}}:(-3,-1)\times (-1,1)\lra \wG$ which transforms the holonomy of $\omega$ into
that of $\omega'$, as long as one restricts oneself to paths which are finite concatenations of paths
which stay on either side of $e$. Since $e$ is geodesic for the auxiliary metric on $M$, piecewise
geodesic paths for this metric have this property. Hence, the assumptions of Proposition \ref{prop:local
jauge} are satisfied on $(-3,-1)\times (-1,1)$ by $\omega$, $\omega'$ and $j$, with $k=1$ and
$p=\infty$. Hence, $j_{\un{e}}$ belongs to $W^{2,\infty}(U;\wG)$. According to the remark made
immediately after Proposition \ref{prop:local jauge}, $j_{\un{e}}$ is smooth outside $e$. It is also
identically equal to $1$ on $e$ by construction.

At the vertex $\ov{e}$, the compatibility conditions $X_{L(e^{\pm 1}),\ov{e}}=\Ad(\tilde
g_{e})X_{L(e^{\pm 1}),\un{e}}$ and the same arguments as above imply that the forms $\omega_{e}$ and
$\omega_{\ov{e}}$ restricted to $U_{\ov{e}}\cap T_{e}$ also satisfy the assumptions of Lemma
\ref{lem:loops to paths}. Thus, we find in the same way $j_{\ov{e}}\in W^{2,\infty}(U_{\ov{e}}\cap
T_{e};\wG)$, smooth outside $e$, identically equal to 1 on $e$ and such that  $j_{\ov{e}}\cdot
\omega_{e}=\omega_{\ov{e}}$ on $U_{\ov{e}}\cap T_{e}$.
  
There remains to extend $j_{\un{e}}$ and $j_{\ov{e}}$ to an element of $W^{2,\infty}(T_{e};\wG)$. The
functions $j_{\un{e}}$ and $j_{\ov{e}}$, which are in particular Lipschitz, extend respectively to
continuous functions on $[-3,-1]\times [-1,1]$ and $[1,3]\times [-1,1]$. We start by interpolating
them by a continuous function $j_{0}:[-3,3]\times [-1,1]\lra \wG$, for example by setting, for $-1\leq x \leq 0$ and
$0\leq y\leq 1$, $j_{0}(x,y)=j_{\un{e}}(-1,(y-x-1)^+)$ and defining $j_{0}$ similarly on the three
other quarters of $[-1,1]^2$.  We still have $j_{0}(x,0)=1$ for every $x\in [-3,3]$.

Let us embed $\wG$ in the linear space ${\mathbb M}_{N}(\R)$ of $N\times N$ real matrices for some
$N\geq 1$. Let us endow this space of matrices with a Euclidean scalar product. As a smooth
Riemannian submanifold of ${\mathbb M}_{N}(\R)$, $G$ admits a tubular neighbourhood. In particular,
there exists $\epsilon>0$ such that, if $G^\epsilon$ denotes the set of matrices at a Euclidean
distance smaller than $\epsilon$ to $G$, then there exists a smooth mapping $pr_{G}:G^\epsilon \lra
G$ which is the identity when restricted to $G$. Since the range of $j_{0}$ is bounded, we can also
assume that $\epsilon$ is small enough to guarantee that any matrix closer than $\epsilon$ to the
range of $j_{0}$ is invertible. 

Take $\epsilon'>0$ and let $J:[-3,3]\times [-1,1]\lra {\mathbb M}_{N}(\R)$ be a smooth function such
that $\|J-j_{0}\|_{\infty} <\epsilon'$. Such a $J$ can be constructed for instance by smoothing
$j_{0}$ by convolution. Replacing $J$ by $J(x,y)J(x,0)^{-1}$, and provided $\epsilon'$ is small
enough, we can assume that $\|J-j_{0}\|_{\infty}<\epsilon$ and $J(x,0)=1$ for all $x\in[-3,3]$. 

Let now $\varphi:[-3,3]\lra [0,1]$ be a smooth function such that $\varphi(x)=0$ if $|x|\geq 2$ and
$\varphi(x)=1$ if $|x|\leq 1$. Set $j_{e}(x,y)=pr_{G}[(1-\varphi(x))j_{0}(x,y) + \varphi(x) J(x,y)]$.
The function defined in this way belongs to $W^{2,\infty}((-3,3)\times (-1,1))$, is smooth
outside the segment $x=0$ and equal to $1$ identically on this segment. Moreover, it coincides
respectively with $j_{\un{e}}$ and $j_{\ov{e}}$ on $(-3,-2)\times (-1,1)$ and $(2,3)\times
(-1,1)$. Hence, $j_{e}$ has the desired properties. \qed

Proposition \ref{prop:arete jauge} applied to each form $\omega_{e}$ produces a new collection
$(j_{e}\cdot \omega_{e},e\in\E^+)$ of forms which, together with $(\omega_{v},v\in\V)$, determine
a unique element $\omega_{\G}\in W^{1,\infty}\Omega^1_{\lg}(U_{\G})$. Let us summarize the
properties of $\omega_{\G}$.

\begin{proposition}\label{prop:graphe} There exists an element $\omega_{\G}\in
W^{1,\infty}\Omega^1_{\lg}(U_{\G})$ such that the following conditions hold.

1. For each edge $e\in \E^+$, $\langle \omega_{\G}, e \rangle_{\wG} = \tilde g_{e}$. Moreover,
$\omega_{\G}$ vanishes in all directions on $e$ inside $B(\un{e},3R/4)\cup B(\ov{e},3R/4)$.

2. For each face $F\in \F$, $\omega_{\G}$ is smooth on $U_{\G}\cap F$. 

3.  For each face $F$ and each edge $e$ bounding $F$, $\omega_{\G}$ is gauge-equivalent in $T_{e}
\cap F$ to $X_{F,v}\lambda$, where $v$ is any vertex on the boundary of $F$ and $\lambda$ is a
a smooth $1$-form such that $d\lambda=\sigma$.

4. For each face $F$ and each vertex $v$ on the boundary of $F$, $\omega_{\G}$ is equal to
$X_{F,v}\lambda$ on $B(v,3R/4)\cap F$, where $\lambda$ is a smooth $1$-form such that $d\lambda=\sigma$.
\end{proposition}

\subsection{The principal bundle}
\label{constr P}

In this paragraph, we construct the principal bundle $P$ on which the minimizing connection is
going to be defined. For this, we start by proving that there exists a family $(z_{e})_{e\in\E^+}$ of
elements of $\Pi$ indexed by $\E^+$ such that, for each face $F$, one has
\begin{equation}\label{eqn:face arete}
z_{F}=\prod_{e\in\E^+:L(e^{-1})=F}z_{e}. 
\end{equation}
This is a simple consequence of the property 1 of Lemma \ref{lem:orientation}. Indeed, the subsets
$\{e\in\E^+ : L(e^{-1})=F\}$ are non-empty and form, as $F$ spans $\F$, a partition of $\E^+$.
Hence, the mapping $\Pi^{\E^+}\lra \Pi^\F$ defined by $(z_{e})_{e\in\E^+}\mapsto
(\prod_{e\in \E^+:L(e^{-1})=F} z_{e})_{F\in\F}$ is onto. 

Let us choose $(z_{e})_{e\in\E^+}$ such that (\ref{eqn:face arete}) holds. Now, for each $z\in\Pi$,
let us choose a smooth curve $\zeta_{z}:(-3,3)\lra \wG$ such that $\zeta_{z}(t)=1$ if $t\leq
-1$ and $\zeta_{z}(t)=z$ if $t\geq 1$. Pick $e\in\E^+$. Consider as usual the coordinates
$(-3,3)\times (-1,1)$ on $T_{e}$ given by Lemma \ref{recouvrement}. Define $\psi_{e}:U_{e}\lra G$ by
setting $\psi_{e}(m)=1$ if $m\in B(\un{e},3R/4)\cup B(\ov{e},3R/4)$ and
$\psi_{e}(m)=\pi(\zeta_{z_{e}}(x))$ if $m\in T_{e}$ and $m=(x,y)$. Finally, extend $\psi_{e}$ on
$U_{\G}$ by setting $\psi_{e}(m)=1$ if $m\notin U_{e}$. Observe that, if $e\neq e'$, then
the subsets $\{m\in U_{\G} : \psi_{e}(m)\neq 1\}$ and $\{m\in U_{\G} : \psi_{e'}(m)\neq 1\}$ are
disjoint, so that $\psi_{e}\psi_{e'}=\psi_{e'}\psi_{e}$ everywhere on $U_{\G}$. For each face $F$,
define $\psi_{F}:U_{\partial F}\lra G$ by setting 
$$ \psi_{F}=\prod_{e\in\E^+:L(e^{-1})=F} \psi_{e}.$$
Finally, if $F$ and $F'$  are two faces which share at least one common vertex, then define $
\psi_{FF'}:U_{\partial F}\cap U_{\partial F'}\lra G$ by $\psi_{FF'}=\psi_{F}^{-1}\psi_{F'}$. 
Since for every pair $(F,F')$ of faces, $U_{\partial F}\cap U_{\partial F'}=U_{F}\cap U_{F'}$,
the collection $(\psi_{FF'})_{F,F'\in\F}$ is a $G$-valued \v{C}ech 1-cochain on $M$, actually a
1-cocycle.  

At first sight, it may seem that this cocycle is actually a coboundary. In fact, the equality
$\psi_{FF'}=\psi_{F}^{-1}\psi_{F'}$ is misleading. It holds on $U_{F}\cap U_{F'}$ which happens to
be equal to $U_{\partial F}\cap U_{\partial F'}$ and the point is that, in general, neither
$\psi_{F}$ nor $\psi_{F'}$ can be extended to smooth or continuous $G$-valued functions on $U_{F}$
or $U_{F'}$.

\begin{definition} Let $P$ be the principal $G$-bundle over $M$ defined by the covering
$(U_{F})_{F\in \F}$ of $M$ and the transition functions $(\psi_{FF'})_{F,F'\in\F}$.
\end{definition}

The bundle $P$ can be described as follows. Consider the disjoint union $\bigsqcup_{F\in\F}
(U_{F}\times G)$. An element of this union is denoted by $(m,g)_{F}$. Declare $(m,g)_{F}$ and
$(m',g')_{F'}$ to be equivalent if $m=m'$ and $g=\psi_{FF'}(m)g'$. Let $\sim$ denote this
equivalence relation. Then $P$ is the manifold $\bigsqcup_{F\in\F} (U_{F}\times G)/\sim$ on which
$G$ acts by right multiplication on the second factor.

In fact, $P$ constructed in this way is endowed with a family of local sections. Indeed for each
face $F$, there is a smooth section $s_{F}$ of $P$ over $U_{F}$ which sends each point $m$ to the
class of $(m,1)_{F}$. If two faces $F$ and $F'$ share at least
one common vertex, then $s_{F}$ and $s_{F'}$ are related on $U_{F}\cap U_{F'}$ by
$s_{F'}=s_{F}\psi_{FF'}$. In particular, if $v$ is a vertex of $\G$ and if $v\in U_{F}\cap
U_{F'}$, then $s_{F}(v)=s_{F'}(v)$.

We turn now to the construction of the connection on $P$.

\subsection{Inside the faces}

Let $\delta>0$ be such that, for each face $F$, the open subset $U^\delta_{\partial F}=\{m \in M :
d(m,\partial F) < \delta \}$ of $M$ is contained in $U_{\partial F}$. Such a $\delta$ exists because
the boundaries of the faces of $\G$ are compact subsets of $M$. Define, for each 
face $F$, $U^\delta_{F}=F\cup U^\delta_{\partial F}$. Define also $U^\delta_{\G}=\cup_{F\in
\F} U^\delta_{\partial F}$. This open subset of $M$ is contained in $U_{\G}=\cup_{F\in\F} U_{\partial
F}$.

The domains of the local sections $(s_{F},U^\delta_{F})$ cover $M$. Hence, according to the remark
\ref{rmk:def connection covering}, a $W^{1,\infty}$ connection on $P$ is specified by the data of a
collection $(\eta_{F})_{F\in\F}$, where for each $F$, $\eta_{F}\in
W^{1,\infty}\Omega^1_{\lg}(U^\delta_{F})$ and, on $U^\delta_{F}\cap U^\delta_{F'}$,
$\eta_{F'}=\psi_{FF'}\cdot \eta_{F}$. In order to construct such a family, let us start by
restricting the form $\omega_{\G}$ given by Proposition \ref{prop:graphe} to each one of the open
subsets $U_{\partial F}$. In this way, we get a collection
$(\omega^0_{\partial F})_{F\in\F}$ of $1$-forms. For each face $F$, set
$$\omega_{\partial F}=\psi_{F}\cdot \omega^0_{\partial F}.$$
Observe that, if $F$ and $F'$ are two faces, then, on $U_{F}\cap U_{F'}$, one has
$\omega_{\partial F'}=\psi_{FF'}\cdot \omega_{\partial F}$. Hence, these locally defined $1$-forms
almost define a connection on $P$. There remains only to extend each $\omega_{\partial F}$ to a
$1$-form defined on $U_{F}$. 

For each face $F$, let $\lambda_{F}\in \Omega^1(F)$ be such that $d\lambda_{F}=\sigma$. Consider
the element $X_{F}=X_{o(F),F}$ of $\LG$. We apply a gauge transformation to the form
$X_{F}\lambda_{F}$ in order to extend $\omega_{\partial F}$ inside $F$. However, as in Proposition
\ref{prop:arete jauge}, we get two forms which do not coincide on $F\cap U_{\partial F}$, but on a
smaller domain, that we arrange to contain $F\cap U^\delta_{\partial F}$.

\begin{proposition} \label{prop:fill face} Let $F$ be a face of $\G$. There exists a smooth $G$-valued
function $j_{F}:F\lra G$ such that $j_{F}\cdot (X_{F}\lambda_{F})=\omega_{\partial F}$ on $F\cap
U^\delta_{\partial F}$.
\end{proposition}

We begin by proving the following result.

\begin{lemma} Let $F$ be a face of $\G$. Let $m$ be a point of $F\cap B(o(F),3R/4)$. Let $l$ be a
smooth loop based at $m$ contained in $F\cap U_{\partial F}$. Then $\langle \omega_{\partial F}, l \rangle_{\wG}=
\widetilde{\exp}(X_{F}\int_{l}\lambda_{F})$.
\end{lemma}

\pf Let $*\Omega_{\partial F}$ be the unique element of $C^\infty(F\cap U_{\partial F})$ such that
the curvature of $\omega_{\partial F}$ is equal to $*\Omega_{\partial F} \sigma$ on this domain.

We begin by treating the case where $l$ is homotopic to a constant loop. Let $L:[0,1]^2\lra F\cap
U_{\partial F}$ be a smooth homotopy $(s,t)\mapsto L(s,t)=l_{s}(t)$ such that for each $s\in
[0,1]$, $l_s$ is a smooth loop based at $m$, $l_0$ is the constant loop and $l_1$ is just $l$. 
Let us pull $\omega_{\partial F}$ back by this smooth homotopy and work on $[0,1]^2$. For each $s\in
[0,1]$, set $h_s=\langle \omega_{\partial F}, l_{s}\rangle_{\wG}=\langle L^*\omega_{\partial F},
 \{s\}\times [0,1] \rangle_{\wG}$. Then, according to \cite{Gross_P}, Theorem 2.2, or Proposition 
\ref{variation}, the mapping $s\mapsto h_{s}$ satisfies the following differential equation:
\begin{equation}\label{eqn:Gross}
h_s^{-1} \frac{\partial h_s }{\partial s}= \int_0^1 {\langle \omega_{\partial F},
l_{s}([0,t])\rangle_{\wG}}^{-1} *\Omega_{\partial F}(l_{s}(t)) \langle\omega_{\partial F}, l_{s}([0,t])
\rangle_{\wG} \Phi^*\sigma_{(s,t)}(\partial_s, \partial_t) \; dt.
\end{equation}
We claim that ${\langle \omega_{\partial F}, l_{s}([0,t])\rangle_{\wG}}^{-1} *\Omega_{\partial
F}(l_{s}(t)) \langle\omega_{\partial F}, l_{s}([0,t]) \rangle_{\wG}$ is identically equal to
$*\Omega_{\partial F}(m)$. More generally, we claim that, whenever a path $c$ in $F\cap U_{\partial
F}$ starts at $m$ and finishes at some point $n$, then 
$$
*\Omega_{\partial F}(m)={\langle\omega_{\partial F} , c \rangle_{\wG}}^{-1} *\Omega_{\partial F}(n)
\langle\omega_{\partial F} , c \rangle_{\wG}.
$$
Indeed, this relation is true if $\omega_{\partial F}$ is replaced by a connection of the form
$X\lambda$, where $X\in \LG$ and $\lambda$ satisfies $d\lambda=\sigma$. Moreover, this relation is
gauge-invariant: if it is true for some connection, it is also satisfied by the image of this
connection by any gauge transformation. Finally, the relation is multiplicative: if it holds for
two paths which one can concatenate, then it holds for their concatenation. Now the result follows
from the fact that any path in $F\cap U_{\partial F}$ can be written as a concatenation of
finitely many shorter paths, each of which is contained in a domain where $\omega_{\partial F}$ is
gauge-equivalent to a connection of the form $X\lambda$. 

Now, (\ref{eqn:Gross}) implies that
$$h_1 = \exp \left( *\Omega_{\partial F}(m) \int_{0}^1 \left\{ \int_{0}^1
L^*\sigma_{(s,t)}(\partial_s,\partial_t) \; dt \right\}\; ds\right).$$
On the other hand,
$$ \frac{\partial}{\partial s} \int_{l_s} \lambda_F =
\int_0^1 \partial_s \left[(L^*\lambda_F)_{(s,t)}(\partial_t)\right] \; dt 
=  \int_0^1 d(L^*\lambda_F)_{(s,t)}(\partial_s, \partial_t) \; dt
= \int_0^1 L^*\sigma_{(s,t)}(\partial_s, \partial_t)\; dt,$$
because $d_{(s,0)}L(\partial_s)=d_{(s,1)}L(\partial_s)=0$ for all $s\in [0,1]$.
Finally, since $*\Omega_{\partial F}(m)=X_{F}$, we get $\langle \omega_{\partial F},l\rangle_{\wG} =
\widetilde{\exp}(X_{F} \int_{l}\lambda_{F})$.

Let us drop the assumption that $l$ is homotopic to a constant loop. Let $\gamma$ be the geodesic
segment from $m$ to $o(F)$. Set $w=\gamma \partial F \gamma^{-1}$. Then $w$ generates the
fundamental group $\pi_{1}(U_{\partial F},m)$, which is isomorphic to $\mathbb{Z}$. Let $r$ be the
unique integer such that $lw^{-r}$ is homotopic to a constant loop. We cannot apply the discussion
above to $lw^{-r}$ because it is not contained in $F\cap U_{\partial F}$. So, let $(w_{n})_{n\geq 0}$
be a sequence of simple loops based at $m$ such that $w_{n}$ converges to $w$ and, for each $n\geq
0$, $lw_{n}^{-r}$ is contained in $F\cap U_{\partial F}$ and homotopic to a constant loop. Then
$\langle \omega_{\partial F}, l \rangle_{\wG} = \langle\omega_{\partial F},w_{n} {\rangle_{\wG}}^r
\langle \omega_{\partial F} , lw_{n}^{-r}\rangle_{\wG}$. 

On one hand, $\langle\omega_{\partial F},w_{n}\rangle_{\wG}$ converges to $\langle \omega_{\partial
F},w\rangle_{\wG}$ because $w_{n}$ converges to $w$ with fixed endpoints. The holonomy of
$\omega_{\partial F}$ along $\partial F$ is equal to $h^\wG_{\partial F}(\tilde
g)z_{F}=\widetilde{\exp}(\sigma(F)X_{F})$ and its holonomy along $\gamma$ commutes to $X_{F}$
because, on $B(o(F),3R/4)$, $\omega_{\partial F}$ takes its values in $\R X_{F}$. So, $\langle
\omega_{\partial F},w\rangle_{\wG}=\widetilde{\exp}(\sigma(F)X_{F})$.

On the other hand, by the discussion of the homotopically trivial case, $\langle \omega_{\partial
F}, lw_{n}^{-r}\rangle_{\wG}=\widetilde{\exp}(X_{F}\int_{l}\lambda_{F})\widetilde{\exp}(-r
X_{F}\int_{w_{n}}\lambda_{F})$. Since $d\lambda_{F}=\sigma$, $\int_{w_{n}}\lambda_{F}$ is equal to
the area enclosed by $w_{n}$. Since $w_{n}$ converges uniformly to $w=\gamma  \partial F
\gamma^{-1}$, this area tends to $\sigma(F)$. The result follows. \qed

\ppf{prop:fill face} Let $F$ be a face of $\G$. Both forms $\omega_{\partial F}$ and
$X_{F}\lambda_{F}$ are smooth on $F\cap U_{\partial F}$. We have just proved that they have the same
holonomy in $\wG$ along all smooth loops
based at $m$. Hence, by Lemma \ref{lem:loops to paths} and Proposition \ref{prop:local jauge}, there
exists a smooth mapping $j^0_{F}:F\cap U_{\partial F}\lra \wG$ such that $j^0_{F}\cdot
(X_{F}\lambda_{F})$ coincides with $\omega_{\partial F}$ on $F\cap U_{\partial F}$. 
Consider a diffeomorphism between $F$ and the unit disk in $\R^2$ with polar coordinates
$(r,\theta)$ such that $F\cap U^\delta_{\G}\subset\{r>\frac{3}{4}\}\subset \{r>\frac{1}{2}\}
\subset F\cap U_{\partial F}$. Consider the restriction of $j^0_{F}$ to $\{r>\frac{1}{2}\}$. 
In order to extend $j^0_{F}$, we start by embedding $\wG$ in a vector space of matrices. The possible
non compactness of $\wG$ is, as always in this paper, not a problem because $\wG$ is the direct
product of a compact group and a group isomorphic to $(\R^m,+)$ for some $m\geq 0$. We may even
assume that, for some $\epsilon>0$, and with the same notation as in the proof of Proposition
\ref{prop:arete jauge}, there is a smooth mapping $pr_{\wG}:\wG^\epsilon\lra \wG$ which restricts to
the identity on $\wG$.  
Now, since $\wG$ is simply connected, we can extend $j^0_{F}$ by continuity on $D(0,1)$. Since
$j_{0}$ is uniformly continuous on $D(0,\frac{7}{8})$, we can also approximate it by a smooth
matrix-valued function $J$ defined on $D(0,\frac{7}{8})$ and such that
$\|J-j_{0}\|_{\infty}<\epsilon$. Finally, let $\varphi:[0,1]\lra [0,1]$ be a smooth function such
that $\varphi(r)=1$ if $r\leq \frac{1}{2}$ and $\varphi(r)=0$ if $r\geq\frac{3}{4}$. We define
$j_{F}:D(0,1)\lra G$ in polar coordinates by
$j_{F}(r,\theta)=\pi[pr_{\wG}((1-\varphi(r))j_{0}(r,\theta) + \varphi(r) J(r,\theta))]$. 
The forms $j_{F}\cdot (X_{F}\lambda_{F})$ and $\omega_{\partial F}$ coincide on the domain of
equation $\{r>\frac{3}{4}\}$, which contains $F\cap U^\delta_{\G}$.   \qed

We can finish the proof of the existence of a minimizing connection.\\

\ppf{prop:mini gene} Let $P$ be the principal $G$-bundle over $M$ constructed in Section
\ref{constr P}. It satisfies  $\oo(P)=\prod_{e\in\E^+}z_{e}=\prod_{F\in\F}z_{F}=z$ (see Appendix A of
\cite{Lawson_Michelsohn}). 

For each face $F$, let us call $\omega_{F}$ the element of $W^{1,\infty} 
\Omega^1_{\lg}(U^\delta_{F})$ which is equal to $j_{F}\cdot (X_{F}\lambda_{F})$ on $F$ and to
$\omega_{\partial F}$ on $U^\delta_{\partial F}$. If $F$ and $F'$ are two faces which share at least
a common vertex, then, on $U^\delta_{F}\cap U^\delta_{F'}=U^\delta_{\partial F}\cap
U^\delta_{\partial F'}$, we have $\omega_{F'}=\omega_{\partial F'}=\psi_{FF'}\cdot \omega_{\partial
F}=\psi_{FF'}\cdot \omega_{F'}$.  Hence, the forms $(\omega_{F})_{F\in \F}$ determine a connection on
$P$. Let $\omega$ denote this connection. By construction, $\omega$ belongs to $W^{1,\infty}\A(P)$
and it is smooth outside $\cup_{e\in \E}e$. 

Let $v$ be a vertex of $\G$. We have observed at the end of Section \ref{constr P} that the local
sections $s_{F}$ determine without ambiguity a point $s_{F}(v)$, which we denote by $p_{v}\in
P_{v}$. Let us compute the holonomy of $\omega$ along the edges of $\G$ with respect to the
reference points $p_{v}$. 

Let $(\tau_{c}, c\in PM)$ be the holonomy induced by $\omega$. Let $e$ be an edge of $\E^+$. By
definition of $\omega$, we have for each face $F$ the equality $\omega_{s_{F}}=\omega_{F}$. In
particular, $\omega_{s_{L(e)}}=\omega_{L(e)}$. Hence, $\tau_{e}(p_{\un{e}})=p_{\ov{e}}\langle
\omega_{L(e)},e\rangle$. Since $\psi_{L(e)}$ is identically equal to $1$ in a neighbourhood of
$e$, $\langle \omega_{L(e)},e \rangle=\langle \omega_{\G},e\rangle$. By Proposition
\ref{prop:graphe}, this is equal to $g_{e}$. Since this holds for every $e\in \E^+$, $\omega$
agrees with $(g_{e})_{e\in\E^+}$ on $\E$.

Finally, let $F$ be a face of $\G$. Inside this face, $\omega_{F}$ is gauge-equivalent to
$X_{F}\lambda_{F}$, so that its curvature is equal to $X_{F}\sigma$. Hence,
$S_{F}(\omega)=\|X_{F}\|^2 \sigma(F)$. The assertion on $S_{F}(\omega)$ follows now from the identity
$\sigma(F)\|X_{F}\|=\tilde \rho(h^\wG_{\partial F}(\tilde g)z_{F})$. \qed

\section*{Appendix: A proof of the energy inequality}
\label{smooth energy}
\renewcommand{\theequation}{A-\arabic{equation}}
\renewcommand{\thetheorem}{A-\arabic{theorem}}
\setcounter{theorem}{0}

In this appendix, we give a proof of the energy inequality (Proposition \ref{ineg fine}), which
consists basically in applying Cauchy-Schwarz inequality in the right context. The proof of a more
general instance of this result can be found in \cite{Sengupta_IE}. However, we are not aware of a
such a compact proof as the one we give here.

Let $D$ be the closed unit disk in $\R^2$ endowed with a smooth volume form $\sigma$. Pick $\omega$
in $\Omega^1_{\lg}(D)$. Let $\Omega \in \Omega^2_{\lg}(D)$ be its curvature and $*\Omega$ the
unique smooth $\LG$-valued function on $D$ such that $\Omega=*\Omega \sigma$.

For all $s,t\in [0,1]$, set 
$$\gamma(s,t)=(1-s+s\cos(2\pi t), s\sin(2\pi t)) \in \R^2. $$
For every $s\in [0,1]$, $t\mapsto \gamma(s,t)$ is a smooth loop based at the point $(1,0)$. The
mapping $\gamma$ realizes a smooth homotopy between the constant loop at $(1,0)$ and the boundary of
$D$ starting at $(1,0)$ with the usual orientation. 

Set $h(s,t)=\langle \omega,\gamma(s,\cdot) \rangle$. By standard results on ordinary differential
equations, $h$ is a smooth mapping from $[0,1]^2$ to $G$. Let us denote by $h_{s}$ and $h_{t}$ the
partial derivatives of $h$ with respect to $s$ and $t$. Similarly, let $\gamma_{s}$ and $\gamma_{t}$
denote the partial derivatives of $\gamma$ with respect to $s$ and $t$.

\begin{proposition} \label{variation}
For all $s\in [0,1]$,
$$h(s,1)^{-1} h_{s}(s,1) =\int_0^1 h(s,t)^{-1} *\Omega(\gamma_{t}(s,t), \gamma_{s}(s,t)) h(s,t)\;
dt.$$
\end{proposition}

\pf For every $s\in [0,1]$, consider the vector field $X_s$ defined on $[0,1]\times G$ by
$$X_s(u,g)=(1,-\omega(\gamma_{t}(s,u))g).$$
Let $(\Phi^t_s)_{t\geq 0}$ denote the flow of $X_s$. The group $G$ acts on $[0,1]\times G$ by
multiplication on the right on the second factor. Since $X_s$ is $G$-invariant, its flow is also
invariant. By definition of $h(s,t)$, one has
\begin{equation} \label{flow}
\Phi^t_s(u,g)=(u+t,h(s,u+t) h(s,u)^{-1} g).
\end{equation}
In particular, $\Phi^1_s(0,1)=(1,h(s,1))$. Since $X_s$ depends smoothly on $s$, a classical result
(see \cite{Duistermaat}, Thm. B.3 for example) asserts that
$$\frac{\partial}{\partial s} \Phi^1_s(0,1)=\int_0^1 d_{\Phi^t_s(0,1)} \Phi^{1-t}_s \left[
\frac{\partial X_s}{\partial s} (\Phi^t_s(0,1)) \right] \; dt.$$ 
The term inside the brackets is equal to $(0,-\frac{\partial}{\partial s} \left[
\omega(\gamma_{t}(s,t)) \right] h(s,t)).$ By differentiating (\ref{flow}), one gets
\begin{eqnarray*}
&& \hskip -2cm 
d_{\Phi^t_s(0,1)} \Phi^{1-t}_s \left[(0,-\frac{\partial}{\partial s} \left[ \omega(
\gamma_{t}(s,t)) \right] h(s,t)) \right] \; = \\ 
&& \hskip 1cm -(0,h(s,1)h(s,t)^{-1} \frac{\partial}{\partial s} \left[ \omega (\gamma_{t}(s,t))
\right] h(s,t)).
\end{eqnarray*}
Hence, 
\begin{equation} \label{1}
h(s,1)^{-1}h_{s}(s,1)= -\int_0^1 h(s,t)^{-1} \frac{\partial}{\partial s} \left[ \omega (
\gamma_{t}(s,t)) \right] h(s,t) \; dt.  
\end{equation} 
Since the vector fields $\partial_t \gamma$ and $\partial_s \gamma$ commute, one has at every point 
\begin{equation} \label{2}
-\frac{\partial}{\partial s} [\omega(\gamma_{t})]=d\omega(\gamma_{t},\gamma_{s})
- \frac{\partial}{\partial t} [\omega(\gamma_{s})].  
\end{equation}
Let us compute $-\int_0^1 h(s,t)^{-1} \frac{\partial}{\partial t} [\omega(\gamma_{s})] h(s,t)
\; dt$ by integration by parts. The boundary terms vanish, because the loops $t\mapsto \gamma(s,t)$
share the same basepoint. There remains
\begin{equation} \label{3}
-\int_0^1 h(s,t)^{-1} \frac{\partial}{\partial t} [\omega(\gamma_{s})] h(s,t) \; dt = \int_0^1
h(s,t)^{-1} [\omega(\gamma_{t}),\omega(\gamma_{s})] h(s,t) \; dt,
\end{equation}
where the last bracket is the Lie bracket of $\LG$. Combining (\ref{1}), (\ref{2}) and (\ref{3}), we
find what we want, that is,
$$h(s,1)^{-1}h_{s}(s,1)= \int_0^1 h(s,t)^{-1} *\Omega(\gamma_{t}(s,t),(\gamma_{s}(s,t)) h(s,t)
\; dt.$$
\qed

\begin{corollary}[Energy inequality] \label{true ei}
The following inequality holds:
$$\rho(\langle \omega,\partial D\rangle)^2 \leq S_D(\omega) \sigma(D).$$
\end{corollary}

\pf Since conjugation preserves the norm in $\LG$,
$$\| h_{s}(s,1) \| \leq \int_0^1 \|  \Omega(\gamma_{t}(s,t),\gamma_{s}(s,t)) \| \; dt .$$
By Cauchy-Schwarz inequality,
\begin{eqnarray*} 
\left| \int_0^1 \| \Omega(\gamma_{t}(s,t),\gamma_{s}(s,t)) \| \; dt \right|^2 &=&  \left| \int_0^1
\| *\Omega(\gamma(s,t)) \| |\sigma(\gamma_{t}(s,t),\gamma_{s}(s,t))| \; dt \right|^2 \\  
&& \hskip -3cm \leq \int_0^1 \| * \Omega(\gamma(s,t)) \|^2 |\sigma(\gamma_{t}(s,t),\gamma_{s}(s,t))|
\; dt \int_0^1 |\sigma(\gamma_{t}(s,t),\gamma_{s}(s,t))| \; dt.
\end{eqnarray*}

Set $A(r)=\int_0^r ds \int_0^1 |\sigma(\gamma_{t}(s,t),\gamma_{s}(s,t))| \; dt $.
It is the area enclosed by the path $\gamma(r,\cdot)$. The function $A$ is a
diffeomorphism of $[0,1]$ onto $[0,\sigma(D)]$. Reparametrize $s\mapsto h(s,1)$ by setting
$$k(u)=h(A^{-1}(\sigma(D) u), 1).$$
The path $u \mapsto k(u)$ is a path in $G$ from $1$ to $h(1,1)$. Hence,
$$\rho(h(1,1))^2 \leq \ell(k)^2 = \left| \int_0^1 \| \dot k(u)
\| \; du \right|^2 \leq \int_0^1 \| \dot k(u) \|^2 \; du.$$

The last term is the energy of the path $k$, which can be estimated as follows.

\begin{eqnarray*}
\int_0^1 \| \dot k(u) \|^2 \; du &=& \int_0^1 \| h_{s}(A^{-1}(\sigma(D) u),1) \|^2
\frac{\sigma(D)^2}{A'(A^{-1}(\sigma(D)u))^2} \; du \\
&=& \int_0^1 \|h_{s}(s,1) \|^2 \frac{\sigma(D)}{A'(s)} \; ds \\
&& \hskip -3cm \leq  \; \sigma(D) \int_0^1 \left(\int_0^1 \| *\Omega(\gamma(s,t)) \|^2
|\sigma(\gamma_{t}(s,t),\gamma_{s}(s,t))| \; dt \right) \; ds \\
&&\hskip -3cm =\; \sigma(D) \int_D \| *\Omega \|^2 \sigma \\ 
&&\hskip -3cm =\; \sigma(D) S_D(\omega).
\end{eqnarray*}

This implies the result because $h(1,1)=\langle \omega, \partial D \rangle$. \qed

\bibliographystyle{plain}
\bibliography{levy-norris-rv}

\begin{thebibliography}{10}

\bibitem{Adams}
Robert~A. Adams.
\newblock {\em Sobolev spaces}.
\newblock Academic Press, New York-London, 1975.
\newblock Pure and Applied Mathematics, Vol. 65.

\bibitem{DZ}
Amir Dembo and Ofer Zeitouni.
\newblock {\em Large deviations techniques and applications}.
\newblock Jones and Bartlett Publishers, Boston, MA, 1993.

\bibitem{Driver}
Bruce~K. Driver.
\newblock Two-dimensional {E}uclidean quantized {Y}ang-{M}ills fields.
\newblock In {\em Probability models in mathematical physics (Colorado Springs,
  CO, 1990)}, pages 21--36. World Sci. Publishing, Teaneck, NJ, 1991.

\bibitem{Duistermaat}
J.~J. Duistermaat and J.~A.~C. Kolk.
\newblock {\em Lie groups}.
\newblock Universitext. Springer-Verlag, Berlin, 2000.

\bibitem{Gray}
Alfred Gray.
\newblock {\em Tubes}.
\newblock Addison-Wesley Publishing Company Advanced Book Program, Redwood
  City, CA, 1990.

\bibitem{Gross_P}
Leonard Gross.
\newblock A {P}oincar\'e lemma for connection forms.
\newblock {\em J. Funct. Anal.}, 63(1):1--46, 1985.

\bibitem{Kobayashi_Nomizu_I}
Shoshichi Kobayashi and Katsumi Nomizu.
\newblock {\em Foundations of differential geometry. {V}ol. {I}}.
\newblock John Wiley \& Sons Inc., New York, 1996.
\newblock Reprint of the 1963 original.

\bibitem{Lawson_Michelsohn}
H.~Blaine Lawson, Jr. and Marie-Louise Michelsohn.
\newblock {\em Spin geometry}, volume~38 of {\em Princeton Mathematical
  Series}.
\newblock Princeton University Press, Princeton, NJ, 1989.

\bibitem{Levy_AMS}
Thierry L{\'e}vy.
\newblock Yang-{M}ills measure on compact surfaces.
\newblock {\em Mem. Amer. Math. Soc.}, 166(790):xiv+122, 2003.

\bibitem{Levy_PTRF}
Thierry L{\'e}vy.
\newblock Discrete and continuous {Y}ang-{M}ills measure for non-trivial
  bundles over compact surfaces.
\newblock {\em Preprint}, math-ph/0501014, 2005.

\bibitem{Li}
Peter Li and Shing-Tung Yau.
\newblock On the parabolic kernel of the {S}chr\"odinger operator.
\newblock {\em Acta Math.}, 156(3-4):153--201, 1986.

\bibitem{Migdal}
A.~A. Migdal.
\newblock Recursion equations in gauge field theories.
\newblock {\em Sov. Phys. JETP}, 42(3):413--418, 1975.

\bibitem{Morita}
Shigeyuki Morita.
\newblock {\em Geometry of differential forms}, volume 201 of {\em Translations
  of Mathematical Monographs}.
\newblock American Mathematical Society, Providence, RI, 2001.

\bibitem{Norris}
James~R. Norris.
\newblock Heat kernel asymptotics and the distance function in {L}ipschitz
  {R}iemannian manifolds.
\newblock {\em Acta Math.}, 179(1):79--103, 1997.

\bibitem{Sengupta_AMS}
Ambar Sengupta.
\newblock Gauge theory on compact surfaces.
\newblock {\em Mem. Amer. Math. Soc.}, 126(600):viii+85, 1997.

\bibitem{Sengupta_IE}
Ambar Sengupta.
\newblock A {Y}ang-{M}ills inequality for compact surfaces.
\newblock {\em Infin. Dimens. Anal. Quantum Probab. Relat. Top.}, 1(1):1--16,
  1998.

\bibitem{Steenrod}
Norman Steenrod.
\newblock {\em The topology of fibre bundles}.
\newblock Princeton Landmarks in Mathematics. Princeton University Press,
  Princeton, NJ, 1999.

\bibitem{Uhlenbeck}
Karen~K. Uhlenbeck.
\newblock Connections with {$L\sp{p}$} bounds on curvature.
\newblock {\em Comm. Math. Phys.}, 83(1):31--42, 1982.

\bibitem{Varadhan1}
S.~R.~S. Varadhan.
\newblock Diffusion processes in a small time interval.
\newblock {\em Comm. Pure Appl. Math.}, 20:659--685, 1967.

\bibitem{Varadhan2}
S.~R.~S. Varadhan.
\newblock On the behavior of the fundamental solution of the heat equation with
  variable coefficients.
\newblock {\em Comm. Pure Appl. Math.}, 20:431--455, 1967.

\bibitem{Varopoulos}
N.~Th. Varopoulos, L.~Saloff-Coste, and T.~Coulhon.
\newblock {\em Analysis and geometry on groups}.
\newblock Cambridge University Press, Cambridge, 1992.

\bibitem{Wehrheim}
Katrin Wehrheim.
\newblock {\em Uhlenbeck compactness}.
\newblock EMS Series of Lectures in Mathematics. European Mathematical Society
  (EMS), Z\"urich, 2004.

\bibitem{Witten}
Edward Witten.
\newblock On quantum gauge theories in two dimensions.
\newblock {\em Comm. Math. Phys.}, 141(1):153--209, 1991.

\end{thebibliography}
\end{document}